# Type Ia Supernova Progenitor Properties and Their Host Galaxies

Sudeshna Chakraborty,[1] Benjamin Sadler,[2] Peter Hoeflich,[1] Eric Y. Hsiao,[1] M. M. Phillips,[3] C. R. Burns,[4] T. Diamond,[1] I. Dominguez,[5] L. Galbany,[6,7] S. A. Uddin,[8] C. Ashall,[9] K. Krisciunas,[10] S. Kumar,[1] T. B. Mera,[1] N. Morrell,[3] E. Baron,[11,12,13] C. Contreras,[3] M. D. Stritzinger,[14] and N. B. Suntzeff[15]

[1]Department of Physics, Florida State University, 77 Chieftan Way, Tallahassee, FL, 32306, USA
[2]Western Governor's University, Department of Physics, Salt Lake City, UT 84107, USA
[3]Las Campanas Observatory, Carnegie Observatories, Casilla 601, La Serena, Chile
[4]Observatories of the Carnegie Institute for Science, 813 Santa Barbara Street, Pasadena, CA 91101, USA
[5]Universidad de Granada 18071, Granada, Spain
[6]Institute of Space Sciences (ICE-CSIC), Campus UAB, Carrer de Can Magrans, s/n, E-08193 Barcelona, Spain
[7]Institut d'Estudis Espacials de Catalunya (IEEC), E-08034 Barcelona, Spain
[8]Texas Agricultural and Mechanical University, College Station, Texas, USA
[9]Virginia Polytechnic Institute and State University, Blacksburg, Virginia, USA
[10]George P. and Cynthia Woods Mitchell Institute for Fundamental Physics and Astronomy, Texas A&M University, Department of Physics and Astronomy, College Station, TX 77843, USA
[11]Planetary Science Institute, 1700 East Fort Lowell Road, Suite 106, Tucson, AZ 85719-2395, USA
[12]Hamburger Sternwarte, Gojenbergsweg 112, D-21029 Hamburg, Germany
[13]Dept of Physics & Astronomy, University of Oklahoma, 440 W. Brooks, Rm 100, Norman, OK USA
[14]Department of Physics and Astronomy, Aarhus University, Ny Munkegade, DK-8000 Aarhus C, Denmark
[15]P. and C. Woods Mitchell Institute for Fundamental Physics and Astronomy, Department of Physics and Astronomy, Texas A&M University, College Station, TX 77843, USA

## ABSTRACT

We present an eigenfunction method to analyze 161 visual light curves (LCs) of Type Ia supernovae (SNe Ia) obtained by the Carnegie Supernova Project to characterize their diversity and host-galaxy correlations. The eigenfunctions are based on the delayed-detonation scenario **(DD)** using three parameters: the LC stretch $s$ being determined by the amount of deflagration-burning governing the $^{56}$Ni production, the main-sequence mass $M_{MS}$ of the progenitor white dwarf (WD) controlling the explosion energy, and its central density $\rho_c$ shifting the $^{56}$Ni distribution. Our analysis tool (SPAT) extracts the parameters from observations and projects them into physical space using their allowed ranges ($M_{MS} \leq 8\ M_\odot,\ \rho_c \leq 7-8\ \times 10^9\ gcm^{-3}$). The residuals between fits and individual LC-points are $\approx 1-3\%$ for $\approx 92\%$ of objects. We find two distinct $M_{MS}$ groups corresponding to a fast ($\approx 40-65\ Myrs$) and a slow($\approx 200-500\ Myrs$) stellar evolution. Most underluminous SNe Ia have hosts with low star formation but high $M_{MS}$, suggesting slow evolution times of the progenitor system. 91T-like SNe show very similar LCs and high $M_{MS}$ and are correlated to star formation regions, making them potentially important tracers of star formation in the early Universe out to $z \approx 4-11$. Some $\sim 6\%$ outliers with 'non-physical' parameters **using DD-scenarios** can be attributed to super-luminous SNe Ia and subluminous SNe Ia with hosts of active star formation. For deciphering the SNe Ia diversity and high-precision SNe Ia cosmology, the importance is shown for LCs covering out to $\approx 60$ days past maximum. Finally, our method and results are discussed within the framework of multiple explosion scenarios, and in light of upcoming surveys.

Corresponding author: Peter Hoeflich
phoeflich@fsu.edu

sc15f@fsu.edu

*Keywords:* Type Ia SNe — Light Curve — Host galaxy — Differential — Main sequence mass — Central density — CSP-I — CSP-II

## 1. INTRODUCTION

Type Ia supernovae (SNe Ia) are thermonuclear explosions of degenerate carbon-oxygen (C/O) white dwarf (WD) stars (Hoyle & Fowler 1960). They are rather homogeneous and can be used as "standardizable candles" within ≈ 0.1 magnitude (mag hereafter) using the classical brightness decline relation (Pskovskii 1977; Phillips 1993; Phillips et al. 1999) to infer cosmological distances which led to the discovery of the dark energy (Perlmutter et al. 1999; Riess et al. 1998). However, to decipher the nature of the dark energy, the precision needs to be improved by an order of magnitude. Though better accuracy may be achieved by larger statistical samples, the control of the systematic shift and spread in the intrinsic SN distribution due to changes with redshift ($z$) in the typical progenitor path and explosion is a major obstacle.

The diversity of SNe Ia can be caused by intrinsic factors such as the progenitor properties and extrinsic factors such as the environment of the exploding WD. Spectral line diagnostics have helped identify different SNe Ia subtypes (Branch et al. 2005, 2009; Wang et al. 2013; Folatelli et al. 2013), and some evidence for variable rise times of SNe Ia with similar overall light curves (LC) has been suggested (Riess et al. 1999). Variations in progenitor systems, explosion scenarios (Hoeflich & Khokhlov 1996; Quimby et al. 2006; Shen et al. 2010; Polin et al. 2019), and/or viewing angle effects (Howell et al. 2001; Wang et al. 2003b; Hoeflich 2006; Motohara et al. 2006; Maeda et al. 2010a; Shen et al. 2018a) can contribute. From theory, variations in peak brightness by the progenitor are expected to be up to ≈ 0.3 mag (e.g. Hoeflich et al. 1998, 2017a), and those have been observed (e.g. Höflich et al. 2010; Gall et al. 2018).

From observations, LCs with high absolute precision are needed and became first available with the Carnegie Supernova Projects I and II (CSP-I and II)[1]. CSP was designed to obtain high-quality data to improve the cosmological use and understanding of SNe Ia. From 2004 to 2009, the Carnegie Supernova Project I (CSP-I, Contreras et al. 2010; Stritzinger et al. 2011; Krisciunas et al. 2017) observed well-calibrated optical and near-infrared (NIR) LCs of multiple types of SNe. From 2011 to 2015, the Carnegie Supernova Project II (CSP-II, Phillips et al. 2019; Hsiao et al. 2019) obtained both LCs and follow-up spectroscopy of SNe, including LCs for 214 SNe Ia and NIR spectra for 157 SNe Ia. The LCs were obtained using the 1 $m$ Henrietta Swope telescope at Las Campanas Observatory (LCO) in Chile. For both CSP-I and II, LCs are placed on the CSP natural filter system (Phillips et al. 2019). CSP-I came mostly from targeted searches, while CSP-II is unbiased and comes from blind searches in the Hubble flow.

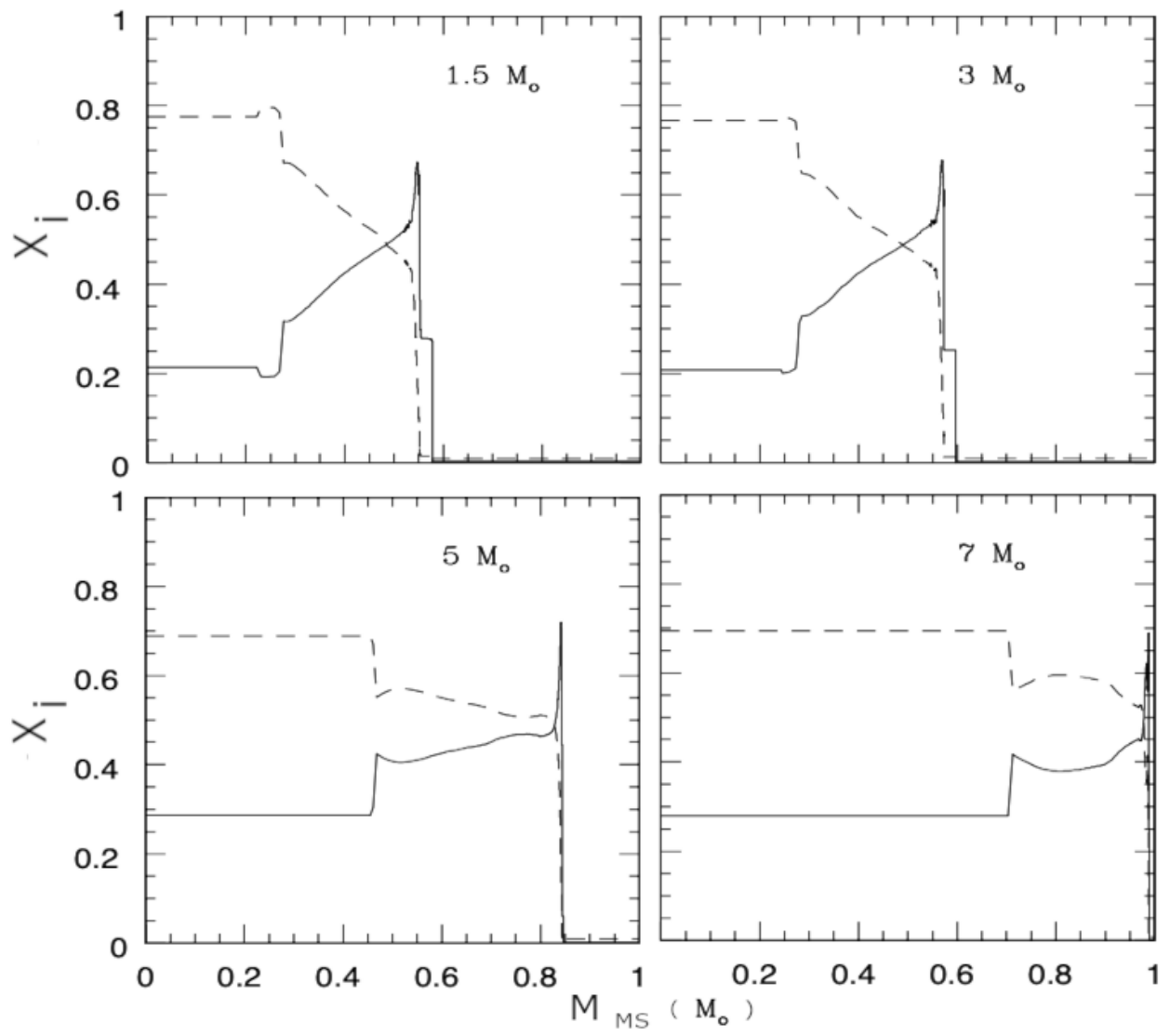

**Figure 1.** Final chemical carbon (solid line) and oxygen (dashed line) profiles in the central region of stars with $M_{MS}$ between 1.5 to 7 $M_\odot$ for solar abundances $Z$ = 0.02 (Pop I) (adapted from Domínguez et al. 2001). Note that low $Z$ (Pop II/III) will affect the size of the central-helium burning core similar to the main sequence mass $M_{MS}$. A change from one to 1/10 $Z_\odot$ is equivalent to a $\delta M_{MS} \approx 1$ $M_\odot$ (Höflich et al. 2000; Domínguez et al. 2001). However, the CSP SNe Ia are local without a widespread in $Z$.

Based on CSP-I data, variations in high-precision monochromatic LCs standardized by the stretch in time (Perlmutter et al. 1997) have been established. In particular, some of the SNe Ia pairs showed time dependence in the differences consistent with one of the individual eigenfunctions predicted by the theoretical models, suggesting that two SNe Ia differ only in one of the variables in the model. Within $M_{Ch}$ mass explosions, the variations were interpreted as differences in the progenitor properties, namely in terms of either the main sequence mass or $M_{MS}$ (Fig. 1), the mass of the initial WD ($M_{WD}$) which, in $M_{Ch}$ mass scenarios, is equivalent to the central density ($\rho_c$) at the time of the explosion of the WD (Fig. 2 [2] ) leading to a central 'hole' in the $^{56}$Ni by electron capture (EC) (Fig. 3) and metallicity ($Z$). This led to the suggestion to describe the differences between all SNe pairs as a linear combination of three theoretical eigenfunctions as a function of time for all SNe Ia (Höflich et al. 2010). This method has been validated using the SNe Ia of CSP-I by Sadler (2012). Though some progenitor properties and relations to the brightness have been found, the analysis has been hampered by the small sample size, a total of 25 SNe Ia, disallowing subsampling.

[1] As discussed in Sect. 8.2, more high-precision, homogeneous data sets will be released and become available in the near future.

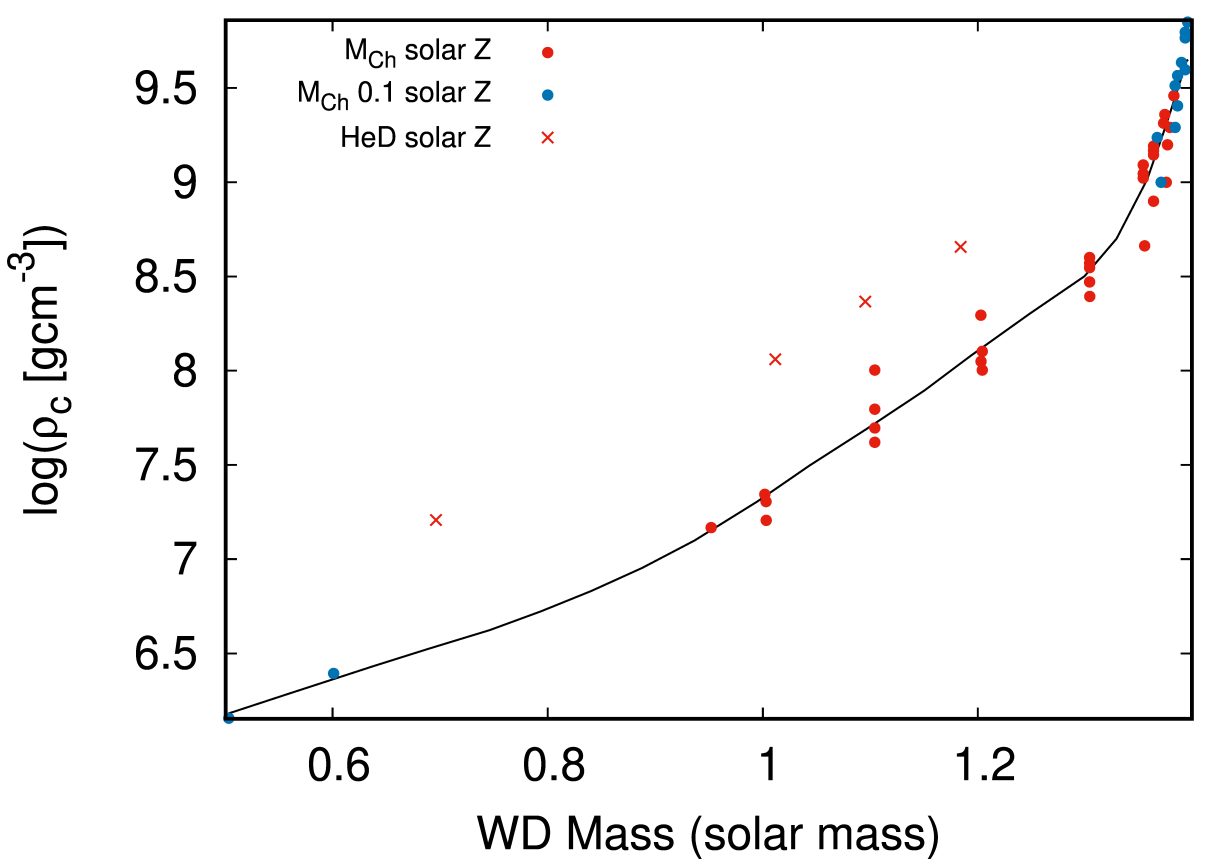

**Figure 2.** $\rho_c$ is shown as a function of the WD mass($M_{WD}$). Models are shown for various densities and solar (red, Pop I) and 1/10 solar (blue, Pop II/III) metallicity (adapted from Hoeflich et al. (2019), and references therein). Pop II/III metallicities can be expected at high redshifts z. The black line indicates the largest density of burning in scenarios starting with a deflagration phase. For He-triggered detonations, the detonation waves compress the material and increase the density (the peak density of burning for central detonation is larger by 0.5 dex). .

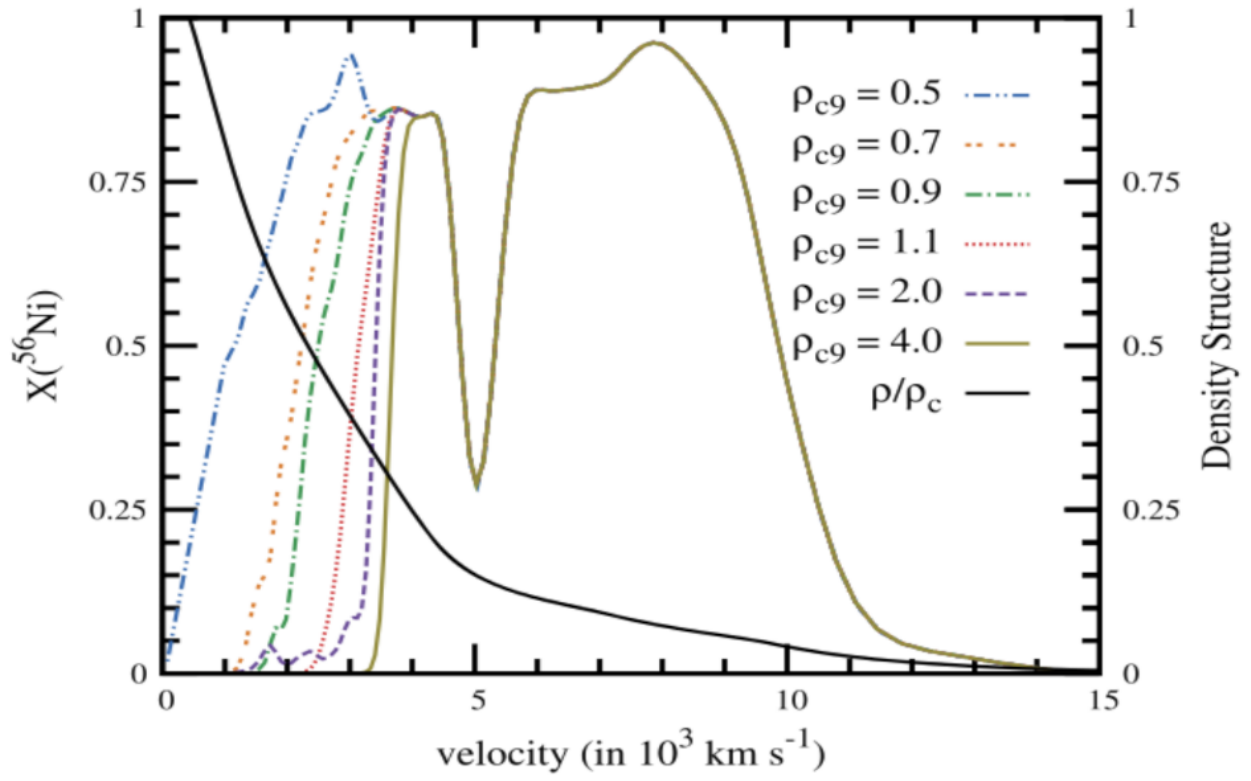

**Figure 3.** The abundance of $^{56}$Ni at $t = 0$ as a function of the expansion velocity for spherical models with various initial central densities $\rho_c$ in units of $10^9$ $g/cm^3$ of the WD (left label). The lack of $^{56}$Ni is due to electron capture (EC), which shifts the nuclear statistical equilibrium (NSE) to stable isotopes of iron group elements. Thus, the size of the $^{56}$Ni hole increases with $\rho_c$. The dip at 5000 $km/s$ is an artifact of spherical delayed detonation models and is caused by the strong reflection wave produced by the DDT. The solid black line ($\rho/\rho_c$) shows the density profile (right label) for the model with $\rho_{c9} = 0.9$ (where $\rho_{c9}$ is $\rho_c$ in units of $10^9$ $g/cm^3$). The variations in the density profiles remain small (adapted from Diamond et al. 2015).

Here, we extend the analysis based on monochromatic LCs to 29 SNe Ia from CSP-I and 226 SNe Ia from CSP-II to probe the number and quality of the eigenfunctions and latency in the observations needed to reduce the dispersion in brightness to ≈ 1%, as required for high-precision SNe Ia cosmology.

[2] e.g., to first order and as an upper limit, we may expect similar electron capture (EC) isotopes in HeD and DDT at 1.2 and 1.3 $M_\odot$, respectively. However, the duration of compression by a detonation is shorter than the WD expansion timescale resulting in about 1/2 the shift in abundance with respect to from partial to NSE burning ($log(\rho_c) \gtrsim -1$), electron capture (EC) isotopes ($log(\rho_c) \geq 8.5$, $^{54}$Fe, $^{57}$Co, $^{58}$Ni) and to increasingly neutron-rich isotopes from $^{55,56}$Fe, $^{55}$Mn to $^{48,52,54}$Cr, $^{59,60}$Fe ($log(\rho_c) \geq 1.87$) which is close to the accretion induced collapse (Höflich et al. 1998b; Brachwitz et al. 2000; Hoeflich et al. 2019).

The main goals of this paper are to study the distribution of the progenitor properties, their correlation to the host galaxies, and to find criteria to recognize subclasses such as 91T-likes and answer whether they are or are not the bright end of normal SNe Ia or a distinct class of objects (Phillips et al. 2022), and to filter out non-standard SNe Ia which may 'contaminate' samples for high-precision cosmology when applying the methods described above **by means of extracting specific information found to be related to the progenitor star and system.**

**We have carefully chosen our methods and observables to be robust against most expected forms of variation and noise. For example, we sample signals from a single photometric band to eliminate most reddening considerations; the V band was carefully chosen, as it is largely free from atomic line-blocking and blending, which complicates U and B band photometry. The method presented can be expected to be stable with respect to variations within the explosion physics because it is based on conserved quantities and nuclear physics (see Sect. 2). As a corollary, it can be expected to be insensitive to variation between and within explosion scenarios.**

**The paper builds upon a series of papers from the CSP collaboration crucial for the analysis:**
**a) LCs corrections for interstellar reddening and k-corrections (Burns et al. 2011; Hsiao et al. 2007; Burns et al. 2014, 2018; Stritzinger et al. 2011; Krisciunas et al. 2017; Hoeflich et al. 2017a) which amount to about twice the secondary effects discussed here.**
**b) Classification of host galaxies to identify possible relation between SNe Ia properties and the hosts (Galbany et al. 2018, 2016; Uddin et al. 2020, 2023).**
**It also draws from previous theoretical works of many groups (see Sects. 2 & 3). Note that color effects and spectra of CSP-SNe Ia have been studied extensively, though, they are not the main subject of this paper which is on secondary parameters in addition to color and LC stretch (Hoeflich et al. 2017b).**

The structure of this paper is as follows:

In Sect. 2, we give some background on SNe Ia and their progenitors to justify the explosion scenario used to reconstruct the eigenfunctions, and lay the groundwork for the discussion of physical parameters in alternative scenarios and for the relation to their environment. In Sect. 3, our method is presented as implemented in our Supernova Parameter Analysis Tool (SPAT). In Sect. 4, it is applied to the observations within the framework of the delayed-detonation scenario, and the number of LC parameters is identified which allows us to characterize the diversity of the majority of SNe Ia in our sample. Distributions of the physical SNe Ia properties are discussed including their correlation with the host-galaxy properties. Subgroups among SNe Ia have been identified by the method and correlated with prior classifications from the literature. In Sect. 6, implications for cosmology and the first generation of SNe Ia are discussed. In Sect. 7, the possible physical parameters are discussed in light of alternative explosion scenarios. In Sect. 8, we summarize the main results, limitations, and future directions.

## 2. BACKGROUND

One of the potential SNe Ia progenitor systems is the single degenerate (SD) system in which a single WD approaches the Chandrasekhar mass ($M_{Ch}$) by accreting, via Roche-Lobe (RL) overflow, from a non-degenerate donor companion which may be a main sequence (MS) star, helium (He) star, or red giant (RG) star (Iben & Tutukov 1984; Webbink 1984; Han & Podsiadlowski 2006; Di Stefano et al. 2011; Nomoto et al. 2003; Branch et al. 1995; Wang & Han 2012). Another potential progenitor system for SNe Ia is a double degenerate (DD) system that consists of two WDs in close orbit, merging via the potential energy loss by gravitational radiation (Iben & Tutukov 1984; Webbink 1984). A triple system with two colliding WDs (Lidov 1962; Rosswog et al. 2009; Thompson 2011; Pejcha et al. 2013; Kushnir et al. 2013; Dong et al. 2015) can also be a possible progenitor system of SNe Ia.

There are three leading scenarios for the explosion physics of SNe Ia. The first is the delayed-detonation scenario (Khokhlov 1991). A WD in a DD or an SD system accretes material from a companion over long time scales (up to $10^8$ years) resulting in a secular merger (Whelan & Iben 1973; Piersanti et al. 2003). The explosion is triggered by compressional heating near the WD center as the WD approaches the $M_{Ch}$. The flame starts as a deflagration (Nomoto et al. 1984) which is then followed by a deflagration-to-detonation transition (DDT). For more detail, see Höflich et al. (2013). The amount of deflagration burning is the physical property governing the brightness-decline rate relation. Within this class of models, most SNe Ia should be normal-bright (Höflich et al. 2002). The C/O ratio affects the explosion energy [3] (Domínguez et al. 2001), so the effects of $M_{MS}$ can be seen in this scenario. Higher $\rho_c$ results in a shift of the abundances from $^{56}$Ni to electron capture (EC) elements close to the center of the WD. At ~ 23 days after the explosion when the central shift to EC elements and loss of $^{56}$Ni becomes apparent as a decrease in luminosity because the central $^{56}$Ni can contribute to the LC, $\rho_c$ contributes to the diversity.

The second explosion scenario is a surface He detonation (HeD) that triggers a central detonation of a sub-$M_{Ch}$ WD with a C/O-core (Woosley et al. 1980; Nomoto 1982a; Livne 1990; Woosley & Weaver 1994; Hoeflich & Khokhlov 1996; Kromer et al. 2010; Sim et al. 2010; Woosley & Kasen 2011; Shen 2015; Tanikawa 2018; Glasner et al. 2018; Polin et al. 2019; Townsley et al. 2019). The C/O detonation may be triggered off-center (Livne et al. 2005). HeD models originate mostly from a C/O-WD with a thin He layer, accreting He from a companion (Woosley & Weaver 1994; Iben & Tutukov 1991). C/O-WD accreting from the wind of a companion RG, namely symbiotic binary stars can also be the origin in this scenario (Munari & Renzini 1992).

The third explosion scenario is the dynamical merging of two WDs, possibly head-on in a triple system, and heating on a dynamical timescale of seconds (Webbink 1984; Iben & Tutukov 1984; Benz et al. 1990; Ra-sio & Shapiro 1994; Hoeflich & Khokhlov 1996; Segretain et al. 1997; Yoon et al. 2007; Wang et al. 2009b,a; Lorén-Aguilar et al. 2009; Pakmor et al. 2010; Isern et al. 2011; Pakmor et al. 2012; Rosswog et al. 2009; Thompson 2011; Pejcha et al. 2013; Kushnir et al. 2013; Dong et al. 2015; García-Berro & Lorén-Aguilar 2017).

[3] Explosion energy is nuclear energy minus binding energy

**Nuclear physics determines the structure of the progenitor WD, the explosion physics (which imprints the burning conditions on the abundance pattern of the ejecta), the average expansion velocities, and the LCs, which are powered by radioactive decay of $^{56}$Ni $\rightarrow^{56}$ Co $\rightarrow^{56}$Fe for the first $2-3$ years. The explosion structures being close to self-similar (Arnett 1980). To first order, this masks the actual diversity of SNe Ia explosions and progenitors - 'stellar amnesia' (Höflich et al. 2003). These differences may be linked to variations in progenitor systems and explosion mechanisms (Hoeflich & Khokhlov 1996; Quimby et al. 2006; Shen et al. 2010; Polin et al. 2019), and/or due to viewing angle effects (Howell et al. 2001; Wang et al. 2003b; Hoeflich 2006; Motohara et al. 2006; Maeda et al. 2010b; Shen et al. 2018b).**

Within a wide range of explosion scenarios, the brightness-decline rate relation and, equivalently, the luminosity time-stretch $s$ (Perlmutter et al. 1999) relation can be understood as a direct effect of the $^{56}$Ni production. Radioactive decay of $^{56}$Ni powers the LCs (Colgate & McKee 1969). Therefore, more $^{56}$Ni means a brighter maximum and more heating, leading to higher opacities, and consequently, a slower decline rate with increasing brightness (Hoeflich et al. 1996; Kasen et al. 2006) [4]. In this work, the stretch $s$ will be the primary parameter for characterizing LCs (see Sect. 1).

Here, the main analysis is based on the framework of spherical delayed-detonation models (Khokhlov 1991) **which has been employed to analyze** many observations of Phillips-normal SNe Ia, i.e., those that follow the luminosity decline relation (Phillips 1993), **including early-time optical and near-infrared spectra, LCs and color-magnitude diagrams** (Wang et al. 2003a; Hoeflich et al. 2017a). Moreover, JWST observations show narrow stable $^{58}$Ni in the nebular spectra (Gerardy et al. (2007); Telesco et al. (2015); Hoeflich et al. (2021); DerKacy et al. (2023, 2024) **(Ashall et al., in preparation)** which all indicate high-mass explosions with little mixing by a passive flow of EC elements in a pre-existing turbulent field prior to the runaway (Höflich & Stein 2002; Zingale et al. 2011) rather than multiple-spot ignitions and Rayleigh-Taylor (RT) instabilities as assumed by Seitenzahl et al. (2013) in their version of delayed-detonation models. **However, from agreement on one explosion scenario with individual objects, one cannot exclude the existence of alternative scenarios (see above). So, we have developed our methods to minimize the influence of these unknowns. For our analysis, we use $V$-band which has been shown to only weakly depend on the scenario and variations within and, thus, has been used as proxy for bolometric LCs (Benetti et al. 2005; Branch et al. 2005, 2009; Wang et al. 2013; Folatelli et al. 2013). E.g. the luminosity decline rate relation is caused by the temperature dependence of the opacity, resulting in similar relations for a wide range of explosion models including delayed-detonations, pulsating delayed-detonations, He-triggered detonations, mergers, and envelope models (see Fig. 2 of Hoeflich et al. 1996).**

An important aspect of this work is the correlation between SNe Ia properties and the evolution time between star formation and the explosion. The correlation between the SNe Ia with the environment has been previously studied (Han & Podsiadlowski 2003, 2004; Blanc & Greggio 2008), and found to be a combination of the evolutionary time to the WD ($t_{stellar}$), and the evolutionary time of the progenitor system ($t_{system}$) (Nomoto & Leung 2019). The $t_{stellar}$ depends mostly on the $M_{MS}$ of the progenitor, ranging from ≈ 60 million to more than 10 billion years (Dominguez et al. 1999). Here, we define the delay time ($t_{delay}$) as the total time required for a SN Ia explosion. Then $t_{delay}$ can be written as:

$$t_{delay} = t_{stellar} + t_{system} \quad (1)$$

The initial WD has a mass somewhere between 0.6 and 1.2 $M_{\odot}$. The Eddington luminosity produced by nuclear burning on top of the WD limits the accretion rate. For H-accretors, stable burning is between $10^{-8}$ to $10^{-6}$ $M_{\odot} yr^{-1}$ (Nomoto & Leung 2019; Hachisu et al. 1996; Li & van den Heuvel 1997; Nomoto & et al. 2000; Nomoto 1982b), resulting in time of accretion to be $\approx 3\times10^{6}$ to $3\times10^{7}$ years. The energy production by He burning is lower by about a factor of 10 resulting in a correspondingly shorter time of accretion (Iben & Tutukov 1994; Nomoto & Iben 1985).

The delay times can range from $\approx 1\times10^{1}$ years, in case of double-degenerate systems, to the age of the Universe (Han & Podsiadlowski 2003, 2004). Binary evolution in the DD scenario happens via gravitational wave radiation, and therefore the dynamical merger scenario can have a long delay time to explosion (Iben & Tutukov 1984). Considering the lifetimes, the delayed detonation models can originate from both single and double-degenerate systems (Whelan & Iben 1973; Piersanti et al. 2003).

[4] Note that it is still under debate which scenario dominates the diverse group of SNe Ia (Höflich et al. 2002; Hoeflich 2017; Shen et al. 2021).

**Table 1.** Regions (see Fig. 4, Sect. 3.4) needed in the LC to distinguish between central density and progenitor mass components.

| Regions | Primary Parameters | Progenitor Mass | Central Density |
|---|---|---|---|
| I | | ✓ | |
| II | ✓ | | |
| III | | ✓ | ✓ |
| IV | | | ✓ |

## 3. METHODOLOGY USED IN SPAT

In this section, the outline is given for the procedural steps of our LC analysis using the Secondary Parameter Analysis Tool (SPAT). Only the LC shapes in specific filters were used without considering colors. Therefore, the method is independent of extinction, as discussed below.

The method relies on uniform data sets and k-corrected LCs. The uniform data sets from CSP-I and CSP-II (see Sect. 1) were used. The k-corrections are based on spectral templates from Burns et al. (2011) and Hsiao et al. (2007).

To characterize the LCs beyond the *classical* brightness-decline relation, the method combines high-precision observations with relations for the LC evolution with time, so-called eigenfunctions, based on theoretical delayed-detonation models. As shown below, combining information allows us to extract progenitor and explosion properties of an individual SN Ia, related to the individual weights to the eigenvalues attributed (Sect. 3.4). Subsequently, the eigenvalues are transformed into physical secondary parameters using prior knowledge of their allowed ranges (Sects. 3.5 and 3.6).

We want to emphasize that early rise times and late time LC coverage up to ≈ 60 days are needed to trace the differential changes (see Sect. 3.3). The suitable LC time coverage required to distinguish between the physical eigenfunction is indicated in Tab. 1.

The primary parameter $\Delta m_{15}(B)$ or equivalently, stretch ($s$) in time (Perlmutter et al. 1999; Jha et al. 2006) is our first parameter, and removes the diversity to first order (Phillips 1993).

For the secondary parameters, differentials of $V$-band light curves are used to boost the accuracy of the analysis (Höflich et al. 2010; Sadler 2012), see Sect. 3.4 and Fig. 4. Note that absolute intrinsic model uncertainties are expected to be a few tenths of a magnitude, i.e., comparable to the differences between two SNe Ia observations. Model uncertainties in differential LC changes are smaller by an order of magnitude. In this work and following (Höflich et al. 2010; Sadler 2012), we identified the main-sequence mass $M_{MS}$ and central density $\rho_c$ as secondary parameters (Sect. 3.4) and will show in Sect. 4 that two parameters are sufficient to model the observations.

### 3.1. *Primary Parameter in Detail*

The first step is to remove the effect that the diffusion time scales become shorter with decreasing brightness. LC templates are used to determine the resulting time-stretch parameter $s$.

#### 3.1.1. *Creation of a Uniform Template*

As a template, a single "fiducial" template is used rather than a set of templates based on observations because those may already contain systematic variations between brightness and tail ratios. A $V$-band "fiducial" template is created by using an average of three templates (Prieto et al. 2006) with $\Delta\bar{m}_{15}(B)$ values of 0.9, 1.0, and 1.1 based on LC-fits using SNooPy (Burns et al. 2011). These templates were adapted because they are indistinguishable from each other in the temporal range $[-3 : 15]$ days where the Höflich et al. (2010) models found no (or little) effect from the "secondary parameters", $M_{MS}$ and $\rho_c$ (Fig. 4). This "fiducial" template is defined to have a stretch $s = 1.0$, and it is normalized such that it peaks at $t = 0$ day with magnitude 0. Each observed visual LC was fitted using this template by $\chi^2$ minimization. This step also determines the peak brightness, which is subtracted from all data to normalize the LC, produces a time of maximum ($t_0$), and the stretch parameter in time ($s$) from each observed LC.

This analysis focuses on the $V$-band rather than $U$ or $B$ because it is least affected by the metallicity (Hoeflich et al. 1998) **and mixing (Aldoroty et al. 2023) by e.g. RT instabilities and passive drag (see Sect. 2)**. From theory, $V$ is least affected by line blending and blanketing and, thus, least susceptible to numerical radiation transport and discretization errors (Höflich et al. 2002). Moreover, as the post-maximum decline is gentler in the $V$-band, knowing the exact time of maximum is less crucial for determining the $s$ parameter.

We use three different 'definitions' for the brightness-decline ratio relations:

1. For reference to modern literature, $\Delta\bar{m}_{15}(B/V)$ defined by template-fitting of the entire observed LC using SNooPy (Burns et al. 2011, 2014);

2. Template fitting of the entire LC using the stretch $s$ in time as parameter (Perlmutter et al. 1999) in combination with the empirical relation $\Delta m_{15}(B) = 3.06 - 2.04 \times s$ (Jha et al. 2006);

3. $\Delta m_{15,s}(B/V)$ defined by the brightness-decline over $\Delta(t) = 15d/s$ past maximum using template fitting for $s$ in region II using the $V$-band LC (Hoeflich et al. 2017a) (see Sect. 3.3). **Namely, to determine s, the boundaries of region II (relative to maximum light) are reduced by $s$ and iterated, $\Delta m(B/V) = m(t_{max})(B/V) - m(t_{max} + 15d/s)(B/V)$ with $\Delta m_{15,s}(B/V) = \Delta m(B/V) \times s_V$.** The limited time range is used because observations outside this region are utilized to constrain secondary LC parameters, and the 's' taken over the entire LC would depend on the LC coverage and the distribution of the actual observation. Note that both $\Delta m_{15,s}(B/V)$ and the color stretch parameter $s_{BV}$ have been introduced to avoid the ambiguity between values of underluminous and transitional SNe Ia. The corresponding LC parameters and host-galaxy properties of all CSP SNe Ia used in our analysis are given in appendices A and B (Tabs. A1 & A2), respectively.

### 3.1.2. *Correction of the Brightness to Peak-to-tail relation*

The time-dependent correction factors to the uniform template are based on theoretical models with all physical parameters fixed except the progenitor properties. More $^{56}$Ni means brighter LCs and higher temperatures which translates to a shallow gradient of opacity (Hoeflich et al. 1996), and therefore the stored energy during the pre-maximum phase is released more slowly in the post-maximum phase for normal-bright SNe Ia than the subluminous SNe Ia. This results in a variation of the ratio of bolometric luminosity and the instantaneous energy release, namely Q (Arnett 1982; Hoeflich & Khokhlov 1996). To remove this effect we use time-dependent magnitude corrections. These corrections are based on models of Hoeflich et al. (2017a) with a range of Q values to reflect the range of observed $\Delta m_{15,s}(V)$. The differences between instantaneous energy release by radioactive decay and the luminosity of SNe Ia were calculated from the models 12, 16, and 25 (Hoeflich et al. 2017a). These differences were interpolated in a time grid between −2 to 20 days with respect to the maximum time and normalized to the model with a brightness decline ratio $\Delta m_{15,s}(V) = 0.68$ (Model 5 - normal bright). The normalized differences were interpolated for a range of $\Delta m_{15,s}(V)$ values that go from normal-bright to subluminous SNe Ia. Finally, these corrections were subtracted from individual LCs according to their $\Delta m_{15,s}(V)$ values (Fig. 5). The largest correction values are of the order of 0.1 mag. Since our goal is to study the more subtle variations caused by secondary parameters, this step is crucial to remove any remaining effects from the primary parameter. It is well separated from the 'classical peak to tail ratio' due to $\rho_c$ because of the different dependence as a function of time (see below). Note that this correction factor only enters for pairs with very different peak brightness.

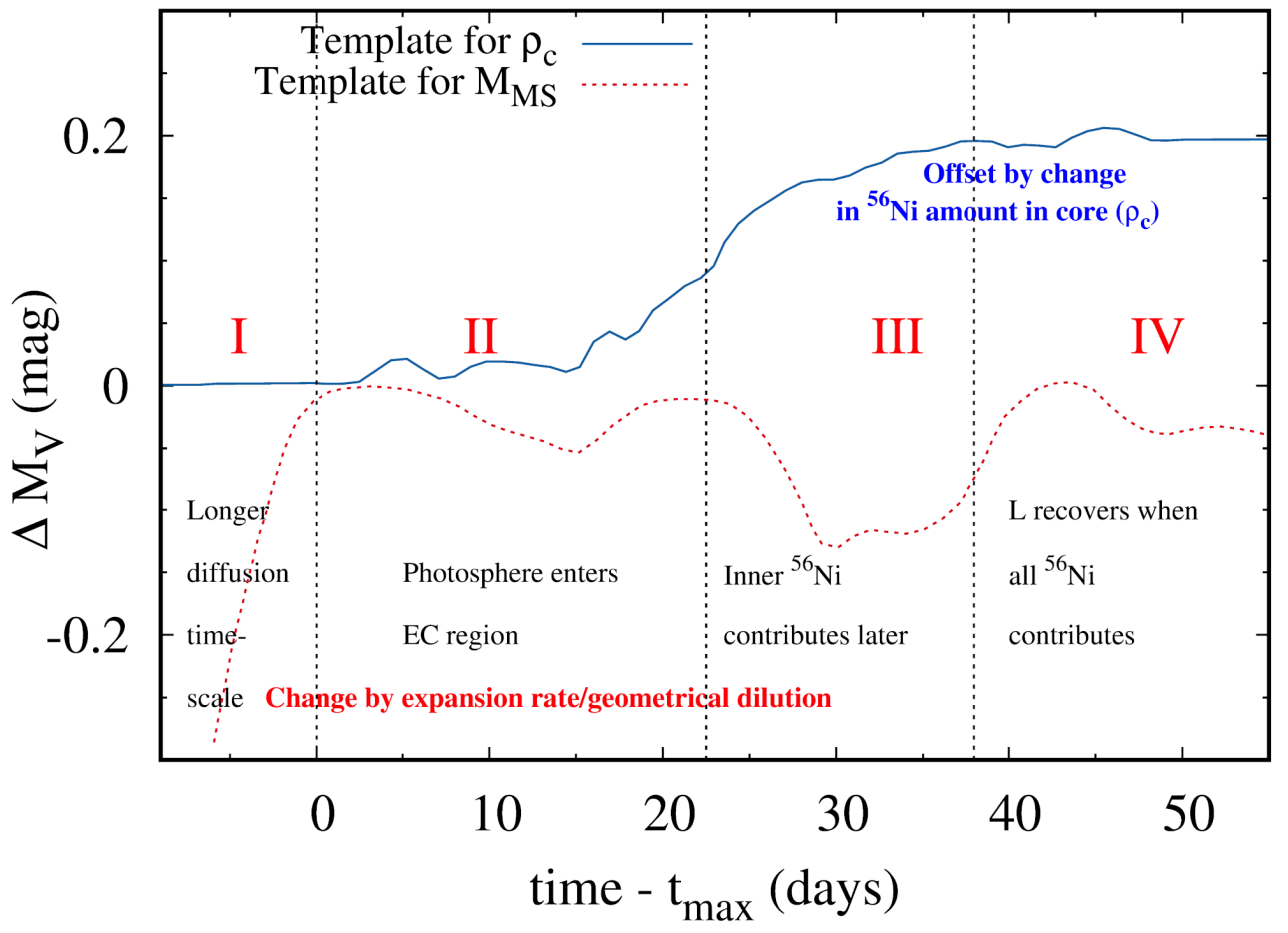

**Figure 4.** Theoretical eigenfunction: theoretical eigenfunctions for variations by $\rho_c$ (blue) and $M_{MS}$ (red), derived from 5 $M_\odot$ and 7 $M_\odot$ theoretical models (Höflich et al. 2010; Hoeflich et al. 2017a) with a $\Delta m_{15}(B)$ of 1.25 and a stretch value of 0.92, corrected for the CSP filter functions (Burns et al. 2011). The four regions are marked as I: pre-maximum, II: maximum to ≈ 22 days after the maximum, III: ~ 22 – 38 days after the maximum, and IV: later than ≈ 38 days after the maximum (see Sect. 3.3 & 3.4). This figure is adapted from Hoeflich et al. (2017a).

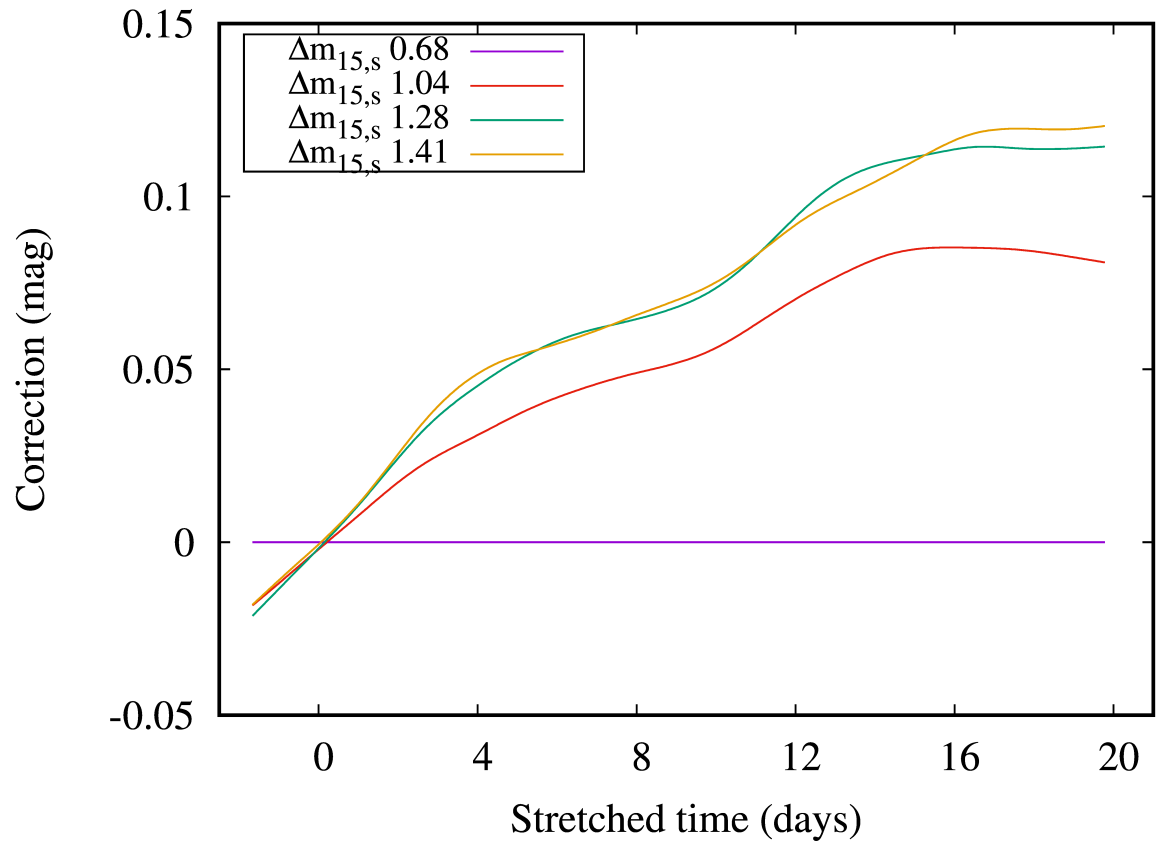

**Figure 5.** Corrections to stretched LCs as a function of the stretched time based on models with different $\Delta m_{15,s}(V)$ normalized to the normal-bright model with $\Delta m_{15,s}(V) = 0.68$, $Q = 1.1$ (Höflich et al. 2002; Hoeflich et al. 2017a)). These corrections are applied to the stretched LCs before constructing the differentials, see Sect. 3.1.2.

### 3.2. *Influence of Uncertainties in the CSM and ISM Reddening on the Differentials*

Reddening will lead to a phase-dependent shift of the flux-averaged wavelength over broadband filters, in particular, due to dust and the varying impact on B and, to a lesser amount, on V, due to the strong line-blending (Phillips et al. 1999). The LC becomes broader ( shift in $s$) and, potentially, leading to a 'distortion' of the differentials by $\Delta V$ which depends on the uncertainty in the extinction coefficient $R_V$ (Fig. 6).

Note that our LCs have been k-corrected using SNooPY fits which takes this into account. However, the uncertainties in reddening do not enter our analysis, but systematics would show up in the residuals of individual differentials and a shift with redshift in our $\rho_c - M_{MS}$ diagrams. The effect of reddening uncertainties can be expected to be the largest in highly reddened SNe Ia. In Fig. 7 (left), the distribution of our SNe Ia is shown. The majority of SNe Ia, some 80% and 96% show low $E_{B-V} \leq 0.2$ and 0.5 mag corresponding to $\approx 0.01$ and 0.025 mag in the differential, respectively. This is well below the error bars of individual data points in the differentials (see Sect. 3.5).

The locations of the mean values in the $\rho_c - M_{MS}$ diagram of the three reddening groups, $0 \leq E(B-V) < 0.1$, $0.1 \leq E(B-V) < 0.2$ and $E(B-V) \geq 0.2$ mag with 55%, 24% and 20% of all SNe, respectively, are consistent with the statistical uncertainties (Fig. 7). The high $E(B-V)$-group is slightly shifted towards the locus of 91T-like SNe because some 16% of all 91T-likes lie in this group vs. 3% in the low-reddening group. We find indeed that the systematic effect on the differentials (Sect. 3.4) and projection in the physical parameter space (Sect. 3.6) remains small (Fig. 7, right), ruling out systematics in the k-corrections applied.

For the few objects with high reddening and large uncertainties in $E(B-V)$ and $R_V$, the possible impact needs to be taken in mind and will be addressed in the corresponding sections.

Note that the size of $\Delta V(t, \Delta E_{B-V}, \Delta R_V, z)(t)$ will become important for high-precision cosmology with unusual SNe Ia, or very large data sample where the statistical error becomes small in the distribution of progenitor properties.

In this context, it may be noticed that the functional form somewhat resembles our eigenfunctions (Fig. 4) only distinguishable beyond some 40+ days as another argument for considering LCs up to 55 – 60 days past maximum light.

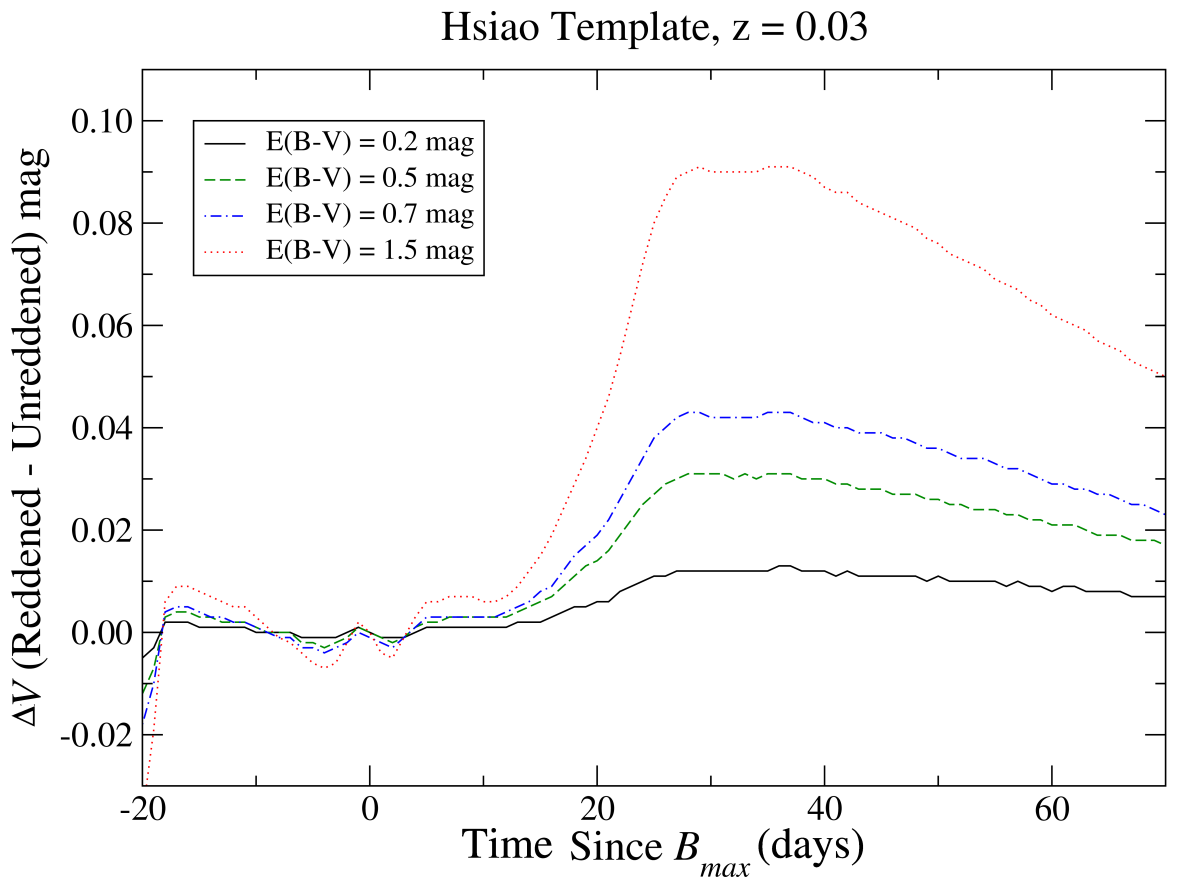

**Figure 6.** Reddening $\Delta V(t, R_V = 3.1, z = 0.03)$ as a function of time for a SNe Ia at a redshift of $z = 0.03$ using the spectral templates by Hsiao et al. (2007) for $E(B-V)$ between 0.2 and 1.5 mag.

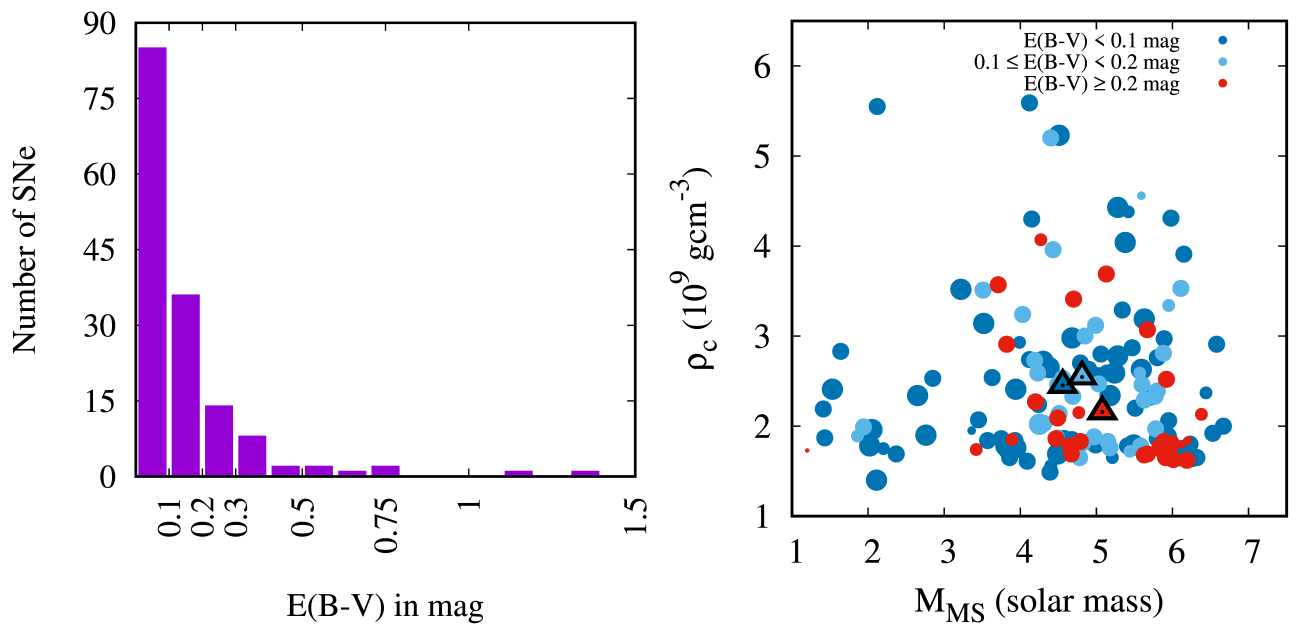

**Figure 7.** Distribution of SNe Ia as a function of $E(B-V)$ (left) and l location of SNe Ia in the $M_{MS} - \rho_c$ plane with low and large reddening(right). The average parameter values for high, medium, and low reddening are indicated by black triangles filled with the same color as the reddening groups.

### 3.3. *Criteria to Select SNe Ia in the SNe Ia Sample*

First, we need to select LCs with proper coverage to extract the primary and secondary parameter effects. For this purpose and based on physical regimes as defined in Fig. 4, the time domain is divided into four regions: region I - the pre-maximum phase, region II - maximum to ~ 22 days after the maximum, region III - 22 – 38 days after the maximum, and region IV - later than ~ 38 days after the maximum. Due to the effects of $M_{MS}$ and $\rho_c$ in different LC regions (see Sect. 2 and Fig. 4), we need data in region II for the primary parameter, the brightness-decline ratio, and III for both the secondary parameter effects. To separate the secondary parameter effects, we need data in regions I or IV (Tab. 1, Fig. 4).

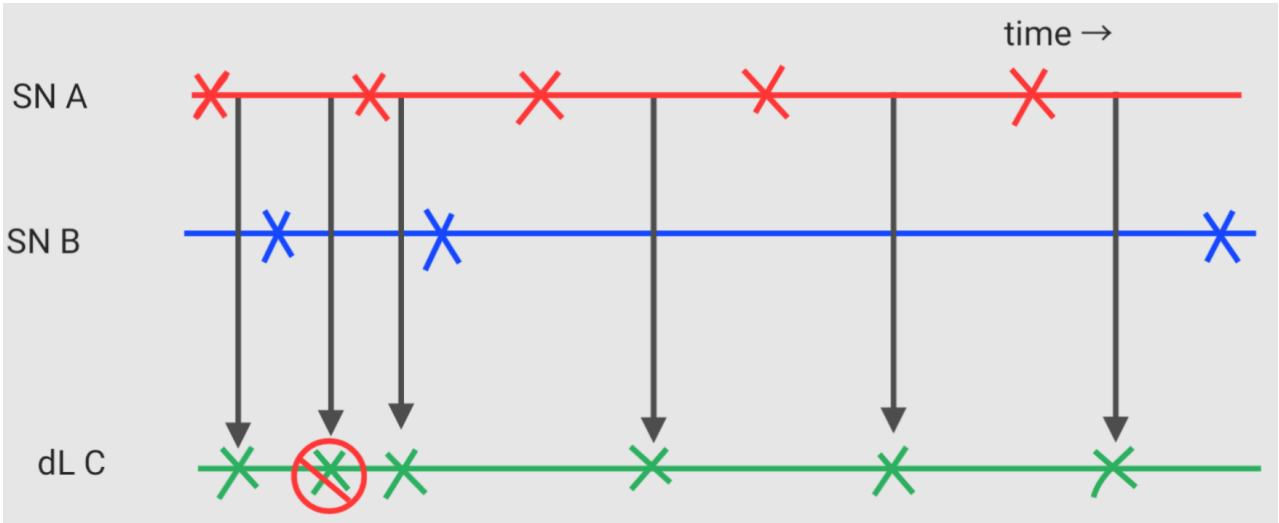

**Figure 8.** Cartoon to show the interpolation scheme for constructing differentials: SNA (red) and SNB (blue) show the times of observations. dLC shows the times at which the differentials are calculated from the interpolated LCs of SNA and SNB (Section 3.4).

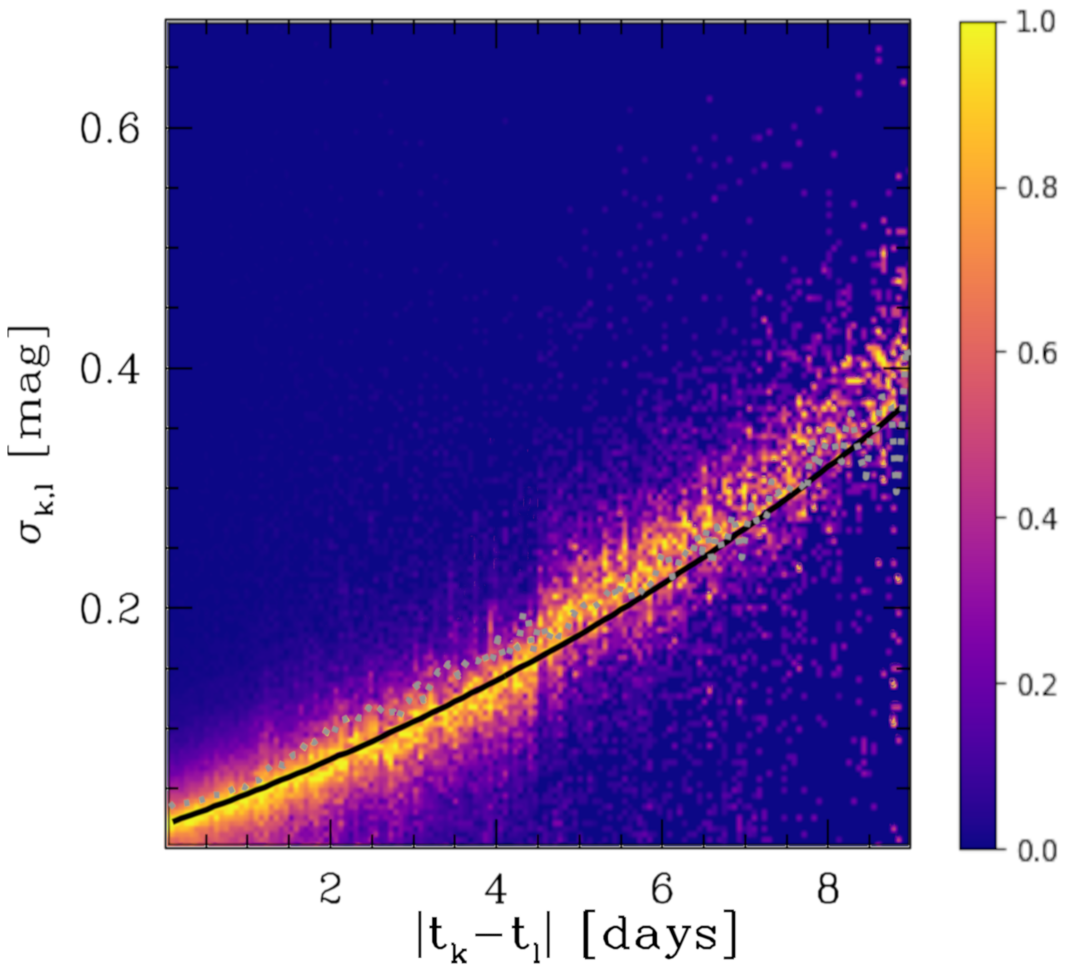

**Figure 9.** Residual deviations for all SNe Ia pairs as a function of the minimum distance in time between observations. The number of pairs has been normalized to their maximum within each time-bin of 0.06 days. In addition, we give the function used as correlation error (Fig. 3) as applied to individual pairs (black line) for solving the overdetermined system (Sect. 3.4).

First, LCs that do not have at least two data points in regions II or III are discarded. Then LCs that do not have at least two data points in either region I or IV are also discarded because the secondary parameter effects cannot be differentiated in those SNe. No cadence requirement was applied. 94 SNe without suitable LC coverage from the 226 SNe of the CSP-II data set were excluded from this analysis. All SNe in the CSP-I sample satisfy the above criteria. The resulting sample is a total of 161 SNe with sufficient LC coverage. The distribution of their hosts is given in Tab. 2.

**Table 2.** Summary of SNe Ia brightness class of their hosts

| Host Type | SNe with Host Info | SNe with SN-brightness | |
|---|---|---|---|
| | | Normal Bright | Underluminous |
| All | 133 | 106 | 27 |
| Spiral - bulge + irr. + dwarf E | 79 | 75 | 4 |
| Spiral bulge | 9 | 6 | 3 |
| S0 | 33 | 16 | 17 |
| Large E | 12 | 9 | 3 |
| unknown | 19 | 17 | 2 |

Note—Normal bright SNe = SNe with $dm_{15}(B) \leq 1.45$, and underluminous SNe= SNe with $dm_{15}(B) > 1.45$. This table shows the numbers of normal bright and underluminous SNe Ia present in different host types. The total numbers of SNe in the corresponding host types are also given. Most of the underluminous are in the spiral bulges or S0 galaxies.

### 3.4. *Construction of Differentials*

The goal of this step is to identify variations in observed SNe Ia pairs and relate them to the theoretical eigenfunctions.

The observed differentials $\Delta F_{i,j}(obs)(t)$ are constructed by taking the differences between the observed LCs of two SNe $i$ and $j$. They are presented as a linear combination of characteristic signals or eigenfunctions $f_l(t)$ for $n$ numbers of physical properties $l$ (see Fig. 4), by determining the weights $\lambda_{i,j,l}$ plus the higher-order terms, $O_{i,j}(t)$, representing contributions from physics not considered.

$$\Delta F_{i,j}(obs)(t) = \sum_{l=1}^{n} \lambda_{i,j,l} f_l(t) + O_{i,j}(t) \tag{2}$$

We use $M_{MS}$ and $\rho_c$ ($n$=2) as the "secondary parameters" $l$ in the framework of delayed detonation models.

In general, photometric observations are not coincident for any two observed LCs of SNe $i$ and $j$, and the observed LCs change differently. Therefore, time interpolations at time $t_{i,j}$ are used to construct the differentials between SNe $i$ and $j$ using a rotated parabola.

Fig. 8 shows the scheme in a cartoon. The differential was calculated at the midpoint, to minimize the interpolation error. When we have a gap in the LC data from one of the SNe in a pair, we take two neighboring LC points from the same SN to create a differential point in the gap and assign larger uncertainties for these points. Note that after the selection criteria, all LC gaps exist in monotonically increasing or decreasing regions.

Photometric uncertainties, $\sigma_i$ and $\sigma_j$, and interpolation uncertainties were summed. We have modeled the interpolation uncertainties in the differentials as exponential in time with a correlation length $t_{cor}$ and size $f_{cor}$. The total uncertainty, $\sigma_{ij}(t)$, of an interpolated differential point at time $t_{j,i}$ is fitted according to (Sadler 2012):

$$\sigma_{ij}(t) = \begin{cases} \dfrac{(\sigma_i(t) + \sigma_j(t)) \;+\; \exp[t_{cor}|(t_j - t_i)|] - 1.0}{f_{cor}} & \text{for } t_{j,i} \leq 9d \\ \infty & \text{otherwise} \end{cases} \quad (3)$$

where $t_i$ and $t_j$ are times of observations in the stretched LCs for SNe $i$ and $j$, with $f_{cor}$ = 4 days and $t_{cor}$ = 0.098, respectively. For the rare case of gaps larger than 9 days (Fig. 8), the corresponding points have been omitted by setting ($\sigma_{ij} = \infty$) to avoid large gaps dominating the uncertainty. As a first step, $t_{cor}$ = 0.09531 and $f_{cor}$ = 4 days have been estimated based on the analysis by Sadler (2012) which was guided by the maximum change of the LC template assuming 0.1 mag/day and ~ 4 days as typical size/amplitude of differential in the CSP-I data. In this work, we optimized the value of $t_{cor}$ = 0.098 to fit the actual distribution of errors in the combined CSP-I and II sample (Fig. 9). The uncertainties in $f_{cor}$ and $t_{cor}$ can be estimated by the scatter of the lines along the x- and y-axes of Fig. 9, respectively, and obtain $\Delta t_{cor} \approx 3\%$ and $\Delta f_{cor} \approx 10\%$, which shows the soundness of the previous analysis.

Fig. 10 gives an example from our data set. Here SN 2011iv (green points) has a gap in the LC data at about ~ 15 to ~ 25 days, while there is data for SN 2011jn. In this case, we created differential points in the region between the neighboring SN2011jn LC data points.

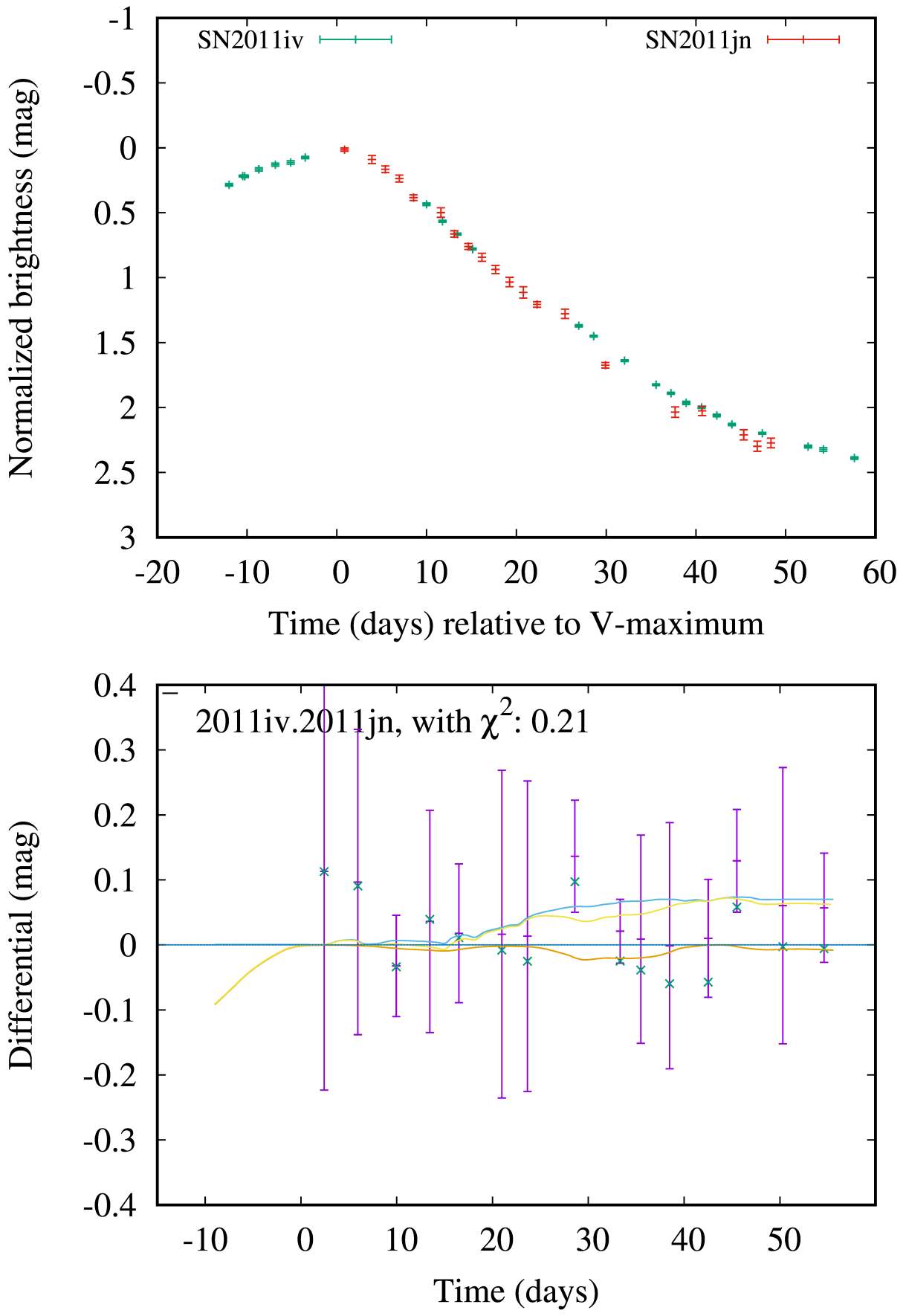

**Figure 10.** We show an example of stretched LC overlay of SN2011iv and SN2011jn (top). SN2011iv has a gap in the LC data between 15 – 25 days. In the bottom image, we show the differential plot of this pair. The orange line is the $M_{MS}$ component, the blue line is the $\rho_c$ component, and the yellow line represents the combined component. The violet points with error bars are the differentials with error bars, and the green crosses are the residuals (Sect. 3.4). Note: as we solve an overdetermined system, this will not affect the solution of our system.

### 3.5. *Determining Generic Progenitor Parameters*

After finding the differences in LC pairs, generic progenitor parameters ($g_i$) of individual SN are determined by solving an overdetermined system of coefficients ($\lambda_{i,j}$), see Equation 2.

$$\lambda_{i,j,l} = (\frac{g_j}{g_i})_l \quad (4)$$

The optimal signal coefficients ($\lambda_{i,j,l}$) in Equation 2 are the relative disparities between the secondary parameters of the progenitors and are found by minimizing the $\chi^2$ of the residuals $O_{i,j}$ in Equation 2. The following equation is minimized (Nelder & Mead 1965), and the optimal scaled signal ($f_l(t) \times \lambda_{i,j,l}$) is subtracted from each differential data point to get the higher order terms in Equation 2. Here $m$ is the number of differential points in a pair of SNe Ia LCs, and $n$ is the number of physical properties $l$.

$$\chi^2(\lambda_1, \lambda_2, ...., \lambda_n) = \sum_{k=1}^{m} \left( \frac{dm_k - \sum_{l=1}^{n} f_l(t_k) \times \lambda_{i,j,l}}{\sigma_k} \right)^2 \quad (5)$$

As there can be secondary parameters affecting the LCs other than the two we are investigating, the residuals ($O_{i,j}(t)$) will not be zero. However, Fig. 11 shows some typical examples where the residuals (green crosses in the left-hand side plots) are well within the error bars, which suggests that the higher-order terms in Equation 2 do not contain any significant amount of information.

In some LC pairs, the $\lambda_{i,j,l}$ coefficients can have a high degree of covariance depending on the availability of the data. To take this into account, the two-dimensional $\lambda_{i,j,l}$ parameter space is scanned to create a probability distribution where the individual dots are Monte-Carlo solutions of the coefficients (see the right-hand side plots in Fig. 11). The quality of the SNe Ia pair and the uniqueness of the solution are determined by the confinement of the signal coefficients in the parameter space. In the example shown in Fig. 11, the top right plot is more constrained in $\rho_c$ than $M_{MS}$, but the middle and bottom right plots are well constrained in both dimensions, although there is some covariance in all these examples. In the example of Fig. 12 (middle right), it is not well constrained in either dimension.

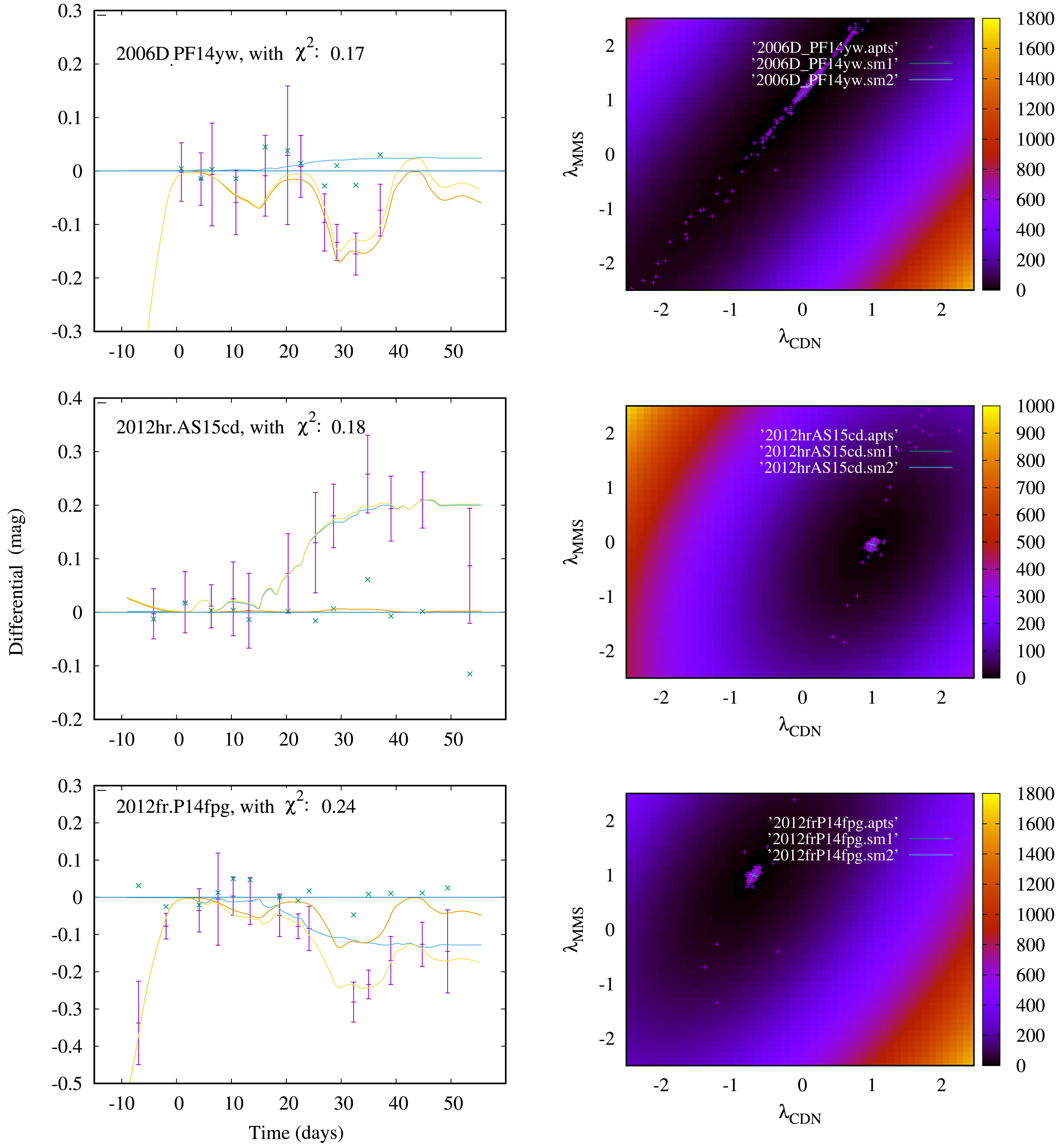

**Figure 11.** Differential (left) and correlation matrix (right) plots for the pairs of: SN2006D and iPTF14yw (top row) - $M_{MS}$ is the dominant component here; SN2012hr and ASASSN-15cd (middle row) - $\rho_c$ is the dominant component here; SN2012fr and iPTF14fpg (bottom row) - comparable contributions from both $M_{MS}$ and $\rho_c$ can be seen in this case. The differential plots show the $M_{MS}$(orange), $\rho_c$(blue), and the combined components(yellow), respectively. In addition, the differentials with error bars (violet) and the residuals (green) are shown. In the correlation plot, the individual dots are the Monte Carlo solution of the coefficients. Lines ".sm1" and ".sm2" represent the semi-major and semi-minor axes of the uncertainty ellipse, respectively. See Sect. 3.5.

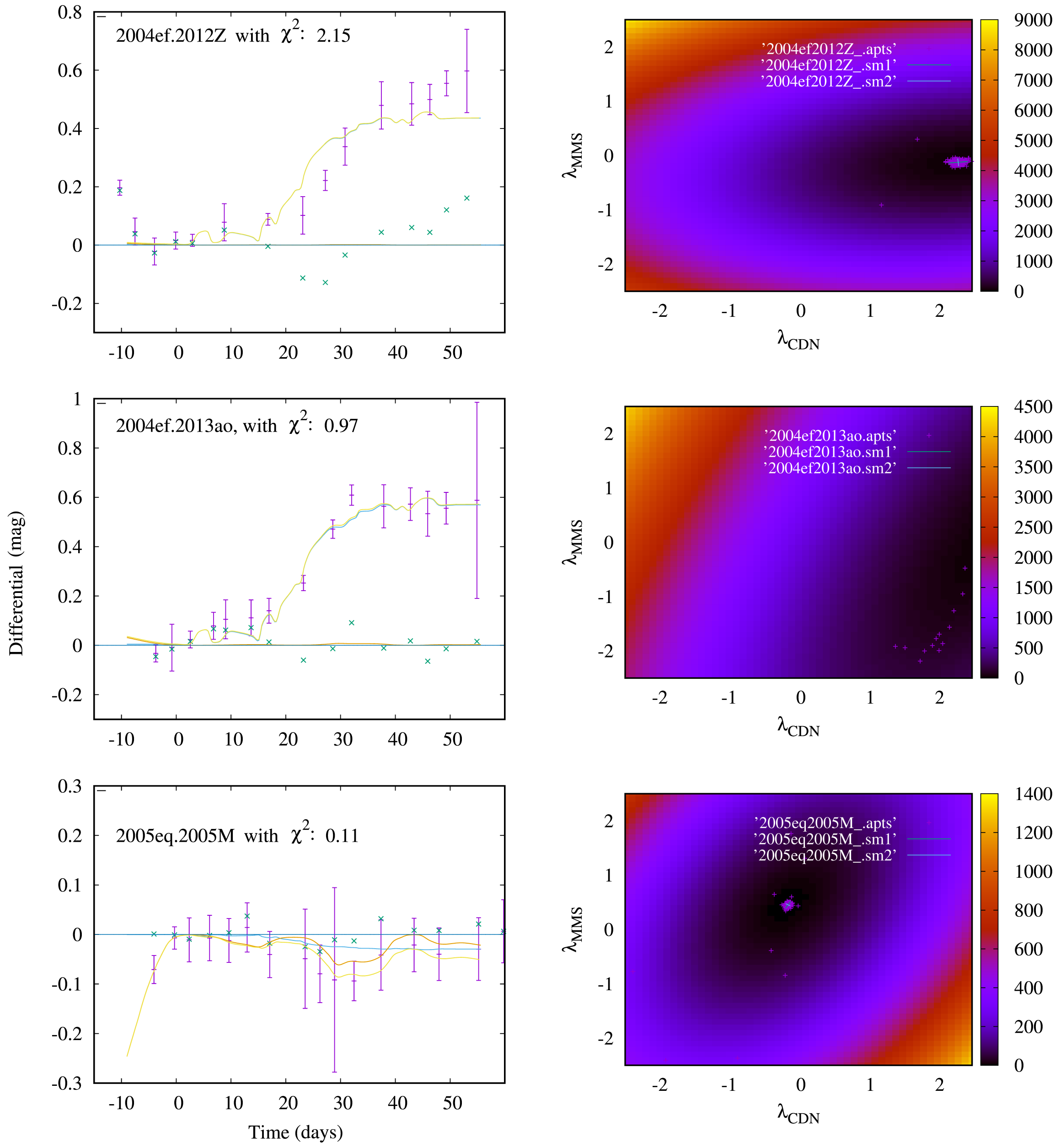

**Figure 12.** Same as Fig. 11 but for pairs with non-standard SNe Ia (see Sect. 4.1). For example, we show SN2004ef and SN2012Z - SN2012Z is a SN Iax; SN2004ef and SN2013ao - SN2013ao is a 03fg-like object, for this pair, the $\chi^2$ minimum is outside the frame, at (-4.5,-0.582); SN2005eq and SN2005M - both are 99aa-like objects.

The coefficients are related to the pairs but do not represent the property of the individual SN, see Equation 4. Therefore, we must solve all SNe simultaneously and determine the $g$ values.

We solve the overdetermined system of linear equations using the algorithm of Businger Golub (Businger & Golub 1965). The algorithm constructs a base of orthogonal solutions using eigenvalues, the so-called Householder algorithm (Householder 1958), and provides the optimal solution for the overdetermined system. The result is the most likely eigenvalue with error bars used for $g$. After solving for the coefficients, we can project them into the physical space.

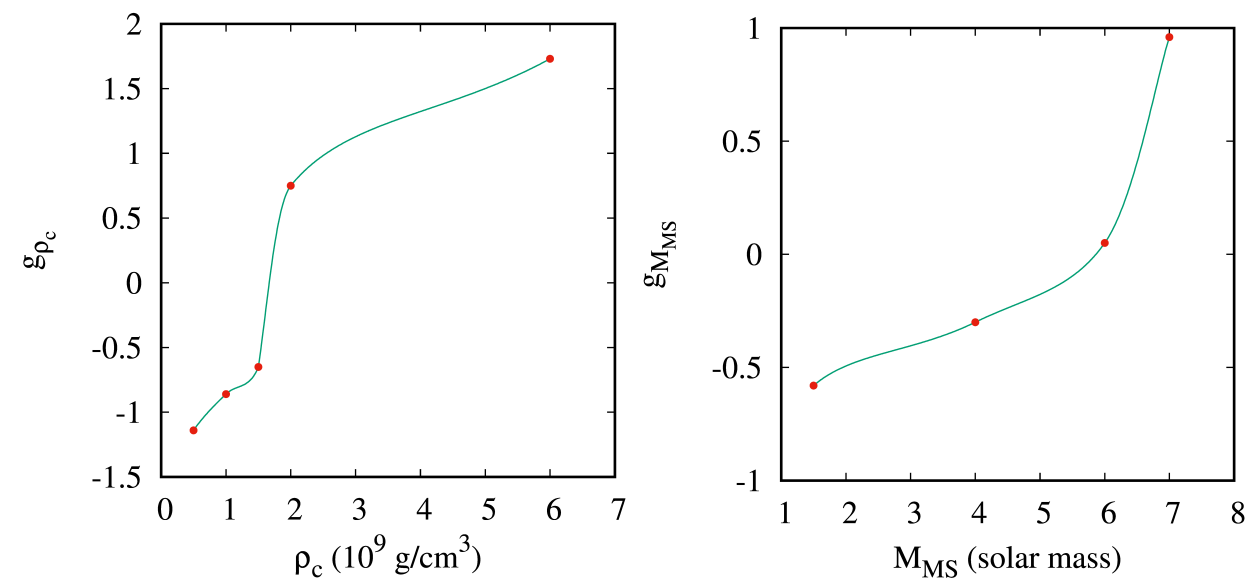

**Figure 13.** Theoretical relationship used to transform generic progenitor parameters into physical space (Sect. 3.5): g values as a function of $\rho_c$ (left) and $M_{MS}$ (right). The red points are based on specific theoretical models (Hoeflich et al. 2017a).

The signal coefficients have been reconstructed and their agreement with the measured signal coefficients has been checked.

### 3.6. *Projection of Generic LC Parameters into Physical Space*

The generic progenitor parameters are projected into the physical space using the theoretical models, see Fig. 13.

The steps for the projection are described below.

1. To transform the generic progenitor parameters into physical $M_{MS}$ and $\rho_c$ parameters, we use our prior knowledge: stars with $M_{MS}$ less than 1 $M_\odot$ have lifetimes longer than the age of the universe, stars with $M_{MS} > \sim 8$ $M_\odot$ become O/Ne WDs and core-collapse SNe, and stars with $\rho_c$ larger than $\approx 7 \times 10^9$ $g/cm^3$ will evolve to be neutron stars through Accretion Induced Collapse or AIC (Höflich et al. 1998b). Note that our limited sample will not span the full theoretical parameter space. Therefore, we use the Monte Carlo method to find the most likely upper and lower physical boundary values. A flat distribution within the error bar has been assumed for determining the most probable ranges. For our sample, we find 1.2 to 7.3 $M_\odot$ for $M_{MS}$ and 0.3 to $5.8 \times 10^9$ $gcm^{-3}$ for $\rho_c$. In Fig. 13, the relations between $g_i$ and the physical parameters are given using the full sample of SNe Ia.

2. The solution of the over-determined system provides the ratios, but $g_i$ can have an arbitrary factor depending on the SN sample or subsample. Here, we use four subsamples with $\approx$ 36 randomly selected SNe and eight common SNe, namely SN2004ef, SN2004gu, SN2005iq, SN2005ke, SN2005na, SN2006ax, SN2006gt, and SN2006X (Figs. 14 & 15). This latter allows testing the sensitivity of the projection from $g_i$ to the physical parameters for individual SNe Ia, as the physical properties of an SN should be sample-independent. We find good agreement (within the error bars) between the physical parameters within the different samples. Both the CSP-I and CSP II follow the same correlation in the full sample showing that our method is overall stable, and different data sets, here CSP-I and II, can be combined. Note that secondary parameter cross-correlations are neglected but, obviously, this does not affect our analysis within the uncertainties. This may hint that the two parameters describe independent physics: within the delayed-detonation scenario, the progenitor mass is a property of the progenitor whereas the central density is mostly given by the accretion rate, e.g. the properties of the binary system and the evolution of the companion star (Höflich et al. 2010).

## 4. RESULTS

There are a total of 161 SNe Ia in our sample (132 from CSP-II and 29 from CSP-I, see Sect. 3.3). The individual properties and parameters for these objects are listed in Tab. A1. **The relation to properties of the host galaxy and, for reference, the color information is given in Tabs. 2 and A3, respectively**.

### 4.1. *Identification of Non-standard Objects*

First, we run an initial analysis for the whole sample of 161 SNe and identify 29 non-standard SNe Ia. In this context, non-standard SNe Ia are given in Tabs. 3 & 4. The two groups are identified as follows:

- All pairs of 91T-like and some of the 99aa-$A$-like objects show both $\lambda_{i,j,l}$ in equation 2 being close to zero (case a). Moreover, all $M_{MS}$ are at the upper end, suggesting classes separate from normal-bright SNe Ia but with a short evolutionary time to form a WD.

- Outliers are defined by non-physical parameters for $\rho_c$ and $M_{MS}$.

#### 4.1.1. *91T and 99aa-like Supernovae*

91T-like SNe (Phillips et al. 1992; Jeffery et al. 1992; Filippenko et al. 1992) are objects at the bright extreme of SNe Ia and defined by their spectroscopic properties, showing very shallow optical Si II features (Branch 2001), with 99aa-like being less extreme (Garavini et al. 2004).

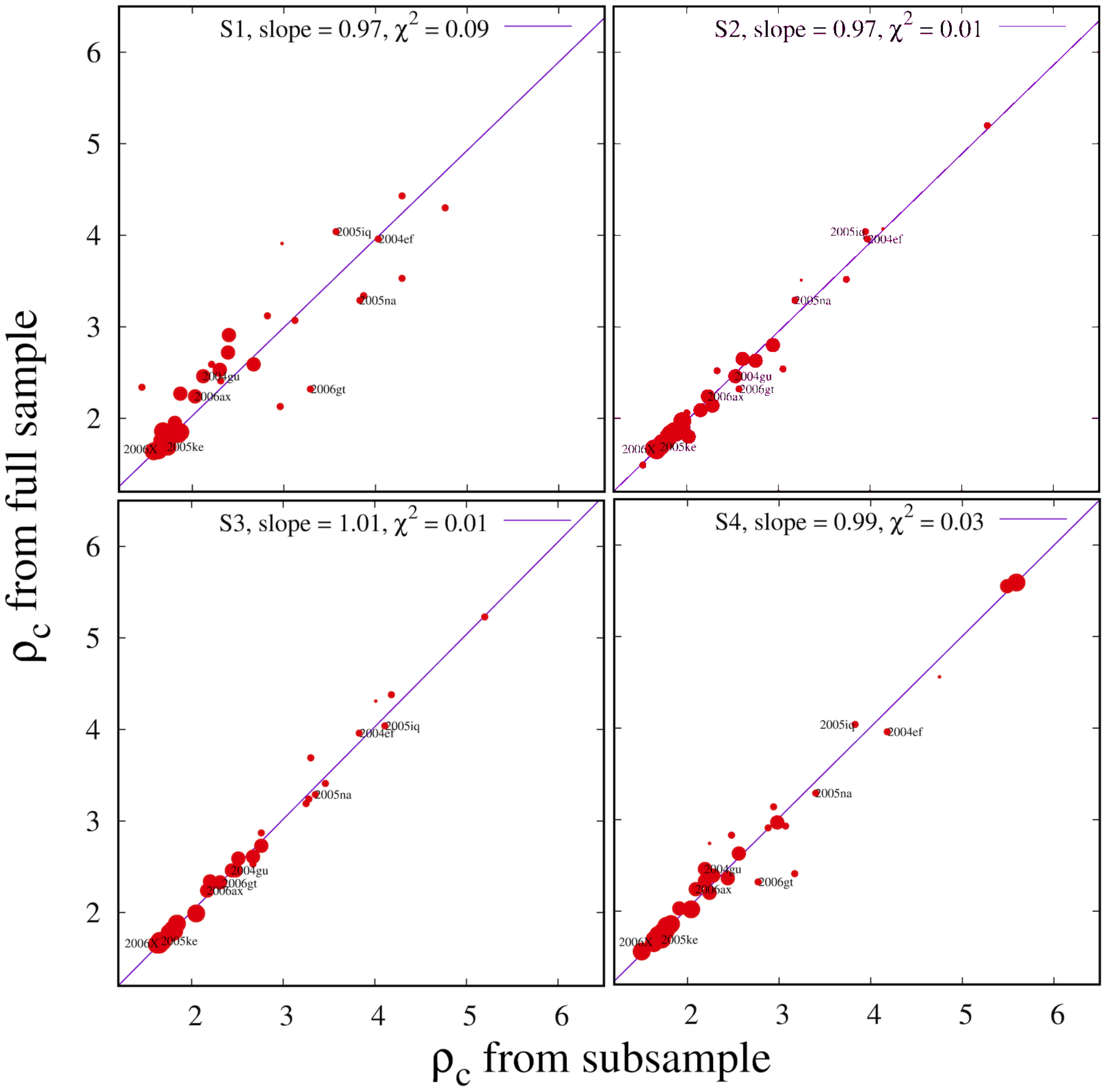

**Figure 14.** Comparison between physical secondary parameter $\rho_c$ values between our full sample and small sub-samples of ≈ 38 objects (red dots). The points with bigger error bars have smaller sizes and vice versa. The smallest and largest sizes correspond to an uncertainty of about 0.1 and 1.5 $\times 10^9$ $gcm^{-3}$, respectively. In addition, the 'common' SNe Ia are identified. The position of the latter and slopes of ≈ 1 validates our method (see text). Subsamples can be used to study correlations between SNe Ia type (Sect. 4) and their host galaxies (Sect. 4.3). Note that the range in $\rho_c$ realized is slightly smaller than the range physically possible because of the limited number of SNe Ia (see Sect. 3.6)).

Two groups with low $s$ have been identified with nearly identical secondary LC parameters (Fig. 16). These two groups are 91T-likes and "99aa-$A$-likes".

(1) Eight 91T-like SNe Ia, namely MASTER OT J093953.18+165516.4, CSS130303:105206-133424, OGLE-2014-SN-107, OGLE-2014-SN-141, SN2013U, ASASSN-14kd, SN2014eg, LSQ12gdj were identified with our method.

(2) A subgroup of five out of 11 total 99aa-likes are termed as "99aa-$A$-like" SNe Ia, namely

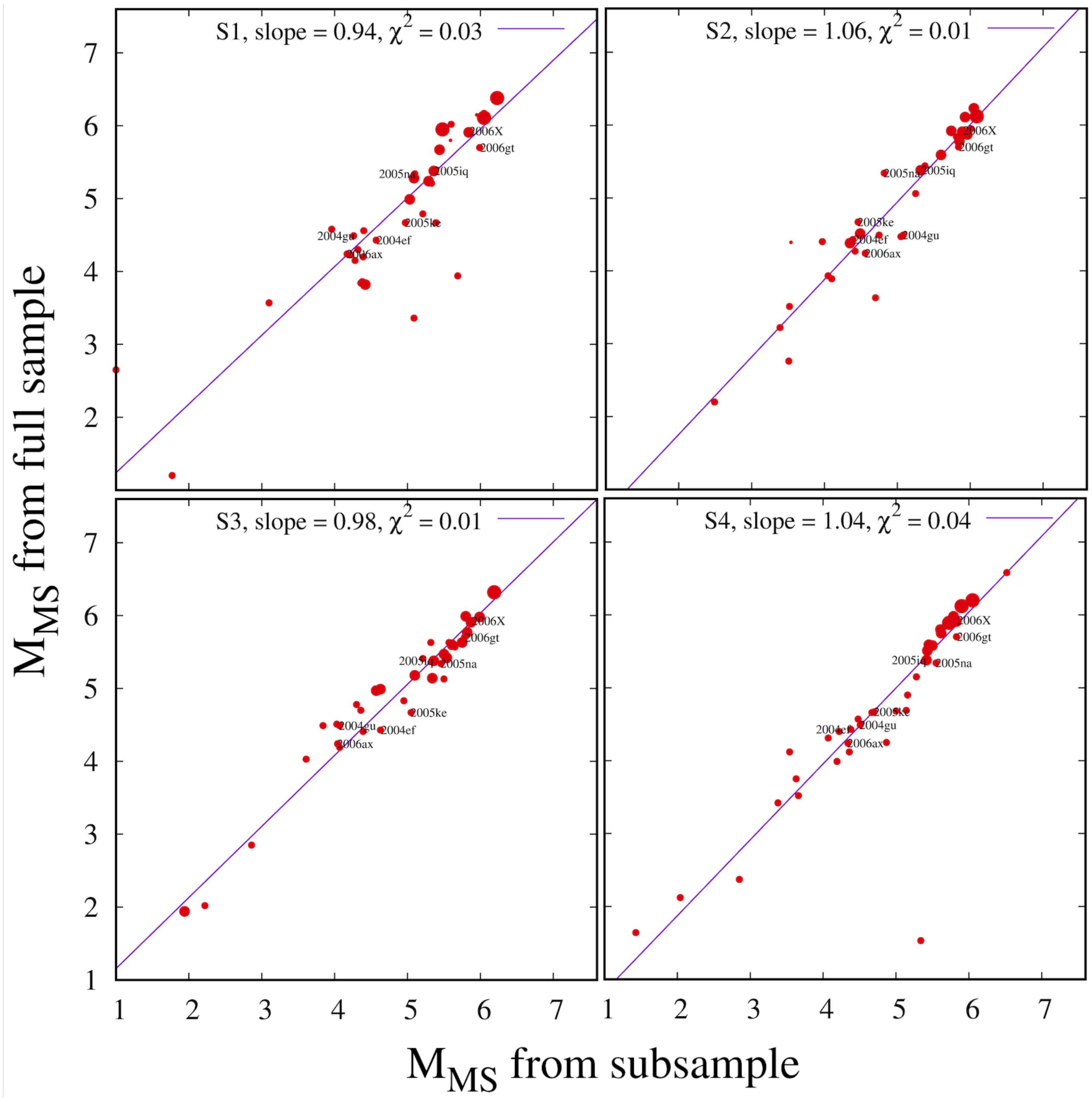

**Figure 15.** Same as Fig. 14, but for $M_{MS}$. The smallest and largest circles correspond to the uncertainty of about 0.4 and 3 $M_{\odot}$, respectively.

SN2004gu, SN2005M, SN2005eq, ASASSN-14lt, LSQ12hzj.

The six 99aa-like SNe Ia in our sample, which are non-99aa-$A$-like, will be subsequently called 99aa-$B$-like. These 99aa-$B$-like objects are SN2007S, SN2012G, ASASSN-14hp, ASASSN-14me, ASASSN-15as, and PS15sv.

91T-likes are clustered at $\rho_c \approx 1.5 \times 10^9$ $gcm^{-3}$ (or lower) and $M_{MS} \approx 6$ $M_{\odot}$ (Fig. 16), i.e. close to the lower and upper edge of the sensitivity and physical range obtained by Monte Carlo in our method, respectively (see Sect. 3.6). 99aa-$A$-like show a systematically higher density, $\gtrapprox 2{\times}10^9$ $gcm^{-3}$. Thus, in nebular MIR spectra, 99a-like SNe should show strong, isolated $Ni$ lines whereas 91T-like SNe should show no or weak $Ni$ lines (Fig. 2).

**Table 3.** The 91T-like (Phillips et al. 2022) and 99aa-like SNe Ia in our sample with their $M_{MS}$ ($M_{\odot}$), $\rho_c$ ($10^9$ $gcm^{-3}$), $s$, $\Delta m_{15,s}$, host galaxy name and type.

| Name | Type | $M_{MS}$ | $\rho_c$ | $s$ | $\Delta m_{15}(B)$ | Host Name | Host Type |
|---|---|---|---|---|---|---|---|
| MASTERJ093953.18+165516.4 | 91T-like | 4.66 | 1.65 | 0.92 | 1.18 | CGCG 092-024 | S |
| CSS130303:105206-133424 | 91T-like | 6.28 | 1.65 | 0.90 | 1.22 | GALEXASC J105206.27-133420.2 | S |
| OGLE-2014-SN-107 | 91T-like | 5.70 | 1.68 | 0.93 | 1.16 | APMUKS(BJ) B004021.02-650219.5 | Low $M_\star$ E |
| OGLE-2014-SN-141 | 91T-like | 4.45 | 1.55 | 1.07 | 0.88 | 2MASX J05371898-7543157 | S (bulge) |
| SN2013U | 91T-like | 5.63 | 1.70 | 1.05 | 0.92 | CGCG 008-023 | S |
| SN2013bz | 91T-like (*) | 5.91 | 1.79 | 0.95 | 1.12 | 2MASX J13265081-1001263 | S |
| ASASSN-14kd | 91T-like | 6.01 | 1.63 | 0.96 | 1.10 | 2MASX J22532475+0447583 | S |
| SN2014eg | 91T-like | 6.19 | 1.62 | 0.8 | 1.43 | ESO 154- G 010 | S |
| LSQ12gdj | 91T-like | 6.26 | 1.64 | 0.87 | 1.29 | ESO 472- G 007 | S |
| SN2004gu | 99aa-like-$A$ | 4.48 | 2.51 | 1.07 | 0.88 | FGC 175A | S |
| SN2005eq | 99aa-like-$A$ | 5.80 | 2.44 | 1.06 | 0.90 | MCG -01-09-006 | S |
| SN2005M | 99aa-like-$A$ | 4.09 | 2.76 | 1.10 | 0.82 | NGC 2930 | S0 |
| ASASSN-14lt | 99aa-like-$A$ | 4.85 | 2.55 | 0.95 | 1.12 | IC 0299 | S0 |
| LSQ12hzj | 99aa-like-$A$ | 4.91 | 2.84 | 1.01 | 1.00 | 2MASX J09591230-0900095 | E |
| SN2007S | 99aa-like-$B$ | 4.48 | 1.89 | 1.03 | 0.96 | UGC 5378 | S |
| SN2012G | 99aa-like-$B$ | 5.86 | 1.87 | 1.10 | 0.82 | IC 0803 NED01 | S |
| ASASSN-14hp | 99aa-like-$B$ | 4.33 | 1.84 | 1.06 | 0.90 | 2MASX J21303015-7038489 (LEDA 127270) | S |
| ASASSN-14me | 99aa-like-$B$ | 2.37 | 1.77 | 1.05 | 0.92 | ESO 113- G 047 | S (bulge) |
| ASASSN-15as | 99aa-like-$B$ | 3.44 | 1.83 | 1.13 | 0.75 | SDSS J093916.69+062551.1 | - |
| PS15sv | 99aa-like-$B$ | 5.89 | 1.80 | 0.95 | 1.12 | GALEXASC J161311.68+013532.2 | - |

NOTE—ASASSN-14kd has only two data points after +20 days, therefore distinguishing this object as a 91T-like from this analysis is difficult. SN2013bz resembles the 99aa-$A$-like SNe from our analysis, although it is a 91T-like object (Phillips et al. 2022). Please see Sect. 4.1.

Our LC-based classification of 91T-likes identification is consistent with that of Phillips et al. (2022) [5] with the possible exception of SN2013bz. Though this object is located close to the 91T-like objects, it has a somewhat larger $\rho_c$. This difference becomes even more obvious when comparing the differential SN2013bz to all other 91T-like objects. They show an offset in the LC tail very similar to the 99aa-$A$- vs. 91T-like objects. As discussed in Sect. 3.2, high-reddening or, here uncertainties, may mimic a change in $\rho_c$, in particular, if there is no observation beyond day 40 after maximum. Indeed, the LC of SN2013bz ends at $\approx$ 35 days with $E(B-V) = 0.257$ mag and $R_V = 2.35$ based on SNooPy fits (Tab. A2). However, $E(B-V)$ is typical for 91T-likes, and with $\Delta V(t, E_{B-V}, \Delta R_V, z)(t) \approx 0.01$ mag, we regard uncertain reddening as an unlikely explanation for this anomaly.

In Fig. 16, 99aa-$A$-like objects are systematically shifted to larger $\rho_c$ suggesting 91T and 99aa-$A$ being separate classes. Both are different from 99aa-$B$-like, which show a low density but a wide spread in $M_{MS}$ and, thus, stellar lifetimes. **We find no dependence of the intrinsic properties of 91T- and 99aa-like objects on the reddening (Tab. 16)** or systematic uncertainties (see Sect. 3.2). Moreover, in the CSP-sample, 91T-like SNe in the CSP-II sample seemed to be slightly more reddened by host dust than the 99aa (Table 1 in Phillips et al. 2022). Our results indicate that the differences in the physical properties are intrinsic to the objects, and not caused by the nearby environment.

From observation and the CSP sample in combination with projecting the 91T- and 91aa-likes to the Hubble flow, Phillips et al. (2022) found two main results: 91T-like are brighter by $\approx$ 0.3mag than 99aa-like, and $\Delta m_{15}(B)$ becomes flat at the bright-extreme of all spectroscopic classes (Figures 7 and 5 of Phillips et al. (2022), respectively). From theory and within the delayed detonation scenario, the latter can be understood as follows: For delayed-detonation models (Hoeflich et al. 2017a; Aldoroty et al. 2023), the brightness-decline-rate relation is dominated by the amount of burning during the deflagration phase because it governs the expansion of the WD. With less burning, the interface between $^{56}$Ni (nuclear statistical equilibrium or

[5] Phillips et al. (2022) who found ten 91T-like objects. SN2014dl was deselected from our sample because of insufficient LC coverage in time (Sect. 3.3 & Sect. 4.1).

NSE) and partially burned material is moved increasingly to the outer, low-mass layers resulting in hardly any gain for $^{56}$Ni production, i.e. the brightness. Decreasing deflagration burning (or increasing transition density $\rho_{tr}$ from deflagration to detonation) does not boost the brightness at the brightness extreme. However, brightness is boosted by the low $\rho_c$ because it adds $^{56}$Ni in the central layers (Fig. 3). To a lesser amount, the larger C-poor core due to high masses (Fig. 1) results in lower explosion energy and, with it less energy loss due to expansion work. Based on delayed-detonation models used in the work, the increase is of the order of $\approx 0.2 - 0.3$ mag (see also Hoeflich et al. (1998, 2017a)) [6]. Note that in Sect. 8.2, late-time nebular spectra are required to test this interpretation (see Sect. 8.2).

**As discussed above, the overall distribution of SNe Ia is weighted towards higher masses and lower densities, including both delayed detonations and most alternative scenarios (see Sect. 7).** Note, e.g. the 'loose' SN cluster of 99aa-A-like, underluminous, and normal-bright SNe Ia near values of $\rho_c$ $2.7 \times 10^9 g/cm^3$ and $M_{MS}$ $4.1 M_\odot$ which have very different colors (Tab. B) and spectroscopic properties. This suggests that very different subtypes have overlapping progenitor properties, but not that the underlying abundance and density structure of the explosion is the same (Figs. 3 and 5 of Höflich et al. (2002)). To first order and within delayed-detonation models, the **brightness and color (B-V) at maximum is determined by the amount of deflagration burning, the primary parameter. For a given $dm_{15,s}V$, the dispersion in these quantities is $\approx 0.2^m$ (see e.g. Fig. 8 in Hoeflich & Khokhlov 1996 and, for CSPI-SNe, Figs. 3-4 of Hoeflich et al. 2017a).** [7]

**It is obvious from the outliers that our models cannot account for all the observed SNe Ia in this sample. For physical reasons, see Sect. 4.1.2. We should also note that just because our models fit the $V$-band LC data used in this analysis does not imply that these SNe are the result of a delayed-detonation explosion. A combination of flux and polarization spectra is needed (see Sect. 2) which is available only for a limited number of objects.**

[6] At the brightness extreme, the LCs are slower rising, giving more time for energy to diffuse out.

[7] Changes in the density profiles by, e.g. pulsational delayed detonation models as suggested for 91T-likes mostly affect deviations from the luminosity decline relation in B (Hoeflich et al. 1994). For most scenarios and to first order, the $^{56}Ni$ mass dominates the decline because V resembles the bolometric LC (see e.g. Fig. 8 in (Hoeflich & Khokhlov 1996)). Spectra are required to probe the diversity.

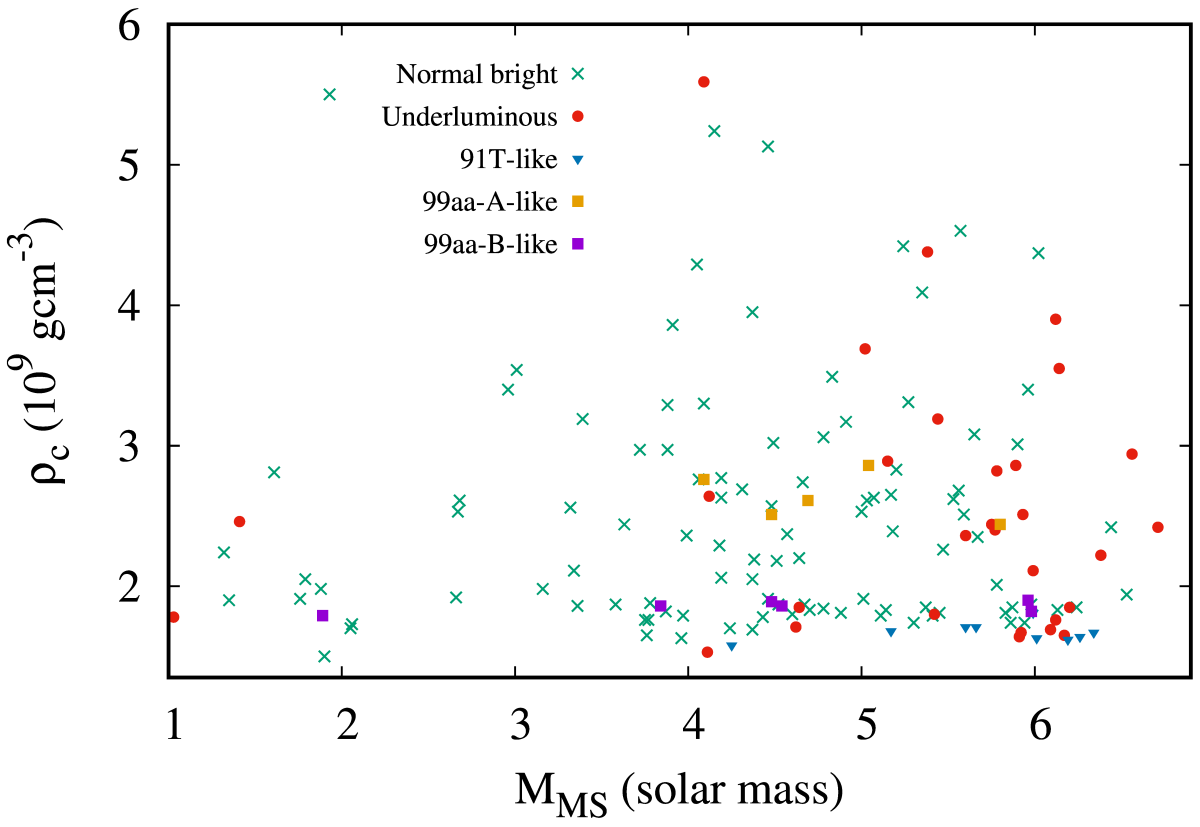

**Figure 16.** $M_{MS}$ and $\rho_c$ correlation plot for all 152 SNe in our sample. Normal-bright, underluminous, 91T-like, and the two groups of 99aa-likes are marked with different symbols. **Note that the location of 91T- and 99aa-like SNe shows no statistically significant difference between those with high and low reddening.** The general trend of SNeIa is towards higher $M_{MS}$ and lower $\rho_c$ (Sect. 4.2.3).

#### 4.1.2. *Outlier Supernovae*

(1) Nine objects among the 29 "non-standard" show values for $M_{MS}$ or $\rho_c$ outside the physical range defined in Sect. 3.6. Only these objects were excluded from further analysis. Note that this may indicate explosion scenarios different from the delayed-detonation scenario used here, or scenarios that produce eigenfunctions non-similar to our models (see Sect. 3.2). **From theory, one cause for 'non-physical' parameters is the presence of a significant amount of additional unburned material, which alters the specific explosion energy similar to $M_{MS}$ but even more significantly, e.g. as found in dynamical mergers, pulsating delayed-detonations and the core-degenerate scenarios (see Sect. 7).**

(2) Among nine outliers in our sample, two have been identified as 03fg-like objects (03fg-like) which require models between 0.1 and $0.7 M_\odot$ of unburned C/O (Hsiao et al. 2020; Lu et al. 2023), and three

**Table 4.** Outlier SNe identified by parameters being non-physical (see Sect. 4.1). The name, classification, and reference for the classification are given in columns 1, 2, and 3, respectively (see text).

| SN Name | Classification | Reference |
|---|---|---|
| KISS15m | 91bg-like | (Ashall et al. 2020) |
| ASASSN-15hy | 03fg-like (03fg-like) | (Ashall et al. 2020) |
| SN2013ao | 03fg-like | (Ashall et al. 2020, 2021) |
| SN2014ek | SN Iax | CSP website |
| SN2012Z | SN Iax | (Ashall et al. 2020) |
| SN2013gr | SN Iax | Ashall, private communication |
| ASASSN-15go | Normal-bright | Physical parameter outside range |
| SN2012bl | Normal-bright | Physical parameter outside range |
| OGLE-2014-SN-019 | Normal-bright (bad-sampling) | (Ashall et al. 2020) |

as underluminous 02cx-like (SNe Iax) (Pakmor et al. 2013). [8]

(3) One SN among these outliers has sparse sampling past 20 days.

(4) Three SNe among the outliers have parameters outside the physical range (Sect. 3.6) possibly suggesting a different explosion scenario: The normal-bright SNe, SN2012bl, and ASASSN-15go have well-sampled LCs; KISS15m, a 91bg-like object, is poorly sampled having only two observations in region two (Tab. 1) and followed by a gap for some 10 days.

### 4.2. *Distribution of the Physical Secondary Parameters - Results from the LC Analysis*

The following analysis of 152 objects is based on $M_{MS}$, $\rho_c$, and $s$ (equivalent to the brightness-decline rate $\Delta m_{15}(B)$) as luminosity indicator for SNe Ia observed. To avoid bias, outliers are excluded from the analysis. Commonly, different brightness classes of SNe Ia are based on the $B$ color (Burns et al. 2018). In this work, we define $\Delta m_{15}(B) > 1.45$ mag separating the normal-bright from underluminous SNe Ia. The relation between integrated SN properties and their hosts is presented by binning the number of realizations in the parameter space.

For quantifying their significance, Gaussian statistics has been employed. **In the text, relations with probabilities less than $2\sigma$ are referred to as "indication", greater than $2\sigma$ as "strong indication", and greater than $3\sigma$ as "evidence".**

Fig. 17 shows the general distribution of normal-bright and underluminous SNe Ia. Most SNe Ia originate from the higher and lower-end of $M_{MS}$ and $\rho_c$, respectively.

None of the residuals in our LC fits of SNe Ia pairs with late-time coverage show a signal expected from uncertainties in the reddening. Any systematics hidden in the k-correction applied would have 'popped up'.

#### 4.2.1. *$M_{MS}$ Distribution*

The SNe distributions indicate two broad maxima around 4 $M_\odot$ and 5.9 $M_\odot$ (Fig. 17) suggesting two populations with typical stellar evolution timescales of $\approx 200-300$ million and $\approx 65$ million years, respectively (see Dominguez et al. 1999). It is strongly indicated that only some $\approx 10\%$ have progenitors with $M_{MS} \lesssim 3\ M_\odot$ and stellar evolution timescales in excess of 500 million years.

There is a strong indication of most underluminous SNe Ia in our sample having a very short stellar evolutionary time (less than 100 million years). **Because underluminous SNe Ia are found in old galaxies (e.g. Uddin et al. 2020), we suggest** that the progenitor system evolution contributes most to the delay time between star formation and explosion (Sect. 2, Equation 1).

#### 4.2.2. *$\rho_c$ Distribution*

**The $\rho_c$ distribution shows evidence of clustering towards the lower end with $\rho_c \lesssim 3 \times 10^9 g/cm^3$ (Fig. 17).** Note that $\rho_c$ depends on the size of the central $^{56}$Ni-free hole and loses its sensitivity for $\rho_c \lesssim 2 \times 10^9\ gcm^{-3}$. Therefore, the low $\rho_c$ SNe Ia may include SNe Ia hydrogen-accretor, He-accretion, and direct C/O accretion for secular mergers (see Sect. 2). Our method does not allow separation between WD masses below $\approx 1.34\ M_\odot$ (Fig. 3). For further separation, we have to rely on nebular spectra in the near IR, namely the [Fe II] at 1.644 $\mu m$ (e.g. (Höflich et al. 2004; Diamond et al. 2015; Graham et al. 2017; Diamond et al. 2018a; Graham et al. 2017) and, in particular, on the direct detection of a significant amount of EC elements such as $^{58}$Ni which requires high-density burning above $\approx 5 \times 10^8$ $gcm^{-3}$ (Fig. 2), and commonly observed in the NIR and MIR of both normal bright and subluminous SNe Ia as discussed in Sect. 2. SNe Ia with $\rho_c$ in the higher end can be seen, but only a few, as expected from Galbany et al. (2019). The fraction of the sample that falls in the tail with $\rho_c > 3 \times 10^9$ $gcm^{-3}$ is $\approx 17\%$.

[8] CSP13abs (MLS140102:120307-010132), no. 89 in Tab. A1 has been marked as peculiar (03fg-like by Ashall et al. (2020); Lu et al. (2023)) and has a maximum brightness of −19.2 mag (CSP-data base). It has not been identified as an outlier because of the lack of early LC points combined with the post-maximum decline (one point but small error bar), resulting in a degeneracy in $M_{MS}$ and $\rho_c$. Note that including this SNe Ia does not affect our analysis.

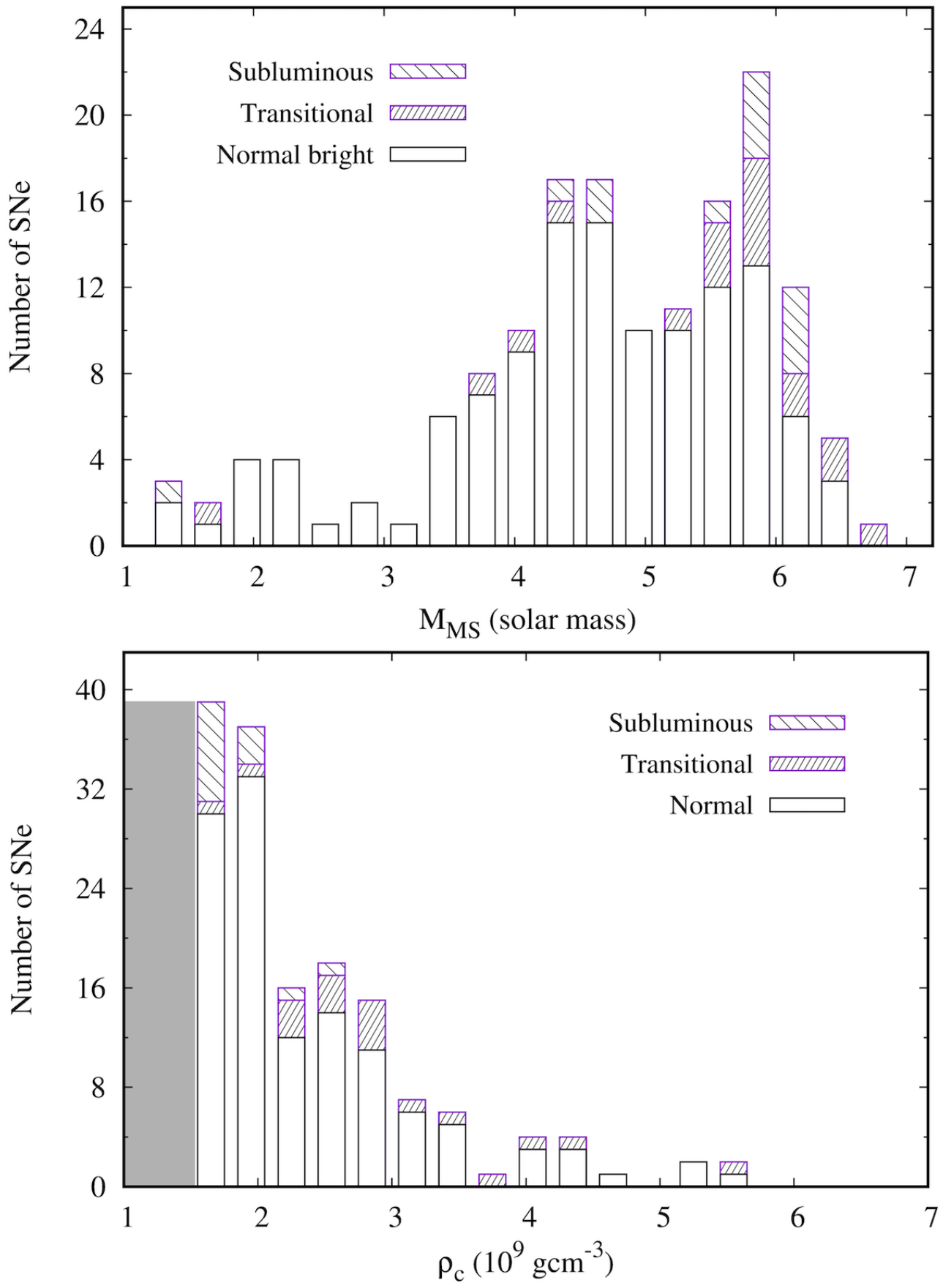

**Figure 17.** Main sequence mass (top) and central density (bottom) distributions for all SNe in our sample, CSP-I, and II combined (Sect. 4.2). The patterns indicate normal ($\Delta m_{15}(B) \leq 1.45$), transitional ($1.45 < \Delta m_{15}(B) < 1.7$) and subluminous ($\Delta m_{15}(B) \geq 1.7$) SNe numbers in each bin. **The confidence level of the overall distribution is $\approx 99\%$, of the two maxima in $M_{MS}$ are $\approx 97$ and $93\%$, separated by a minimum on a $\approx 80\%$ confidence level (upper plot). The shaded area (lower plot) at the low end of $\rho_c$ indicates the region where the variation in differentials becomes comparable to the data accuracy because little EC elements are produced. Note that this region covers both the lower end of delayed-detonation models (e.g. (Diamond et al. 2015; DerKacy et al. 2023) and many other scenarios without high-density burning (Sect. 7).**

#### 4.2.3. *Other Correlations*

Fig. 16 shows the correlation of $M_{MS}$ and $\rho_c$ as well as SNe-brightness. The ratio between normal-bright and underluminous SNe Ia is $\approx 4$. We find no evidence for a correlation between $M_{MS}$ and $\rho_c$ for both normal-bright and underluminous SNe Ia in our sample. However, we find that 91T-like and 99aa-A-like SNe Ia show very similar $M_{MS}$ and $\rho_c$ and $s$ for many objects in the sample as discussed in Sect. 4.1 [9].

#### 4.2.4. *Stability of the trends of the secondary parameter distributions*

**The trends and distributions discussed are stable within 1 to 3% when restricting the sample to CSP-II only. The number of objects in the CSP-I subsample is only 29 and details are lost.**

### 4.3. *Including Additional Information: Host Galaxies*

Additional information about the host galaxy allows for defining subgroups. We subdivide our SN sample into hosts with active star formation, namely spiral, irregular, S0, and small ellipticals, and without active star formation, namely giant ellipticals , and perform the LC analysis on these subsamples. In the presence of host-SNe correlations, we may expect some of the signatures in the distribution (Sect. 4.2) may be enhanced in one sample and weaker in the other, e.g., if the timescales towards the explosion are dominated by the stellar evolution, the peak at high $M_{MS}$ can be expected to become more prominent in spirals than in giant ellipticals.

Information about the morphological classification of the host galaxies has been used mainly from NASA/IPAC Extragalactic Database (NED), SIMBAD Astronomical Database [10], IRSA and SDSS images, HyperLEDA [11], and white images obtained from collapsing integral field spectroscopy from the PISCO (Galbany et al. 2018) and AMUSING (Galbany et al. 2016) compilations. The host stellar mass information is obtained from Uddin et al. (2020, 2023).

Among the 152 SNe in our sample, we were able to acquire morphological information of the host galaxies for 133 SNe Ia. See Tab. A2 for a detailed breakdown of the numbers of different SN-brightness and host galaxy types.

[9] Note that, by chance, some individual SNe Ia pairs are expected to have similar parameters, but not for a significant fraction of the sample.

[10] https://simbad.u-strasbg.fr/simbad/sim-fid

[11] http://leda.univ-lyon1.fr/

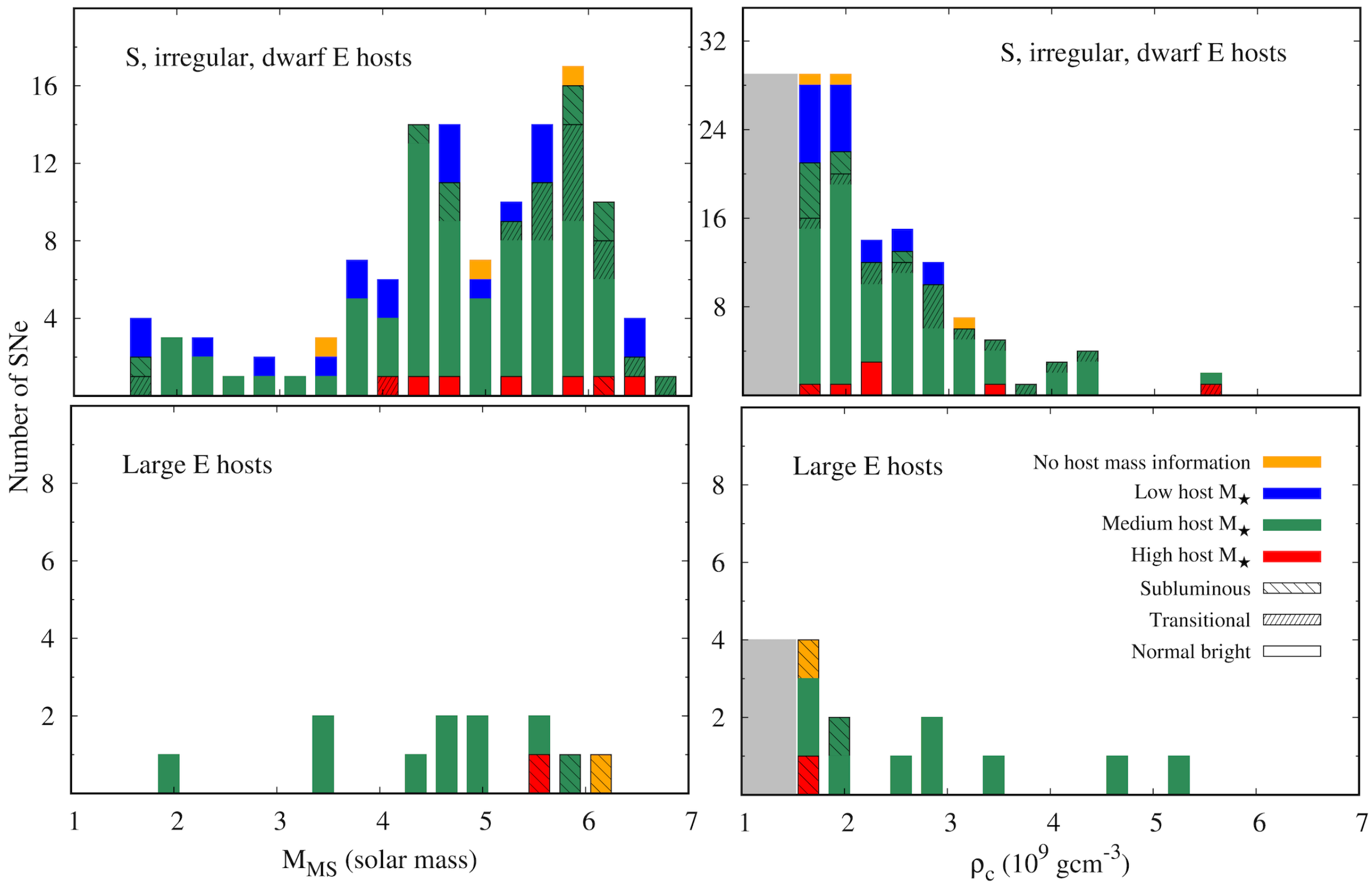

**Figure 18.** Main sequence mass (left-hand side) and central density (right-hand side) distributions for all SNe Ia used from CSP I and II (Sect. 4.3). The upper row shows SNe in star-forming (spirals, irregulars, and dwarf ellipticals) hosts, and the bottom row shows SNe in older hosts (large ellipticals). Different colors refer to the stellar mass ($M_\star$) ranges of the host galaxies, and the patterns indicate whether the SNe are normal ($\Delta m_{15}(B) \leq 1.45$), transitional ($1.45 < \Delta m_{15}(B) < 1.7$) or subluminous ($\Delta m_{15}(B) \geq 1.7$). The shaded parts in the $\rho_c$ distributions indicate the number of SNe Ia with central densities in the region insensitive to our method (see text). Note that, based on NIR and MIR line profiles, a fraction of $M_{Ch}$ models has low densities (see Diamond et al. 2015; Hoeflich et al. 2021; DerKacy et al. 2023).

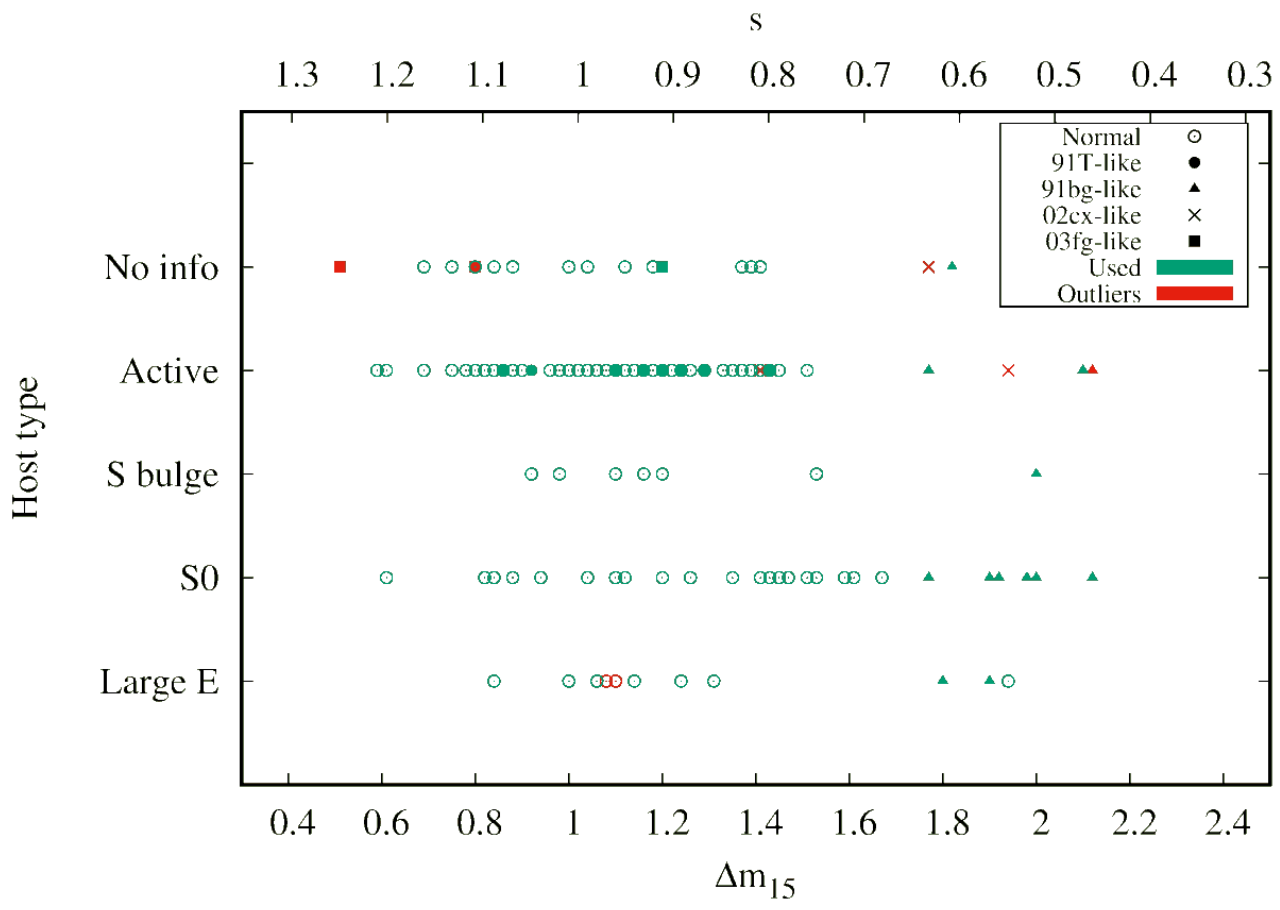

**Figure 19.** We show our $\Delta m_{15}(B)$ values and host galaxy types of all SNe Ia in our sample, including the outliers (used SNe Ia in green, outliers in red).

### 4.3.1. *SNe Ia in Hosts with Ongoing Star Formation*

All relations with respect to $M_{MS}$ and $\rho_c$ found in Sect. 4.2 are confirmed (upper plot, Fig. 18), but the existence of two peaks in $M_{MS}$ rises to ≈ 95%. This strengthens the existence of two populations with the implication with regard to the peak brightness (see Sect. 5). About 11% of the objects have a $M_{MS} < 3\ M_\odot$ and about 18% of the objects have $\rho_c > 3 \times 10^9\ gcm^{-3}$, which is consistent with Sect. 4.2.

### 4.3.2. *SNe Ia in Hosts with Potentially Little to no Ongoing Star Formation*

The $M_{MS}$ distribution is flat with a ≈ 15% probability for 2 peaks, though, low progenitor masses and, thus, long stellar evolutionary times are rare but still present (lower plots, Fig. 18). The presence of SNe Ia with high $M_{MS}$ shows the presence of systems with long evolution times of the progenitor system, i.e., from the formation of the WD to the explosion.

Therefore, in old progenitor systems, the delay time is dominated by the progenitor system (see Equation 1). Assuming that the number of systems of low mass is about the same in all hosts, the reduction of the ratio between high/low-mass $M_{MS}$ from ≈ 1 to 2.5 in giant ellipticals and other galaxies may indicate progenitor evolutionary times in about 20 to 30% of all SN Ia systems.

We see some indication for a shift in $\rho_c$ towards larger $\rho_c$ but with albeit low significance of ≈ 60% (Fig. 18).

More progenitors may be found in the galactic bulge. However, we found that the resulting $M_{MS}$ and $\rho_c$ distributions to be consistent with the total distribution (Fig. 17) because the number of SNe Ia projected on the bulge outnumbers those in the bulge.

#### 4.3.3. *Underluminous SNe Ia*

Our sample includes 12 subluminous, 91bg-like, and 15 transitional SNe Ia (Tab. A2, Fig. 18). Most underluminous SNe Ia have **progenitors with high $M_{MS}$ compared to our total sample indicating long evolution times for the system on a 98% level**. As discussed in the next section (Sect. 5), this is consistent with the number of underluminous SNe Ia decreasing with redshift (González-Gaitán et al. 2011). For transitional SNe Ia, typical high $\rho_c$ are observed making up the majority in our entire sample. This is consistent with detailed analyses in the literature, e.g. SN2016hnk and SN2020qxp (Galbany et al. 2019; Hoeflich et al. 2021). Note that our two transitional SNe Ia in actively star-forming galaxies are the only ones showing low $\rho_c$, which may suggest two distinct classes (please see Sect. 5). Underluminous SNe Ia seem to be clustered towards low $\rho_c$ in contrast to the transitional objects.

#### 4.3.4. $\Delta m_{15}(B)$ *and* $s$ *vs. the Host Galaxy Morphological Groups*

$\Delta m_{15}(B)$ from Krisciunas et al. 2017; Phillips et al. 2019 and $s$ vs. the host galaxy morphological groups of all SNe Ia in our sample, including the non-standard SNe, is shown in Fig. 19.

No particular trend between the $\Delta m_{15}(B)$ and $s$ values vs. the host galaxy morphological groups was observed for normal-bright SNe Ia.

23 among the total 27 underluminous SNe Ia (including 91bg-likes) used for the physical parameter distributions appear in potentially low star-forming galaxies (confirming our observation in Sect. 4.3.3). Therefore they have a long delay time.

91T-like SNe Ia are at the bright side of $\Delta m_{15}(B)$ range. They come mostly from hosts with ongoing star formation (Fig. 19), so their progenitor system evolution is fast.

Among the outliers, 03fg-likes are in the normal bright $\Delta m_{15}(B)$ range. They come from young hosts or active parts of older hosts (private communication with Jing Lu, also see Lu et al. 2021), see Tab. A2. For the three 02cx-likes, host information was unavailable for one and the other two are in active host galaxies. All three of these objects have higher $\Delta m_{15}(B)$ and $s$ values.

## 5. GENERAL DISCUSSION AND LIMITATION

We present fits of SNe Ia observations obtained by the CSP I and II. Monochromatic LCs are commonly fitted by templates using the stretch $s$ or $\Delta m_{15}$ as parameter (Perlmutter et al. 1999; Jha et al. 2006). Our method to construct secondary corrections is based on eigenfunctions related to the physical parameters of specific explosion models. We find and discuss many relations between the physical properties and the characteristics of the LCs, and the sample size required to improve probabilities for the relation [12].

Note that our method is mostly independent of reddening because it makes use of individual LCs. We use the V-band as it is less affected by metallicity than the $u$ or $B$-bands (Hoeflich et al. 1998; Fisher 2000), see Sect. 3). Based on the $V$-band, our method allows detecting non-standard SNe Ia such as 91T-like, 02cx-like, and 03fg-like in the sample (Sect. 4.1). Note that 91T-like supernovae not showing correction by secondary effects in either $\rho_c$ or $M_{MS}$ suggests a narrow path to and homogeneous group of, likely, small $\rho_c$ and little EC elements. This is also supported by the peaked 1.644 $\mu m$ forbidden [Fe II] feature (Meikle et al. 1996). Though the outer region is consistent with the bright range of classical delayed-detonation models with a short period of deflagration burning only (Hoeflich et al. (2017a) and Phillips et al. in preparation), it points towards a separate class of explosions.

### 5.1. *Analysis in the Framework of* $M_{Ch}$ *Explosions*

In the framework of the delayed detonation models, the secondary eigenfunctions have been related to two physical properties, namely $M_{MS}$ and $\rho_c$ (Fig. 4, Sect. 2 and 3.3). $M_{MS}$ causes variations in the explosion energy and provides a measure of the stellar evolutionary times. $\rho_c$ is related to $M_{WD}$ at the time of the explosion and depends on the accretion rate rather than mixing during the runaway (Sect. 2).

Our work shows that the regions of early and late LC data are key in determining the secondary parameters (Tab. 1), therefore photometry data up to ~ 60 days relative to the V-maximum are necessary for this purpose (Sect. 3.3). This is important for high-precision cosmology.

Using the entire sample, we see evidence that most of the progenitors come from high $M_{MS}$ with, over-all, short stellar evolutionary time. However, there was an indication for two maxima in the $M_{MS}$ distribution which implies two populations, a prompt and those with a delay-time of $\approx 2\times10^8$ years (Sect. 4.2.1). Even though the majority of the SNe Ia have low $\rho_c$, the $\rho_c$ distribution extends up to the high densities at which WDs approach the regime of AIC (Nomoto & Kondo 1991; Nomoto 1982b; Nomoto et al. 1984), see Sect. 4.2.2.

[12] **For a comprehensive list, see Chakraborty (2023).**

### 5.2. *Addition of Host Galaxy Information*

Taking the host galaxy into account allows us to make a connection between the location of the SNe Ia and the properties of the progenitors and, to improve the relations by adding information.

We found our results from LC analysis are stable both with respect to $M_{MS}$ and $\rho_c$ , in particular with respect to galaxies with ongoing star formation Sect. 4.3.1.

For SNe Ia in host galaxies without star formation, the results are consistent but hampered by the low number of objects. However, SNe Ia in host galaxies without ongoing star formation allowed us to separate the stellar evolution time from the system evolution time when considering the presence or absence of objects in the $M_{MS}/\rho_c$ space (Sect. 4.3.2).

#### 5.2.1. *Underluminous SNe Ia*

In our sample, most underluminous SNe Ia have hosts with low star formation (Sect. 4.3.3), consistent with earlier studies based on SN/galaxy relation (Filippenko 1989; Branch & van den Bergh 1993; Gallagher et al. 2005, 2008; Howell et al. 2009; Crocker et al. 2017; Panther et al. 2019; Hamuy et al. 1996; Howell et al. 2001; Wang et al. 2007; Patat et al. 2012). Only some 15% of the underluminous objects have hosts with active star formation (Sect. 4.3.3) and many of those may have different secondary LC parameters.

Most subluminous SNe Ia show $\rho_c$ similar to the SNe Ia population. However, most of the transitional SNe Ia have high $\rho_c$. Our finding is consistent with the spectral analysis in the literature (e.g. Galbany et al. 2019). As an exception, the two transitional SNe Ia in active galaxies have lower $\rho_c$. This might suggest at least two classes of transitional SNe Ia with different explosion or progenitor channels. Combining the above information, underluminous SNe Ia are a diverse group.

Our $M_{MS}$ distribution leans towards the higher end with a short stellar evolution time (Sect. 4.2.1) but being in hosts with low star formation, strongly suggests long evolution times of the system. This implies that, in general, a non-degenerate donor should have low mass, or, underluminous SNe Ia originating from secular mergers (Blanc & Greggio 2008; Han & Podsiadlowski 2003, 2004; Greggio 2005), see Sect. 4.3.3.

### 5.3. *Non-standard SNe Ia*

SNe Ia that showed no improvement to the LC fits after implementing the secondary parameter variations are 91T-likes, and their $M_{MS}$ is at the upper end of WDs SNe Ia. They come mostly from hosts with ongoing star formation (Sect. 4.3.4, Fig. 19), which is in agreement with studies such as Hamuy et al. (2000); Howell et al. (2001); Phillips et al. (2022); Taubenberger (2017), and Phillips et. al, in preparation. In summary, 91T-likes are a separate class of objects with short evolutionary time scales for both the progenitor and the system evolution.

In literature, 99aa-like SNe are often linked with 91T-like objects (Greggio 2005; Phillips et al. 2022) raising whether both classes are related. From our analysis (Tab. A1) and based on the differentials, we find two distinct subclasses: a) 99aa-$A$-like SNe Ia (SN2004gu, SN2005M, SN2005eq, SN2013bz, ASASSN-14lt, LSQ12hzj show high $M_{MS}$ similar to 91T-likes but, systematically higher central densities; b) 99aa-$B$-like (SN2007S, ASASSN-14hp, ASASSN-14me, ASASSN-15as, PS15sv, SN2012G) which are bright but with spread in $M_{MS}$ and $\rho_c$ not distinguishable from SNe Ia at the bright end of SNe Ia. Please see Tab. 3 for a list of our 91T-like and 99aa-like SNe, along with their secondary parameters, and $\Delta m_{15}(B)$ values.

Though there may be some overlap between 99aa-$A$- and 91T-like SNe, our analysis suggests them being separate classes. 99aa-$A$-like are a homogeneous group with a $s$ and $\Delta m_{15}(B)$ indicating lower brightness and higher $\rho_c$ than 91T, hinting towards different progenitor systems. 99aa-$B$-like SNe are an inhomogeneous group among 1991aa-likes with a broad spread in properties. As will be discussed in Sect. 8.2, the two groups may be distinguishable by NIR and MIR nebular spectra.

For some SNe Ia, we found the physical parameters to be inconsistent with the range expected from theoretical models (Tab. 4 and Tab. A1, Sect. 4.1). Among these outliers, 02cx-likes are underluminous. Around 67% of the 02cx-likes came from active galaxies. We did not have host information for the other 33%. 03fg-likes are in the normal bright $\Delta m_{15}(B)$ and $s$ range. They come from young hosts or active parts of older hosts (private communication with Jing Lu [13], and Lu et al. (2021). For details, see Tab. A2, Fig. 19, Sect. 4.3.4. The outliers might suggest other explosion scenarios.

## 6. IMPLICATIONS FOR COSMOLOGY AND THE FIRST GENERATION OF SNE IA

[13] iD

Our method demonstrates the importance of acquiring LC data long beyond the maximum light in high-precision cosmology. Therefore, this work strongly suggests the coverage should be extended up to ~ 60 days after the maximum light in order to get the secondary parameter effects rather than the limited coverage to 20 days past the maximum as planned for the Nancy Grace Roman Space Telescope (Fakhouri et al. 2015). For reconstruction of differentials, the current latency of ≈ 5 days for the general survey is insufficient (Foley et al. 2019), though the intensive field may be promising to significantly increase the number of SNe Ia. More promising ongoing surveys are POISE (Burns et al. 2021) (see Sect. 8.2), etc.

91T-like and 99aa-*A*-like SNe are bright, and their high masses and low densities suggest potentially evolutionary time-scales as short as 50 Myrs, and they are often connected with star formation (see Sect. 4.1). Though they amount to only ≈ 6% (nine among 152) of all local SNe Ia they amount to ≈ 12% of SN Ia with $M_{MS} \geq 4.5\ M_{\odot}$. From our analysis, these SNe Ia may dominate the SNe Ia population at high redshift. With a brightness of ≈ −19.5 mag they will outshine high-redshift, $z = 4 - 11$, galaxies and their globular clusters (Mowla et al. 2022), and may allow tracing the star formation history at moderate red-shifts, $z = 1 - 4$, by observing galactic clusters as found with JWST (Noirot et al. 2023). Within the $M_{Ch}$ mass scenarios and high $^{56}$Ni masses (Hoeflich & Khokhlov 1996; Hoeflich et al. 2017a; Aldoroty et al. 2023), and the low $\rho_c$ found here, the ejecta will pollute the surroundings with Fe-rich and very low electron-capture elements (e.g. Co), little intermediate mass elements (Si/S), and fewer products of explosive carbon-burning (e.g. Mg, Ne).

## 7. ALTERNATIVE EXPLOSION SCENARIOS

### 7.1. *Standard SNe Ia in Light of Explosion Scenarios*

We want to see our eigenvalue method in the context of other explosion scenarios. As seen in Fig. 4, at least 2 eigenfunctions are needed. We have not studied the other explosion scenarios but will discuss possible equivalent physical parameters. **Note that we need two parameters for $V$-band but at least two and one additional secondary parameters to fit $B$ and $U$, respectively (Hoeflich et al. 2017c; Sadler 2012).**.

One of our secondary parameters, the $M_{MS}$, is directly related to the initial abundance structure and the specific nuclear energy produced (see e.g. Domínguez et al. 2001). However, for dynamical mergers, $M_{MS}$ does not reflect the main sequence mass of one of the WDs but the change of the C/O ratio of the combined WDs which cannot directly be translated to lifetimes (Domínguez et al. 2003). Still, the combined change of the C/O ratio will influence the LCs in a similar way. For all explosion scenarios, the effect of this parameter and the resulting eigenfunction can be expected to be similar (Hoeflich et al. 1998; Shen et al. 2018a).

Our other secondary parameter, $\rho_c$, does not apply to the other explosion scenarios, as there are little to no EC elements in these other classes (see Sect. 2, and Figs. 3 & 2).

In HeD models $M_{WD}$ produces the luminosity decline relation (Shen et al. 2021; Blondin et al. 2018) but no significant amount of EC elements are produced **for models with a WD mass $M_{WD}$ less than ≈ $1.2M_{Ch}$. Those may be related to the bright end of normal-bright SNe Ia. For those, the actual $\rho_c$ will result in a shift of the $^{56}$Ni distribution**. Having shown that two terms are needed **for the entire brightness range**, the obvious free parameter is the amount of He on top. The mass of He has to be low for consistency with observations (Woosley & Weaver 1994; Hoeflich & Khokhlov 1996; Shen & Bildsten 2009). However, thin-shell ignition has been shown to produce asymmetry in the outer layers which undergo partial burning whereas the inner, nuclear statistical equilibrium (NSE) layers remain almost spherical (Livne 1999; Woosley & Kasen 2011; Boos et al. 2021; Gronow et al. 2021). However, He may trigger burning in the C/O core not in the center but, likely, off-center (Boos et al. 2021) which may lead to asymmetric outer layers and asymmetric luminosity during the optically thick, photospheric phase (Höflich 1991; Yang et al. 2018). Some 1 to 3 weeks after the maximum, the envelope becomes mostly optically thin, and the luminosity becomes isotropic. The resulting eigenfunction may be expected to resemble the offset produced by $\rho_c$ (Fig. 4). Though the $^{58}$Ni observed in the MIR (Gerardy et al. 2007; Telesco et al. 2015; Kumar et al. 2023; DerKacy et al. 2023; Kwok et al. 2023) favors $M_{Ch}$ mass explosions, HeDs may well be in our sample.

For the merger scenario, we would expect an overall asymmetry even reaching further inwards because of the variation of masses (Pakmor et al. 2012; Bulla et al. 2016).

In principle, the offset between anisotropic emission and isotropic emission should work in a similar fashion but may be gradual. However, note that we cannot rule out such a gradual change because of insufficient sampling in the current data set.

For normal-bright SNe Ia, large asymmetries in the overall density distribution can be ruled out because the continuum polarization for SNe Ia is ≤ 0.1% (Cikota et al. 2019), and little C/O is observed in the surface layers (Höflich et al. 2002). Note that the narrowness of the brightness decline relation of ≈ 0.1 mag may be consistent with the HeD scenario with low polarization but inconsistent with the merger.

### 7.2. *Remarks on non-standard and Underluminous SNe Ia in Light of Explosion Scenarios*

*For 91T-like SNe:* Our findings in combination of similar $s$, $M_{MS}$ and $\rho_c$ and low polarization (Cikota et al. 2019) favors explosion scenarios with very similar mass consistent with $M_{Ch}$ explosions which are triggered by compression using a high-accretion rate in a single degenerate or secular accretion in a double degenerate system as discussed in Hoeflich et al. 2019. The peculiarities in the color-magnitude diagram can be understood within the framework of 'classical' delayed-detonation scenarios (Aldoroty et al. 2023). High-amplitude pulsational delayed detonation models previously suggested (Mueller et al. 1991) can be ruled out by the upper limit on the unburned outer layers observed. Si has been observed as early as 13 days before the maximum light (Phillips et al. 2022) which, in combination with Fig 11 in (Hoeflich et al. 2023), sets an upper limit for C/O products of $10^{-2}$ $M_{\odot}$. Any model with a significant amount of unburned material or with a strong asymmetry can be ruled out (see Sect. 2).

*Underluminous SNe Ia* have different polarization properties than normal-bright SNe Ia. The observed continuum polarization is 0.5 – 0.7% indicating an overall asymmetry of 10 to 20% (Howell et al. 2001; Patat et al. 2012; Patat 2017). Spectra indicate a thick layer of a significant amount of unburned C/O (Höflich et al. 2002; Patat et al. 2012; Kumar et al. 2023). **These observations may be understood in terms of the delayed detonation of a very fast-rotating WD or dynamical mergers of two WDs (Patat et al. 2012).**

## 8. CONCLUSIONS

### 8.1. *Main Results*

1. We presented a method (Sect. 3) to link the progenitor and progenitor system properties to visual LCs. That requires a time-coverage of ∼ 60 days (Fig. 4, Sect. 3.3) and a precision of ≈ 0.02 mag. The method allows combining different surveys (Figs. 14 & 15). **The correction to the brightness amounts to ±0.3 mag, i.e. ≈ 50% of k- and reddening corrections suggesting the importance in high-precision cosmology.**

2. The presence of secondary LC parameters has been shown (Sect. 4) for ≈ 75% in our sample (Sect. 4.1). The variables are the $M_{MS}$ of the progenitor and the central density (or mass) of the initial WD which govern the explosion energy and $^{56}Ni$ distribution within the framework of near $M_{Ch}$ delayed detonation models. Alternative scenarios may produce similar relations because of their dependence on the explosion energy, though more eigenfunctions may be needed (Sect. 7). The code is provided on GIT-HUB (see below) which allows application of the system to datasets beyond the CSP survey and eigenfunctions by being used based on alternative models.

3. Independent of whether a galaxy prior is used, the basic properties can be separated by using LCs. General agreement in the results was found between not using and using a galaxy prior. Additional information about the host galaxies, if used, consolidates the trends, but it is not necessary (Sect. 5.2).

4. In galaxies with ongoing star-formation, we find evidence of two SNe Ia populations having a short and a long stellar evolution time, respectively (Fig. 17, Sects. 4.2 & 4.3). Most of our SNe Ia have a shorter stellar evolutionary time, so delay time can mostly be attributed to the progenitor system evolution as discussed in Sect. 4.

5. Our method can identify peculiar SNe Ia which such as 91T-like, 02cx-like, and 03fg-like. The 02cx-like and 03fg-like classes can be deselected by their 'non-physical' corrections.

   One cause for 'non-physical' parameters is the presence of a large amount of unburned material **e.g. C and O**. Partial explosive burning of a WD strongly alters the average specific explosion energy (Sect. 4.1).

6. **The 91T-like objects can be identified within a sample because the differentials are zero for pairs of 91T-likes. SNe Ia appear in actively star-forming hosts, potentially introducing bias in high-redshift surveys. Our method allows us to identify and deselect SNe Ia in cosmological samples, or use them as a homogeneous subclass in high-precision cosmology (Sect.4.1).**

7. We find very similar $s$ and secondary parameters among the 91T-like (and 99aa-$A$-like). Our analysis favors the explosion scenarios of similar mass, and sufficiently low central density to avoid the production of EC elements in the center or, alternatively, strong mixing in the inner layers, and have large $M_{MS}$ (Sect. 4.1). Note that this statement is strictly true only in configurations where the C/O ratio is dominated by the stellar evolution of a single component, e.g. in SD systems or DD systems with secular accretion (Sect. 7). From our analysis, 91T-like SNe Ia are expected to become contributors to the SNe Ia population with increasing redshift because of their shorter lifetimes. Due to their brightness, they can be detected at moderate redshifts, $z = 1 - 4$, and, high red-shifts, $z = 4 - 11$, in lensed galaxies and their globular clusters and may imprint their peculiar abundance structure on the early abundances (see Sect. 6).

8. 99aa-like SNe consist of two separate classes. 99aa-$A$-like are a homogeneous group with a $s$ or $\Delta m_{15}(B)$ indicating lower brightness but systematically higher $\rho_c$ compared to 91T-like SNe Ia. The difference between these groups is intrinsic and not caused by differences in their environment (see Sect. 4.1). 99aa-$B$-like SNe Ia are an inhomogeneous group with a brightness similar to the 99aa-like SNe. Our analysis suggests them being the bright extension of normal bright SNe Ia (Sect. 4.1).

9. Underluminous SNe Ia mostly appear in hosts with little to no star formation. Therefore, the progenitor system dominates the delay time. We found about 15% underluminous SNe Ia in our sample in active host galaxies, and transitional SNe Ia in active hosts have higher $\rho_c$. This suggests that underluminous SNe Ia may be an inhomogeneous group and have different explosion scenarios.

### 8.2. *Limitations and Future Directions*

From the current analysis, we showed that two secondary parameters are needed, though, for low $\rho_c$ LCs become insensitive to $\rho_c$ (see Sect. 2). Though effects of $M_{MS}$ can be expected to play a role within all scenarios, other physical parameters may be found to be specific for alternative scenarios (see Sect. 7). Those will be considered in the future. The effect of the secondary underlying physics on the peak brightness has been discussed. However, future data sets at larger redshifts are needed to verify that this leads to a reduction to the level indicated by the residuals, i.e. 0.02 mag, in brightness dispersion down in the Hubble flow. For large redshifts and Pop II/III, variations in $Z$ must be and can be included with rest frame $u$.

91T-like and 99aa-like SNe Ia have been identified as separate classes with, at least, two subgroups among the 99aa-likes. 91T late-time NIR and MIR show a peaked [Fe II] at 1.644 $\mu m$ (Mueller et al. 1991) in contrast to typical SNe Ia showing round profiles for [Fe II] and the [Co III] at 11.8 $\mu$m (Höflich et al. 2004; Gerardy et al. 2007; Telesco et al. 2015; Diamond et al. 2015, 2018b; Hoeflich et al. 2021; DerKacy et al. 2023; Kumar et al. 2023) (see Fig. 2 ).

As $V$-band is insensitive to the metallicity of the individual SN, our analysis only considers the $V$-band LCs. We have considered the effects of $M_{MS}$ and $\rho_c$, but not $Z$. However, from Hoeflich et al. (2017a) models it can be seen that for metallicity ($Z$), a variation from 0.01 $Z_\odot$ to 0.1 $Z_\odot$ changes the absolute $V$ and $B$ magnitudes $\approx$ 0.01 mag and $\approx$ 0.03 mag, respectively, and the Z results in an offset of early LCs, i.e. some days before maximum, relative to maximum and later phases of the LC. (Höflich et al. 1998a; Lentz et al. 2000; Baron et al. 2015).

In Sadler 2012, this early offset has been used as an eigenfunction with the $u$-band, although for a smaller sample (CSP-I). A broad peak around 1/3 of $Z_\odot$ is seen **with the amplitude of the differentials $\Delta A$ of $\approx$** $0.05^m$, but the tail goes up to $log(Z) = -4$ corresponding to $\Delta A \approx 0.2^m$.

Eigenfunctions used in our method are based on and can be understood by our models but, as a next step for higher accuracy, they may be optimized using big data sets and low-time-resolution adjustments, e.g. by using a few adjustment points over LCs spanning some 60 days.

CSP-II has only around 65 $u$-band LCs. Therefore, in the future, our work will be extended to include the $u$-band and the $Z$ effect when there is better data. **Currently, the variations in $B$ are being studied to try to identify the additional eigenfunctions in $B$ in order to probe the additional physics evading $V$.**

One of the main limitations of this work is the small sample size. While there is some strong evidence of $3\sigma$ significance, most of the trends are only indications with a significance of $\sim 1\sigma$. Higher significance requires more objects obtained with the same instrumental setup and low latency. In particular, the number of underluminous SNe Ia in our sample is too small to study the diversity among those.

**Another current limitation is the use of morphology as a proxy for star formation in the host galaxies. The intrinsic color of the host is a better indicator, based on the passive evolution of the stellar population from the host-galaxy color from blue to red. Based on stellar evolution (e.g. Chieffi et al. 2002), the corresponding time-scales**

**are** $\approx 10 - 30$ **Myrs, which is too short for significant statistics in our sample but, again, will be feasible with future surveys.**

An example of ongoing surveys is ATLAS (Asteroid Terrestrial-impact Last Alert System), which expects to find 300 SNe Ia per year with $V$-magnitude $< 17$ mag [14] (Tonry et al. 2018). Another such survey is the Zwicky Transient Facility (ZTF) which (phase I) discovered $>$ 3000 SNe Ia with a 3 day cadence and median redshift of 0.057 in a period of $\approx$ 3 years (Smith et al. 2014; Bellm et al. 2019; Dhawan et al. 2022) in combination with high-precision photometry such as POISE. These surveys will be able to solidify **the trends.**

**The galaxy morphology is a first-order criterion for ongoing star formation. For higher accuracy, likely a combination of galaxy colors, morphology, and, maybe, masses of galaxies are needed, in combination with detailed studies of the specific star formation rate (sSFR) (e.g. Sullivan et al. 2006).**

Several surveys span SNe up to high redshifts and, thus, probe the redshift evolution of the progenitor system. A requirement for their use is k-corrections (Hsiao et al. 2007; Lu et al. 2023) and, in some cases, construction of eigenfunctions for different filters such as $g$ instead of V which is straightforward. One ground-based survey is the Hyper Suprime-Cam (HSC) at Subaru Telescope with about 6000 SNe Ia LCs up to redshifts of $\approx 1.5$ (Suzuki 2017; Miyazaki et al. 2018; Aihara et al. 2019; Yasuda et al. 2019; Rubin et al. 2019). Another step forward is the Nancy Grace Roman Space Telescope (RST) which is expected to be launched in 2025 [15] (Spergel et al. 2015; Dore et al. 2019). RST will obtain data for $\geq 100$ SNe Ia per $\Delta z = 0.1$ bin over $0.2 \leq z \leq 1.7$ with a cadence of around 5 days (Scolnic et al. 2019; Hounsell et al. 2023). More than 10,000 SNe Ia with a large fraction having $z > 1$ (Hounsell et al. 2018), going up to $z = 3$ will be obtained.

The Large Synoptic Survey Telescope (Vera Rubin Telescope) will pursue another strategy by obtaining low-latency LCs for the whole sky (and the deep drilling field) namely 10(3) days, but for hundreds of thousands of SNe Ia [16] (Ivezic et al. 2008; LSST Science Collaboration et al. 2009; Ivezić et al. 2019; Hambleton et al. 2022). It will observe $> 300,000$ SNe Ia around $z = 0.3 - 0.4$ in its wide survey and $> 10,000$ around $z = 0.9 - 1.0$ in deep fields (Rose et al. 2021).

This work is based on the PhD thesis of S.C. who would like to thank her fellow graduate students for their support. P.H. acknowledges support by NSF grants AST-1715133 and AST-2306395. The observations have been obtained as part of the Carnegie Supernovae Project I and II, funded by the NSF under grants AST-0306969, AST-0607438, AST-1008343, AST-1613426, AST-1613455, and AST-1613472. S.C. would like to thank her husband, Arka Santra, for his support. L.G. acknowledges financial support from the Spanish Ministerio de Ciencia e Innovación (MCIN), the Agencia Estatal de Investigación (AEI) 10.13039/501100011033, and the European Social Fund (ESF) "Investing in your future" under the 2019 Ramón y Cajal program RYC2019-027683-I and the PID2020-115253GA-I00 HOSTFLOWS project, from Centro Superior de Investigaciones Científicas (CSIC) under the PIE project 20215AT016, and the program Unidad de Excelencia María de Maeztu CEX2020-001058-M. I.D. is partially supported by the project PID2021-123110NB-I00 financed by MCIN/AEI /10.13039/501100011033 / FEDER, UE. **M.D. Stritzinger is funded by the Independent Research Fund Denmark (IRFD, grant number 10.46540/2032-00022B).**

*Software:* The secondary parameter analysis tool (SPAT) has been developed as part of the PhD theses by B. Sadler and S. Chakraborty and is available at https://github.com/sudeshnafsu/SPAT. The plot package Gnuplot was used.

*Data:* Plots of all differential and correlations are available on request.

*Facilities:* Magellan, du Pont, Swope, Beowulf system of the Astrophysics group at Florida State University.

[14] https://atlas.fallingstar.com/exploding_stars.php

[15] https://roman.gsfc.nasa.gov/

[16] https://roman.gsfc.nasa.gov/science/workshop112021/presentations/Thu_SN/Roman_SNIa_Wood-Vasey_synergies_20211118.pdf

# APPENDIX

## A. PROPERTIES OF ALL SNE IA IN OUR SAMPLE

**Table A1**. Results for individual SNe. We give the names, g values, $M_{MS}[M_{\odot}]$, and $\rho_c$ [$10^9$ $gcm^{-3}$] values, their corresponding uncertainties and error range with a 95% (2 $\sigma$) confidence level are in brackets, and stretch ($s$). In the last column, we give $\Delta m_{15}(B)$ based on region 2 (see Sect. 3.1.1). For the actual names of the CSP SNe Ia and their host galaxy information, see the notes below Tab. A2. Values of 1.0 and 0.1 for $M_{MS}$ and $\rho_c$ indicate **'out of physical range'. For the physical reason, see Sects. 4 & 8.**

| No. | Name | $g_\rho$ | $g_m$ | $\rho_c$ | $M_{MS}$ | $s$ | $\Delta m_{15}(B)$ |
|---|---|---|---|---|---|---|---|
| | | | All SNe Ia used in analysis | | | | |
| 1 | SN2004ef | 1.55 (8.85E-02) | -2.52E-01 (1.01E-01) | 3.95 (3.1-4.9) | 4.37 (1.6-6.0) | 0.82 | 1.39 |
| 2 | SN2004eo | 1.39 (8.66E-02) | 3.31E-02 (7.86E-02) | 3.4 (2.7-4.3) | 5.96 (4.8-6.4) | 0.83 | 1.37 |
| 3 | SN2004ey | 1.11 (8.1E-02) | -2.75E-01 (1.11E-01) | 2.63 (2.2-3.3) | 4.19 (1.4-6.0) | 1.00 | 1.02 |
| 4 | SN2004gs | 1.48 (1.2E-01) | -1.75E-01 (1.27E-01) | 3.69 (2.6-5.0) | 5.02 (1.6-6.3) | 0.72 | 1.59 |
| 5 | SN2004gu | 1.06 (9.97E-02) | -2.39E-01 (1.26E-01) | 2.51 (2.0-3.3) | 4.48 (1.4-6.2) | 1.07 | 0.88 |
| 6 | SN2005al | 1.43 (1.29E-01) | -4.07E-01 (2.44E-01) | 3.54 (2.5-5) | 3.01 (1.0-6.5) | 0.87 | 1.29 |
| 7 | SN2005am | 1.64 (1.19E-01) | -2.93E-01 (1.56E-01) | 4.29 (3.1-5.5) | 4.05 (1.0-6.2) | 0.80 | 1.43 |
| 8 | SN2005A | 8.75E-01 (1.06E-01) | -2.37E-01 (1.56E-01) | 2.18 (1.9-2.8) | 4.51 (1.1-6.3) | 0.89 | 1.24 |
| 9 | SN2005el | 1.20 (9.32E-02) | -1.49E-01 (1.05E-01) | 2.83 (2.2-3.7 ) | 5.2 (2.3-6.2) | 0.80 | 1.43 |
| 10 | SN2005eq | 1.02 (1.13E-01) | -2.04E-02 (1.27E-01) | 2.44 (1.9-3.3) | 5.8 (3.1-6.5) | 1.06 | 0.90 |
| 11 | SN2005hc | 1.27 (6.55E-02) | 9.70E-03 (7.89E-02) | 3.01 (2.5- 3.6) | 5.9 (4.6-6.4) | 0.97 | 0.98 |
| 12 | SN2005iq | 1.59 (1.11E-01) | -1.26E-01 (9.28E-02) | 4.09 (3-5.3) | 5.35 (3-6.2) | 0.83 | 1.37 |
| 13 | SN2005ir | 1.91 (1.24E-01) | -2.79E-01 (1.75E-01) | 5.24 (3.9-5.6) | 4.15 (1.0-6.3) | 0.86 | 1.31 |
| 14 | SN2005kc | 9.01E-01 (1.81E-01) | 2.66E-01 (1.01E-01) | 2.22 (1.8-3.6) | 6.38 (5.7-6.7) | 0.79 | 1.34 |
| 15 | SN2005ke | 8.34E-02 (1.43E-01) | -2.23E-02 (1.73E-01) | 1.71 (1.6-1.8) | 4.62 (1-6.4) | 0.52 | 2.00 |
| 16 | SN2005ki | 1.33 (1.03E-01) | -1.09E-01 (1.42E-01) | 3.19 (2.4- 4.3) | 5.44 (1.7 -6.4) | 0.78 | 1.47 |
| 17 | SN2005M | 1.17 (6.92E-02) | -2.89E-01 (8.25E-02) | 2.76 (2.3 -3.4) | 4.09 (1.7 -5.7) | 1.10 | 0.82 |
| 18 | SN2005na | 1.36 (1.06E-01) | -1.38E-01 (1.71E-01) | 3.31 (2.5-4.5) | 5.27 (1.3-6.5) | 0.93 | 1.16 |
| 19 | SN2006ax | 9.43E-01 (1.06E-01) | -2.76E-01 (1.26E-01) | 2.29 (1.9-3.0) | 4.18 (1.3-6.1) | 0.99 | 1.04 |
| 20 | SN2006bh | 1.67 (8.56E-02) | -1.44E-01 (9.05E-02) | 4.42 (3.5 -5.3) | 5.24 (2.9-6.2) | 0.83 | 1.37 |
| 21 | SN2006gt | 9.86E-01 (1.96E-01) | -7.56E-02 (2.68E-01) | 2.36 (1.8 -4.1) | 5.60 (1.00-6.8) | 0.61 | 1.82 |
| 22 | SN2006kf | 1.66 (1.39E-01) | -1.21E-01 (1.18E-01) | 4.38 (3.0-5.6) | 5.38 (2.2-6.3) | 0.76 | 1.51 |
| 23 | SN2006ob | 2.49 (1.53E-01) | -2.88E-01 (1.66E-01) | 5.59 (5.58 -5.6) | 4.09 (1.0-6.3) | 0.75 | 1.53 |
| 24 | SN2007af | 1.32 (8.4E-02) | -1.89E-01 (7.6E-02) | 3.17 (2.5-4.0) | 4.91 (2.9 -6.0) | 0.95 | 1.12 |
| 25 | SN2007ba | 1.06 (1.69E-01) | 2.02E-02 (1.21E-01) | 2.51 (1.9-4.0) | 5.93 (3.7-6.5) | 0.56 | 1.92 |
| 26 | SN2007bd | 1.98 (1.04E-01) | -5.08E-01 (1.39E-01) | 5.5 (4.4-5.6) | 1.93 (1.0-5.5) | 0.93 | 1.16 |
| 27 | SN2007S | 5.90E-01 (1.42E-01) | -2.40E-01 (1.17E-01) | 1.89 (1.7-2.4) | 4.48 (1.4 -6.1) | 1.03 | 0.96 |
| 28 | SN2011iv | -6.03E-02 (2.28E-01) | 6.49E-02 (1.61E-01) | 1.67 (1.4-1.9) | 6.04 (2.9-6.7) | 0.54 | 1.56 |
| 29 | SN2011jh | 1.23 (8.22E-02) | -3.51E-01 (8.42E-02) | 2.92 (2.4 -3.7) | 3.59 (1.4 -5.5) | 0.84 | 1.35 |
| 30 | SN2011jn | -2.00E-01 (1.32E-01) | 1.20E-01 (2.98E-01) | 1.63 (1.5 -1.7) | 6.15 (1.0-7.0) | 0.52 | 1.65 |
| 31 | SN2012ar | 1.25 (1.27E-01) | -3.20E-01 (1.76E-01) | 2.97 (2.2 -4.3) | 3.85 (1.0-6.3) | 0.82 | 1.39 |
| 32 | SN2012bo | 7.88E-01 (1.74E-01) | -2.50E-01 ( 1.28E-01) | 2.05 (1.8-3.1) | 4.39 (1.3-6.2) | 1.07 | 0.98 |
| 33 | SN2012fr | 3.49E-02 (1.79E-01) | 3.80E-03 (1.41E-01) | 1.69 (1.6 -1.9) | 5.88 (2.9-6.6) | 0.96 | 1.10 |
| 34 | SN2012G | 5.74E-01 (1.56E-01) | -3.35E-03 (2.32E-01) | 1.87 (1.7 -2.5) | 5.86 (1.1-6.8) | 1.10 | 0.82 |
| 35 | SN2012hd | 1.29 (1.41E-01) | -8.38E-02 (1.03E-01) | 3.09 (2.2-4.6) | 5.56 (3.2-6.3) | 0.88 | 1.26 |

| No. | Name | $g_\rho$ | $g_m$ | $\rho_c$ | $M_{MS}$ | $s$ | $\Delta m_{15}(B)$ |
|---|---|---|---|---|---|---|---|
| 36 | SN2012hr | 1.14 (8.13E-02) | -2.36E-01 (8.31E-02 ) | 2.68 (2.2-3.4) | 4.51 (2.1-5.9) | 0.98 | 1.06 |
| 37 | SN2012ht | 1.03 (1.06E-01) | -1.69E-01 (1.02E-01) | 2.45 (2.0 -3.3) | 5.06 (2.2-6.2) | 0.81 | 1.41 |
| 38 | SN2012ij | 8.40E-01 (2.33E-01) | 2.76E-02 (2.15E-01) | 2.12 (1.7-3.9) | 5.95 (1.4 -6.8) | 0.62 | 1.80 |
| 39 | SN2013aa | 8.90E-01 (9.77E-02) | -1.39E-01 (9.69E-02) | 2.20 (1.9-2.8) | 5.27 (2.7-6.2) | 0.96 | 1.10 |
| 40 | SN2013aj | 1.18 (1.04E-01) | -5.05E-02 (9.46E-02) | 2.78 (2.2- 3.7) | 5.70 (3.7-6.3) | 0.75 | 1.53 |
| 41 | SN2013bz | 3.50E-01 (1.45E-01) | 1.38E-02 (1.53E-01) | 1.79 (1.7-2.1) | 5.91 (2.5-6.6) | 0.95 | 0.92 |
| 42 | SN2013E | 2.33E-01 (1.30E-01) | 1.30E-01 (-1.53E-01) | 1.75 (1.7-1.9) | 5.39 (2.8-6.3) | 0.98 | 0.96 |
| 43 | SN2013fy | 3.35E-01 (1.81E-01) | -2.10E-01 (1.44E-01) | 1.78 (1.6-2.2) | 4.73 (1.3-6.3) | 1.03 | 0.96 |
| 44 | SN2013fz | 7.38E-01 (9.91E-02) | -3.02E-01 (1.62E-01) | 1.99 (1.8-2.5) | 3.99 (1.0-6.3) | 1.03 | 0.96 |
| 45 | SN2013gy | 1.11 (8.93E-02) | -1.65E-01 (7.43E-02) | 2.61 (2.1-3.3) | 5.09 (3.2-6.0) | 0.92 | 1.18 |
| 46 | SN2013hh | 5.48E-01 (1.30E-01) | 2.31E-02 (1.66E-01) | 1.86 (1.7-2.3) | 5.93 (2.2-6.7) | 0.89 | 1.14 |
| 47 | SN2013H | 4.93E-01 (1.32E-01) | -3.59E-01 (9.52E-02) | 1.84 (1.7 -2.2) | 3.52 (1.3-5.6) | 1.03 | 0.96 |
| 48 | SN2013M | 7.12E-01 (9.37E-02) | -5.41E-02 (9.77E-02) | 1.96 (1.8-2.4) | 5.68 (3.6-6.3) | 0.91 | 1.20 |
| 49 | SN2013U | 4.61E-02 (1.16E-01) | -6.79E-02 (1.16E-01) | 1.70 (1.6 -1.8) | 5.63 (2.9-6.4) | 1.05 | 0.92 |
| 50 | SN2014ao | 1.35 (1.09E-01) | -1.96E-01 (1.71E-01) | 3.25 (2.4-4.4) | 4.85 (1.1-6.4) | 0.87 | 1.29 |
| 51 | SN2014at | 9.80E-01 (1.15E-01) | -1.61E-01 (1.20E-01) | 2.35 (1.9-3.2) | 5.12 (1.8-6.3) | 0.96 | 1.10 |
| 52 | SN2014dn | 4.74E-01 (1.77E-01) | -2.52E-01 (1.96E-01) | 1.83 (1.7-2.4) | 4.38 (1.0-6.5) | 0.57 | 1.90 |
| 53 | SN2014eg | -2.65E-01 (9.39E-02) | 1.41E-01 (1.07E-01) | 1.62 (1.5-1.7) | 6.19 (5.0-6.6) | 0.80 | 1.13 |
| 54 | SN2014I | 1.12 (1.07E-01) | -2.74E-01 (9.31E-02) | 2.65 (2.1-3.6) | 4.20 (1.6-5.9) | 0.96 | 1.10 |
| 55 | SN2015F | 1.06 (1.14E-01) | -1.09E-01 (1.64E-01) | 2.51 (2.0-3.4) | 5.44 (1.4-6.5) | 0.86 | 1.31 |
| 56 | SNhunt281 | 1.22 (1.26E-01) | -1.71E-01 (1.31E-01) | 2.88 (2.1-4.1) | 5.05 (1.6-6.3) | 0.71 | 1.61 |
| 57 | ASASSN-14ad | 1.02 (1.21E-01) | -4.27E-01 (1.40E-01) | 2.43 (1.9-3.4) | 2.77 (1.0-5.8) | 1.09 | 0.84 |
| 58 | ASASSN-14hp | 4.79E-01 (1.26E-01) | -2.57E-01 (1.38E-01) | 1.84 (1.7-2.1) | 4.33 (1.2-6.2) | 1.06 | 0.90 |
| 59 | ASASSN-14hr | 1.00 (1.32E-01) | -3.39E-02 (1.13E-01) | 2.39 (1.9-3.4) | 5.76 (3.4-6.4) | 0.78 | 1.47 |
| 60 | ASASSN-14hu | 2.11E-01 (1.41E-01) | -1.32E-01 (1.39E-01) | 1.74 (1.6-1.9) | 5.31 (1.7-6.4) | 0.97 | 1.08 |
| 61 | ASASSN-14jc | 8.85E-01 (1.02E-01) | -2.30E-01 (1.09E-01) | 2.19 (1.9-2.8) | 4.57 (1.6-6.1) | 0.89 | 1.24 |
| 62 | ASASSN-14jg | 5.25E-02 (2.29E-01) | -1.15E-01 (2.80E-01) | 1.70 (1.5-2.0) | 5.41 (1.0-6.8) | 1.02 | 0.98 |
| 63 | ASASSN-14kd | -1.91E-01 (1.65E-01) | 5.38E-02 (1.23E-01) | 1.63 (1.5-1.8) | 6.01 (3.9-6.6) | 0.96 | 1.03 |
| 64 | ASASSN-14kq | 9.01E-01 (1.16E-01) | -6.22E-01 (1.31E-01) | 2.22 (1.9-3.0) | 1.35 (1.0-4.6) | 1.15 | 0.81 |
| 65 | ASASSN-14lp | 3.61E-01 (1.19E-01) | -1.12E-02 (1.40E-01) | 1.79 (1.7-2.0) | 5.83 (2.7-6.6) | 0.90 | 1.12 |
| 66 | ASASSN-14lt | 1.08 (9.16E-02) | -1.96E-01 (1.34E-01) | 2.55 (2.1-3.3) | 4.85 (1.4-6.3) | 0.95 | 1.02 |
| 67 | ASASSN-14lw | -3.18E-01 (1.62E-01) | -2.92E-01 (1.17E-01) | 1.60 (1.2-1.7) | 4.06 (1.3-6.0) | 1.07 | 0.88 |
| 68 | ASASSN-14me | 3.02E-01 (1.17E-01) | -4.61E-01 (1.79E-01) | 1.77 (1.7-1.9) | 2.37 (1.0-6.1) | 1.05 | 0.92 |
| 69 | ASASSN-14mf | 1.11 (1.79E-01) | -5.26E-01 (2.33E-01) | 2.61 (1.9-4.3) | 1.80 (1.0-6.2) | 1.07 | 0.88 |
| 70 | ASASSN-14mw | 6.13E-01 (1.34E-01) | -4.35E-01 (1.16E-01) | 1.90 (1.7-2.4) | 2.67 (1.0-5.5) | 1.08 | 0.86 |
| 71 | ASASSN-14my | 1.04 (1.06E-01) | -1.03E-01 (9.78E-02) | 2.47 (2.0-3.3) | 5.47 (3.1-6.3) | 0.91 | 1.20 |
| 72 | ASASSN-15aj | 1.00 (1.33E-01) | -2.05E-01 (1.97E-01) | 2.39 (1.9-3.4) | 4.78 (1.0-6.5) | 0.81 | 1.41 |
| 73 | ASASSN-15al | 5.58E-01 (2.22E-01) | -2.91E-01 (3.07E-01) | 1.87 (1.7-2.9) | 4.07 (1.0-6.7) | 1.11 | 0.80 |
| 74 | ASASSN-15as | 4.50E-01 (1.72E-01) | -3.67E-01 (2.43E-01) | 1.83 (1.7-2.3) | 3.44 (1.0-6.5) | 1.13 | 0.75 |
| 75 | ASASSN-15ba | 1.04 (1.10E-01) | -2.26E-01 (1.65E-01) | 2.46 (2.0-3.3) | 4.60 (1.1-6.4) | 1.02 | 0.98 |
| 76 | ASASSN-15be | 3.50E-01 (1.48E-01) | -2.13E-01 (8.55E-02) | 1.79 (1.7-2.1) | 4.72 (2.3-6.0) | 1.01 | 1.00 |
| 77 | ASASSN-15bm | 8.66E-01 (1.73E-01) | -2.92E-01 (1.26E-01) | 2.16 (1.8-3.4) | 4.06 (1.2-6.1) | 1.02 | 0.98 |
| 78 | ASASSN-15cd | 3.59E-01 (9.85E-02) | -1.33E-01 (1.07E-01) | 1.79 (1.7-1.9) | 5.30 (2.4-6.3) | 0.95 | 1.12 |
| 79 | ASASSN-15da | -6.79E-01 (2.57E-01) | -2.54E-01 (4.46E-01) | 1.48 (1.0-1.7) | 4.36 (1.0-7.0) | 0.63 | 1.47 |
| 80 | ASASSN-15db | 1.36 (8.94E-02) | -3.29E-01 (1.06E-01) | 3.30 (2.6-4.3) | 3.78 (1.3-5.8) | 0.96 | 1.10 |
| 81 | ASASSN-15dd | 1.26 (1.13E-01) | -2.50E-01 (1.93E-01) | 3.01 (2.3-4.2) | 4.39 (1.0-6.5) | 0.85 | 1.33 |
| 82 | ASASSN-15eb | 1.55 (1.69E-01) | 6.20E-02 (3.35E-01) | 3.96 (2.5-5.6) | 6.03 (1.0-7.0) | 0.78 | 1.47 |
| 83 | ASASSN-15ga | 2.89E-01 (1.99E-01) | -7.54E-01 (2.89E-01) | 1.77 (1.6-2.2) | 1.00 (1.0-6.1) | 0.46 | 2.12 |
| 84 | ASASSN-15gr | 4.98E-01 (1.24E-01) | -2.59E-01 (1.58E-01) | 1.84 (1.7-2.2) | 4.31 (1.0-6.3) | 1.00 | 1.02 |

| No. | Name | $g_\rho$ | $g_m$ | $\rho_c$ | $M_{MS}$ | $s$ | $\Delta m_{15}(B)$ |
|---|---|---|---|---|---|---|---|
| 85 | ASASSN-15hf | 1.07 (9.02E-02) | -1.99E-01 (1.33E-01) | 2.52 (2.1-3.2) | 4.83 (1.4-6.3) | 0.90 | 1.22 |
| 86 | ASASSN-15hx | 5.29E-01 (2.76E-01) | -3.76E-01 (1.31E-01) | 1.86 (1.6-3.3) | 3.34 (1.0-5.9) | 1.06 | 0.90 |
| 87 | OGLE-2013-SN-126 | 8.39E-01 (2.84E-01) | -3.85E-01 (1.98E-01) | 2.12 (1.7-4.5) | 3.25 (1.0-6.3) | 1.11 | 0.80 |
| 88 | PSN J03055989+0432382 | 6.47E-01 (1.46E-01) | -3.61E-01 (1.25E-01) | 1.92 (1.7-2.6) | 3.49 (1.0-5.9) | 1.08 | 0.86 |
| 89 | MLS140102:120307-010132 | 6.85E-01 (1.97E-01) | 3.93E-01 (1.61E-01) | 1.94 (1.7-3.0) | 6.53 (5.5-6.9) | 0.91 | 1.20 |
| 90 | MOT J0939534+ | -1.19E-01 (1.04E-01) | -2.19E-01 (2.20E-01) | 1.65 (1.6-1.7) | 4.66 (1.0-6.6) | 0.92 | 1.18 |
| 91 | CSS130303:105206-133424 | -1.42E-01 (1.34E-01) | 1.97E-01 (1.10E-01) | 1.65 (1.5-1.8) | 6.28 (5.3-6.7) | 0.90 | 1.12 |
| 92 | OGLE-2014-SN-021 | 1.01 (2.45E-01) | 3.05E-01 (2.34E-01) | 2.41 (1.8-4.7) | 6.43 (3.1-7.0) | 0.95 | 1.12 |
| 93 | CSS140914-010107-101840 | 9.56E-01 (2.00E-01) | -3.60E-01 (3.49E-01) | 2.31 (1.8-4.0) | 3.51 (1.0-6.8) | 0.92 | 1.18 |
| 94 | OGLE-2014-SN-107 | -3.49E-02 (1.06E-01) | -5.08E-02 (1.80E-01) | 1.68 (1.6-1.8) | 5.70 (1.5-6.6) | 0.93 | 1.16 |
| 95 | SN2014du | 1.54 (1.92E-01) | -3.13E-01 (2.68E-01) | 3.94 (2.3-5.6) | 3.91 (1.0-6.6) | 0.85 | 1.33 |
| 96 | OGLE-2014-SN-141 | -5.08E-01 (1.13E-01) | -2.43E-01 (1.64E-01) | 1.55 (1.0-1.6) | 4.45 (1.0-6.4) | 1.07 | 0.88 |
| 97 | SN2015bo | 3.33E-01 (1.09E-01) | -1.50E-01 (1.22E-01) | 1.78 (1.7-1.9) | 5.20 (1.9-6.3) | 0.57 | 1.90 |
| 98 | PSN J13471211-2422171 | 1.11 (1.61E-01) | -2.88E-01 (1.35E-01) | 2.61 (1.9-4.1) | 4.09 (1.1-6.1) | 0.79 | 1.45 |
| 99 | LSQ11bk | 8.44E-03 (1.56E-01) | -2.65E-01 (2.66E-01) | 1.69 (1.6-1.8) | 4.27 (1.0-6.7) | 0.99 | 1.04 |
| 100 | LSQ11ot | 3.91E-01 (9.59E-02) | 2.05E-01 (9.36E-02) | 1.8 (1.7-1.9) | 6.12 (5.0-6.5) | 0.88 | 1.26 |
| 101 | LSQ11pn | 2.39E-01 (1.18E-01) | 9.21E-02 (1.04E-01) | 1.75 (1.7-1.9) | 6.09 (4.6-6.5) | 0.52 | 1.86 |
| 102 | LSQ12agq | 1.42E-01 (1.96E-01) | -3.87E-01 (3.34E-01) | 1.73 (1.6-2.0) | 3.24 (1.0-6.7) | 0.95 | 0.99 |
| 103 | LSQ12bld | 1.35 (3.03E-01) | -2.66E-01 (1.84E-01) | 3.28 (1.8-5.6) | 4.26 (1.0-6.4) | 0.97 | 1.08 |
| 104 | LSQ12fuk | 9.14E-01 (1.59E-01) | -3.95E-02 (2.42E-01) | 2.24 (1.8-3.0) | 5.67 (1.1-6.7) | 1.07 | 0.88 |
| 105 | LSQ12fxd | 8.00E-01 (1.00E-01) | -2.45E-01 (8.57E-02) | 2.07 (1.8-2.6) | 4.43 (2.0-5.9) | 1.08 | 0.86 |
| 106 | LSQ12gdj | -1.57E-01 (1.19E-01) | 1.86E-01 (8.81E-02) | 1.64 (1.5-1.7) | 6.26 (5.6-6.6) | 0.87 | 1.19 |
| 107 | LSQ12gxj | 1.36E-01 (1.25E-01) | -3.26E-02 (1.11E-01) | 1.72 (1.6-1.8) | 5.77 (3.5-6.4) | 0.98 | 1.06 |
| 108 | LSQ12hzj | 1.20 (8.35E-02) | -1.89E-01 (1.53E-01) | 2.84 (2.3-3.6) | 4.91 (1.3-6.4) | 1.01 | 1.00 |
| 109 | LSQ13cwp | 1.76 (2.37E-01) | -1.02E-01 (1.31E-01) | 4.73 (2.5-5.6) | 5.47 (2.0-6.4) | 0.93 | 1.16 |
| 110 | LSQ13dby | 4.54E-01 (2.77E-01) | 1.42E-01 (1.44E-01) | 1.83 (1.6-3.1) | 6.19 (4.1-6.7) | 1.01 | 1.00 |
| 111 | LSQ13dhj | 4.44E-01 (2.56E-01) | -1.87E-01 (2.15E-01) | 1.82 (1.6-2.9) | 4.93 (1.0-6.6) | 1.11 | 0.80 |
| 112 | LSQ13dpm | 7.46E-01 (8.12E-02) | -5.24E-01 (9.20E-02) | 2.00 (1.8-2.4) | 1.82 (1.0-4.4) | 1.12 | 0.78 |
| 113 | LSQ13dsm | 1.11 (8.94E-02) | -1.06E-01 (1.05E-01) | 2.61 (2.1-3.4) | 5.45 (2.8-6.3) | 0.95 | 1.12 |
| 114 | LSQ13lq | 2.33E-01 (1.81E-01) | -3.45E-01 (1.06E-01) | 1.75 (1.6-2.0) | 3.64 (1.2-5.8) | 1.15 | 0.71 |
| 115 | LSQ13ry | 1.09 (8.81E-02) | -1.84E-01 (8.71E-02) | 2.58 (2.1-3.3) | 4.95 (2.6-6.1) | 0.88 | 1.26 |
| 116 | LSQ13vy | 1.22 (1.01E-01) | -1.87E-01 (9.55E-02) | 2.89 (2.2-3.9) | 4.93 (2.2-6.1) | 0.93 | 1.16 |
| 117 | LSQ14age | 9.70E-02 (2.71E-01) | -4.82E-01 (2.21E-01) | 1.71 (1.4-2.2) | 2.16 (1.0-6.3) | 1.11 | 0.80 |
| 118 | LSQ14ahc | 2.70E-01 (2.19E-01) | -2.30E-01 (1.30E-01) | 1.76 (1.6-2.3) | 4.57 (1.4-6.2) | 1.09 | 0.84 |
| 119 | LSQ14ahm | 2.44E-01 (1.95E-01) | -2.93E-01 (1.74E-01) | 1.75 (1.6-2.1) | 4.05 (1.0-6.3) | 1.16 | 0.69 |
| 120 | LSQ14ajn | 1.23 (1.44E-01) | 3.85E-01 (3.21E-01) | 2.91 (2.1-4.4) | 6.52 (1.5-7.0) | 0.72 | 1.59 |
| 121 | LSQ14asu | 1.69 (2.15E-01) | 4.17E-02 (1.79E-01) | 4.48 (2.5-5.6) | 5.98 (2.0-6.7) | 0.81 | 1.41 |
| 122 | LSQ14auy | -6.63E-02 (1.77E-01) | -2.76E-01 (2.08E-01) | 1.67 (1.5-1.8) | 4.18 (1.0-6.5) | 1.07 | 0.88 |
| 123 | LSQ14fms | 2.99E-01 (1.74E-01) | -3.61E-03 (1.46E-01) | 1.77 (1.6-2.1) | 5.86 (2.6-6.6) | 0.82 | 1.39 |
| 124 | LSQ14foj | 3.81E-01 (1.45E-01) | -2.07E-01 (1.23E-01) | 1.80 (1.7-2.1) | 4.76 (1.5-6.2) | 0.96 | 1.10 |
| 125 | LSQ14ghv | 6.85E-01 (2.16E-01) | -5.06E-01 (2.26E-01) | 1.94 (1.7-3.2) | 1.95 (1.0-6.2) | 1.00 | 1.02 |
| 126 | LSQ14gov | -7.42E-01 (4.05E-01) | -5.12E-01 (1.58E-01) | 1.40 (0.1-1.8) | 1.91 (1.0-5.7) | 1.20 | 0.81 |
| 127 | LSQ14ie | -1.89E-01 (2.55E-01) | -3.27E-01 (1.46E-01) | 1.63 (0.82-1.9) | 3.79 (1.0-6.1) | 1.16 | 0.69 |
| 128 | LSQ14jp | 1.46 (1.53E-01) | 7.86E-02 (1.03E-01) | 3.63 (2.4-5.3) | 6.07 (4.6-6.5) | 0.72 | 1.59 |
| 129 | LSQ14mc | 1.16 (7.63E-02) | -2.83E-01 (1.26E-01) | 2.73 (2.3-3.4) | 4.12 (1.2-6.1) | 1.03 | 0.96 |
| 130 | LSQ14q | 1.09 (2.03E-01) | -4.31E-01 (1.23E-0 | 2.57 (1.8-4.5) | 2.72 (1.0-5.7) | 1.06 | 0.90 |
| 131 | LSQ14wp | 2.90E-01 (2.39E-01) | -3.19E-01 (1.95E-01) | 1.77 (1.6-2.4) | 3.87 (1.0-6.4) | 1.20 | 0.81 |
| 132 | LSQ14xi | 3.25E-01 (1.86E-01) | -2.36E-01 (1.06E-01) | 1.78 (1.6-2.2) | 4.51 (1.6-6.1) | 1.05 | 0.92 |
| 133 | LSQ15aae | 5.39E-01 (1.63E-01) | -5.14E-01 (1.28E-01 ) | 1.86 (1.7-2.4) | 1.89 (1.0-5.3) | 1.21 | 0.59 |

| No. | Name | $g_\rho$ | $g_m$ | $\rho_c$ | $M_{MS}$ | $s$ | $\Delta m_{15}(B)$ |
|---|---|---|---|---|---|---|---|
| 134 | LSQ15agh | 5.35E-01 (1.61E-01) | -2.48E-01 (1.39E-01) | 1.86 (1.7-2.4) | 4.41 (1.2-6.2) | 1.11 | 0.80 |
| 135 | LSQ15aja | 1.09E-03 (1.93E-01) | -4.78E-01 (1.14E-01) | 1.68 (1.5-1.9) | 2.20 (1.0-5.3) | 1.07 | 0.88 |
| 136 | LSQ15alq | 1.08 (2.03E-01) | -2.91E-01 (1.62E-01) | 2.56 (1.8-4.5) | 4.06 (1.0-6.3) | 0.97 | 1.08 |
| 137 | LSQ15bv | 4.87E-01 (1.42E-01) | -1.38E-01 (3.42E-01) | 1.84 (1.7-2.2) | 5.27 (1.0-6.9) | 0.92 | 1.18 |
| 138 | PS1-14ra | 1.07 (1.17E-01) | -3.76E-01 (2.24E-01) | 2.53 (2.0-3.5) | 3.34 (1.0-6.4) | 0.8 | 1.43 |
| 139 | PS1-14rx | 9.62E-01 (3.81E-01) | 5.50E-01 (2.32E-01) | 2.32 (1.6-5.6) | 6.68 (5.2-7.0) | 0.75 | 1.25 |
| 140 | PS15sv | 3.90E-01 (9.26E-02) | 5.69E-03 (1.35E-01) | 1.80 (1.7-1.9) | 5.89 (3.1-6.6) | 0.95 | 1.12 |
| 141 | iPTF11pbp | 6.08E-01 (1.47E-01) | -1.99E-01 (9.97E-02) | 1.90 (1.7-2.5) | 4.83 (2.0-6.1) | 1.02 | 0.98 |
| 142 | iPTF13anh | 1.91 (1.87E-01) | -2.41E-01 (1.96E-01) | 5.27 (3.3-5.6) | 4.47 (1.0-6.5) | 1.08 | 0.86 |
| 143 | iPTF13ebh | 1.02 (1.35E-01) | -4.61E-02 (1.08E-01) | 2.43 (1.9-3.5) | 5.72 (3.4-6.4) | 0.67 | 1.69 |
| 144 | iPTF14fpg | 5.50E-01 (1.59E-01) | -6.20E-01 (1.09E-01) | 1.86 (1.7-2.4) | 1.35 (1.0-4.0) | 1.11 | 0.80 |
| 145 | iPTF14gnl | 5.51E-01 (1.28E-01) | -3.58E-01 (1.41E-01) | 1.86 (1.7-2.3) | 3.53 (1.0-6.0) | 1.09 | 0.84 |
| 146 | iPTF14w | 1.03 (1.64E-01) | -5.93E-01 (3.95E-01) | 2.44 (1.9-3.9) | 1.45 (1.0-6.7) | 0.76 | 1.53 |
| 147 | iPTF14yw | 1.35 (1.01E-01) | -3.59E-01 (1.50E-01) | 3.26 (2.5-4.3) | 3.52 (1.0-6.1) | 0.88 | 1.26 |
| 148 | iPTF14yy | 1.41 (3.15E-01) | -4.24E-01 (1.45E-01) | 3.45 (1.8-5.6) | 2.80 (1.0-5.9) | 0.92 | 1.18 |
| | Underluminous in active hosts | | | | | | |
| 149 | SN2006D | 1.21 (9.93E-02) | 7.27E-03 (1.12E-01) | 2.86 (2.2- 3.8 ) | 5.89 (3.8 -6.5) | 0.76 | 1.05 |
| 150 | SN2006X | -5.45E-02 (1.05E-01) | 1.54E-02 (9.08E-02) | 1.67 (1.6-1.8) | 5.92 (4.3-6.4) | 0.78 | 1.27 |
| 151 | iPTF11pra | -1.91E-01 (1.59E-01) | 2.98E-02 (2.17E-01) | 1.63 (1.5-1.8) | 5.95 (1.4-6.8) | 0.47 | 2.10 |
| 152 | iPTF14aje | 4.00E-01 (9.53E-02) | 1.26E-01 (1.01E-01) | 1.80 (1.7-1.9) | 6.16 (5.0-6.6) | 0.63 | 1.77 |
| | Outliers | | | | | | |
| 153 | SN2013ao | -4.5 (2.10E-01) | -5.82E-01 (1.89E-01) | 0.1 (0.1-0.1) | 1.49 (1.0 -5.8) | 1.10 | 0.82 |
| 154 | ASASSN-15hy | -6.32 (3.19E-01) | -4.35E-01 (2.06E-01) | 0.1 (0.1-0.1) | 2.67 (1.0-6.3) | 1.25 | 0.51 |
| 155 | SN2012Z | -3.73 (2.00E-01) | -6.17E-01 (1.58E-01) | 0.1 (0.1-0.1) | 1.36 (1.0-5.2) | 0.81 | 1.41 |
| 156 | SN2013gr | -4.43 (2.02E-01) | -2.01E-01 (3.46E-01) | 0.1 (0.1-0.1) | 4.81 (1.0-6.9) | 0.55 | 1.94 |
| 157 | SN2014ek | -4.39 (1.94E-01) | -7.18E-01 (1.90E-01) | 0.1 (0.1-0.1) | 1.08 (1.0-5.2) | 0.63 | 1.77 |
| 158 | KISS15m | -4.14 (1.49E-01) | -7.78E-01 (1.71E-01) | 0.1 (0.1-0.1) | 1.0 (1.0-4.3) | 0.47 | 2.10 |
| 159 | ASASSN-15go | -4.6 (3.51E-01) | -1.12 (3.85E-01) | 0.1 (0.1-0.1) | 1.0 (1.0-6.0) | 0.96 | 1.10 |
| 160 | SN2012bl | -4.27 (2.72E-01) | -6.3E-01 (2.43E-01) | 0.1 (0.1-0.1) | 1.32 (1.0-6.1) | 0.98 | 1.06 |
| 161 | OGLE-2014-SN-019 | -9.78E-02 (4.97E-01) | -4.17E-01 (2.93E-01) | 1.66 (0.1-3.4) | 2.88 (1.0-6.6) | 0.95 | 1.12 |

$^{+}$ MASTER OT J09311953.18+165516.4.

## B. SUPERNOVAE AND THEIR HOST GALAXY

**Table A2**. SNe Ia in our sample with the host names, star-forming activities indicated by the galaxy type, and the CSP (SNooPy) values for $z_{hel}$, $s_{BV}$, $\Delta m_{15}(B)$, are from Uddin et al. (2023). For error bars, see (Uddin et al. 2023). $E(B-V)$, $R_V$ are from Burns (private communication). Values for 91T-like SNe are from Phillips et al. (2022). The last column gives the special type of the object, if any, in our analysis. The numbers in brackets beside the name of the object are their order in Tab. A1.

| Name | Host | Host Type | $z_{hel}$ | $s_{BV}$ | $\Delta\bar{m}_{15}(B)$ | $E(B-V)$ † | $R_V$ Host † | Type |
|---|---|---|---|---|---|---|---|---|
| | | | SNe Ia in spiral galaxies | | | | | |
| SN2004ef (1) | UGC 12158 | S | 0.03096 | 0.852 | 1.353 | 0.175 | 2.131 | |
| SN2004eo (2) | NGC 6928 | S | 0.01569 | 0.836 | 1.389 | 0.124 | 2.582 | |
| SN2004ey (3) | UGC 11816 | S | 0.01577 | 1.10 | 0.954 | 0.052 | 3.476 | |

| Name | Host | Host type | $z_{hel}$ | $s_{BV}$ | $\Delta\bar{m}_{15}(B)$ | $E(B-V)^{\dagger}$ | $R_V$ Host $^{\dagger}$ | Type |
|---|---|---|---|---|---|---|---|---|
| SN2004gu (5) | FGC 175A | S | 0.0458 | 1.149 | 0.853 | 0.197 | 1.857 | 99aa-A |
| SN2005am (7) | NGC 2811 | S | 0.00789 | 0.813 | 1.524 | 0.118 | 2.716 | |
| SN2005A (8) | NGC 958 | S | 0.01912 | 0.963 | 1.115 | 1.096 | 2.065 | |
| SN2005eq (10) | MCG -01-09-006 | S | 0.02895 | 1.152 | 0.835 | 0.135 | 2.809 | 99aa-A |
| SN2005hc (11) | MCG +00-06-003 | S | 0.0459 | 1.113 | 0.884 | 0.092 | 3.866 | |
| SN2005iq (12) | MCG -03-01-008 | S | 0.03402 | 0.924 | 1.244 | 0.041 | 3.894 | |
| SN2005kc (14) | NGC 7311 | S | 0.01511 | 0.932 | 1.222 | 0.33 | 2.564 | |
| SN2005na (18) | UGC3634 | S | 0.0263 | 1.014 | 0.98 | 0.079 | 3.837 | |
| SN2006ax (19) | NGC 3663 | S | 0.0167 | 1.019 | 1.038 | 0.046 | 3.827 | |
| SN2006bh (20) | NGC 7329 | S | 0.01084 | 0.856 | 1.418 | 0.06 | 3.051 | |
| SN2006D (149) | MCG -01-33-34 | S | 0.00852 | 0.831 | 1.414 | 0.15 | 1.82 | |
| SN2006X (150) | NGC 4321 | S | 0.005237 | 0.971 | 1.057$^{\dagger}$ | 1.349 | 1.906 | |
| SN2007af (24) | NGC 5584 | S | 0.00546 | 0.944 | 1.183 | 0.18 | 2.696 | |
| SN2007S (27) | UGC 5378 | S | 0.01387 | 1.173 | 0.833 | 0.468 | 2.395 | |
| SN2011jh (29) | NGC 4682 | S | 0.00778 | 0.789 | 1.462 | 0.456 | 2.55 | |
| SN2012bo (32) | NGC 4726 | S | 0.02541 | 1.154 | 1.154 | 0.104 | 4.015 | |
| SN2012fr (33) | NGC 1365 | S | 0.00545 | 1.009 | 0.802 | 0.074 | 2.329 | |
| SN2012G (34) | IC 0803 NED01 | S | 0.0258 | 1.154 | 0.909 | 0.038 | 3.699 | 99aa-B |
| SN2012hd (35) | IC 1657 | S | 0.01194 | 0.888 | 1.291 | 0.218 | 2.55 | |
| SN2012hr (36) | ESO 121- G 026 | S | 0.00756 | 0.973 | 1.075 | 0.085 | 3.344 | |
| SN2013aa (39) | NGC 5643 | S | 0.00399 | 1.097 | 0.924 | 0.035 | 3.618 | |
| SN2013bz (41) | 2MASX J13265081-1001263 | S | 0.0192 | 1.12 | 0.76 | 0.225 | 2.769 | 91T |
| SN2013E (42) | IC 2532 | S | 0.00941 | 1.129 | 0.872 | 0.151 | 2.644 | |
| SN2013fy (43) | ESO 287- G 040 | S | 0.03085 | 1.17 | 0.904 | 0.089 | 3.55 | |
| SN2013fz (44) | NGC 1578 | S | 0.02059 | 1.016 | 0.927 | 0.099 | 3.058 | |
| SN2013gy (45) | NGC 1418 | S | 0.01401 | 0.911 | 1.234 | 0.084 | 3.562 | |
| SN2013hh (46) | UGC 06483 | S | 0.01298 | 1.19 | 1.036$^{\dagger}$ | 0.706 | 2.658 | |
| SN2013H (47) | ESO 036- G 019 | S | 0.01548 | 1.048 | 0.909 | 0.299 | 2.924 | |
| SN2013M (48) | ESO 325- G 043 | S | 0.03493 | 0.941 | 1.124 | 0.11 | 2.769 | |
| SN2013U (49) | CGCG 008-023 | S | 0.0345 | 1.25 | 1.02 | 0.204 | 3.716 | 91T |
| SN2014ao (50) | NGC 2615 | S | 0.01407 | 0.888 | 1.252 | 0.715 | 2.129 | |
| SN2014at (51) | NGC7119 | S | 0.03222 | 0.95 | 1.112 | 0.052 | 3.891 | |
| SN2014eg (53) | ESO 154- G 010 | S | 0.0186 | 1.17 | 0.92 | 0.316 | 2.2 | 91T |
| SN2015F (55) | NGC 2442 | S | 0.00488 | 0.887 | 1.254 | 0.165 | 2.878 | |
| ASAS14ad (57) | KUG 1237+183 | S | 0.0264 | 1.01 | 0.761 | 0.062 | 3.517 | |
| ASAS14hp (58) | 2MASX J21303015-7038489 | S | 0.03889 | 1.074 | 0.801 | 0.022 | 3.759 | 99aa-B |
| ASAS14hu (60) | ESO 058- G 012 | S | 0.02159 | 1.052 | 0.903 | 0.036 | 3.284 | |
| ASAS14jc (61) | 2MASX J07353554-6246099 | S | 0.01132 | 0.915 | 1.182 | 0.538 | 2.444 | |
| ASAS14jg (62) | 2MASX J23331223-6034201 | S | 0.01482 | 1.285 | 1.003 $^{\dagger}$ | 0.01 | 3.475 | |
| ASAS14kd (63) | 2MASX J22532475+0447583 | S | 0.0243 | 1.13 | 0.79 | 0.292 | 3.15 | 91T |
| ASAS14kq (64) | 2MASX J23451480-2947009 | S | 0.03358 | 1.147 | 0.958 | 0.059 | 3.972 | |
| ASAS14lp (65) | NGC 4666 | S | 0.0051 | 1.029 | 0.845 | 0.351 | 2.224 | |
| ASAS14lw (67) | GALEXASC J010647.95-465904.1 | S | 0.02089 | 1.25 | 0.687 | 0.05 | 3.02 | |

| Name | Host | Host type | $z_{hel}$ | $s_{BV}$ | $\Delta\bar{m}_{15}(B)$ | $E(B-V)^{\dagger}$ | $R_V$ Host $^{\dagger}$ | Type |
|---|---|---|---|---|---|---|---|---|
| ASAS15aj (72) | NGC 3449 | S | 0.01091 | 0.831 | 1.440 | 0.208 | 2.666 | |
| ASAS15al (73) | GALEXASC J045749.46-213526.3 | S | 0.03378 | 1.077 | 0.844 $^{\dagger}$ | 0.124 | 2.482 | |
| ASAS15ba (75) | SDSS J140455.12+085514.0 | S | 0.02312 | 0.967 | 1.062 | 0.086 | 3.167 | |
| ASAS15cd (78) | CGCG 064-017 | S | 0.03429 | 1.003 | 0.937 | 0.046 | 3.971 | |
| ASAS15gr (84) | ESO 366- G 015 | S | 0.02428 | 1.035 | 0.924 | 0.077 | 3.738 | |
| PSN J03055989+0432382 (88) | SDSS J030559.63+043246.0 | S | 0.02818 | 1.129 | 0.944 | 0.069 | 3.462 | |
| MASTER OT J093953.18+165516.4 (90) | CGCG 092-024 | S | 0.0478 | 1.12 | 0.84 | 0.034 | 3.497 | 91T |
| CSS130303:105206-133424 (91) | GALEXASC J105206.27-133420.2 | S | 0.0789 | 1.19 | 1.08 | 0.02 | 3.473 | 91T |
| LSQ11ot (100) | CGCG 421-013 | S | 0.02732 | 0.982 | 1.008 | 0.474 | 3.072 | |
| LSQ12fxd (105) | ESO 487- G 004 | S | 0.03122 | 1.16 | 0.948 | 0.083 | 4.183 | |
| LSQ12gdj (106) | ESO 472- G 007 | S | 0.0303 | 1.14 | 0.73 | 0.029 | 3.459 | 91T |
| LSQ13dpm (112) | GALEXASC J102908.61-170654.2 | S | 0.05086 | 1.054 | 0.871 | 0.107 | 4.777 | |
| LSQ13dsm (113) | APMUKS(BJ) B033105.19-262232.9 | S | 0.04237 | 0.909 | 1.260 | 0.068 | 4.934 | |
| LSQ13lq (114) | SDSS J134410.77+030345.3 | S | 0.07555 | 1.069 | 0.856 | 0.033 | 4.099 | |
| LSQ13vy (116) | 2MASX J16065563+0300046 | S | 0.04177 | 0.891 | 1.270 | 0.175 | 2.88 | |
| LSQ14age (117) | GALEXASC J132408.58-132629.0 | S | 0.08054 | 1.115 | 0.789 | 0.046 | 2.767 | |
| LSQ14ahc (118) | 2MASX J13434760-3254381 | S | 0.05086 | 1.222 | 0.896 | 0.008 | 3.552 | |
| LSQ14foj (124) | GALEXASC J002634.59-324825.5 | S | 0.04607 | 1.012 | 0.889 | 0.153 | 2.093 | |
| LSQ14mc (129) | SDSS J090213.35+170335.4 | S | 0.05662 | 0.963 | 1.103 | 0.062 | 3.803 | |
| LSQ14q (130) | SDSS J085357.19+171942.6 | S | 0.06695 | 0.957 | 1.146 | 0.05 | 4.079 | |
| LSQ14wp (131) | SDSS J101405.83+064032.5 | S | 0.06945 | 1.134 | 0.805 | 0.023 | 3.904 | |
| LSQ15aae (133) | 2MASX J16301506+0555514 | S | 0.05156 | 1.218 | 0.895 | 0.138 | 3.111 | |
| LSQ15agh (134) | 2MASX J10525434+2335518 | S | 0.05996 | 1.134 | 1.002 | 0.055 | 3.628 | |
| LSQ15alq (136) | ESO 508- G 016 | S | 0.04707 | 0.921 | 1.243 | 0.072 | 3.321 | |
| iPTF11pbp (141) | NGC 7674 | S | 0.0289 | 1.15 | 0.94 | 0.152 | 3.516 | |
| iPTF11pra (151) | NGC 881 | S | 0.01754 | 0.439 | | 0.436 | 2.167 | |
| iPTF14aje (152) | SDSS J152512.43-014840.1 | S | 0.02767 | 0.684 | 1.589 $^{\dagger}$ | 0.654 | 2.508 | |
| iPTF14gnl (145) | LCSB S0066P | S | 0.05369 | 1.021 | 0.915 | 0.055 | 4.342 | |
| iPTF14yw (147) | NGC 3861 | S | 0.01697 | 0.856 | 1.299 | 0.011 | 3.409 | |
| iPTF14yy (148) | SDSS J122608.78+095847.1 | S | 0.04297 | 0.837 | 1.368 | 0.356 | 2.898 | |
| SNe Ia in low-mass elliptical galaxies | | | | | | | | |
| OGLE-2014-SN-107 (94) | APMUKS(BJ) B004021.02-650219.5 | E | 0.0664 | 1.19 | 1.12 | 0.13 | 3.548 | 91T |

| Name | Host | Host type | $z_{hel}$ | $s_{BV}$ | $\Delta\bar{m}_{15}(B)$ | $E(B-V)^{\dagger}$ | $R_V$ Host $^{\dagger}$ | Type |
|---|---|---|---|---|---|---|---|---|
| LSQ12fuk (104) | GALEXASC J045815.88-161800.7 | E | 0.02059 | 1.004 | 1.036 | 0.1 | 3.932 | |
| LSQ13dhj (111) | GALEXMSC J021234.60-372019.1 | E (?) | 0.0935 | 1.167 | 0.777 | 0.163 | 2.977 | |
| SNe Ia in S0 galaxies | | | | | | | | |
| SN2004gs (4) | MCG +03-22-020 | E/S0 | 0.02663 | 0.705 | 1.626 | 0.231 | 2.394 | |
| SN2005al (6) | NGC 5304 | E/S0 | 0.01239 | 0.865 | 1.193 | 0.014 | 3.549 | |
| SN2005el (9) | NGC 1819 | S0 | 0.0149 | 0.865 | 1.341 | 0.016 | 3.57 | |
| SN2005ki (16) | NGC 3332 | E/S0 | 0.01919 | 0.839 | 1.246 | 0.037 | 3.225 | |
| SN2005M (17) | NGC 2930 | S0 | 0.022 | 1.204 | 0.871 | 0.077 | 3.235 | 99aa-A |
| SN2006kf (22) | UGC 2829 | S0 | 0.02129 | 0.771 | 1.581 | 0.083 | 3.68 | |
| SN2007ba (25) | UGC 9798 | S0 | 0.03849 | 0.556 | 1.82 | 0.175 | 1.799 | |
| SN2011jn (30) | 2MASX J12571157-1724344 | E/S0 | 0.04744 | 0.641 | 1.572 $^{\dagger}$ | 0.071 | 3.826 | |
| SN2014dn (52) | IC 2060 | E/S0 | 0.02217 | 0.466 | 1.716 $^{\dagger}$ | 0.228 | 1.585 | |
| SN2014I (54) | ESO 487-G36 | S0 | 0.02999 | 0.913 | 1.203 | 0.059 | 3.402 | |
| SNhunt281 (56) | NGC 5839 | S0 | 0.00407 | 0.693 | 1.563 | 0.073 | 3.986 | |
| ASAS14hr (59) | 2MASX J01504127-1431032 | S0 | 0.0336 | 0.802 | 1.409 | 0.129 | 3.45 | |
| ASAS14lt (66) | IC 0299 | S0 | 0.03202 | 0.944 | 0.833 | 0.05 | 3.415 | 99aa-A |
| ASAS14mf (69) | GALEXASC J000454.54-322615.3 | E/S0 | 0.03108 | 0.984 | 1.064 | 0.097 | 4.084 | |
| ASAS14mw (70) | AM 0139-655 NED02 | S0 | 0.02739 | 1.063 | 0.808 | 0.046 | 3.387 | |
| ASAS15da (79) | 2MASX J05235106-2442201 | E/S0 | 0.0487 | 0.853 | 1.122 $^{\dagger}$ | 0.012 | 3.558 | |
| ASAS15eb (82) | ESO 561- G 012 | S0 | 0.01647 | 0.821 | 1.112 $^{\dagger}$ | 0.013 | 3.6 | |
| ASAS15ga (83) | NGC 4866 | S0 | 0.00663 | 0.496 | 2.132 | 0.218 | 2.97 | |
| ASAS15hf (85) | ESO 375- G 041 | S0 | 0.00617 | 0.943 | 1.087 | 0.127 | 4.866 | |
| ASAS15hx (86) | GALEXASC J134316.80-313318.2 | S0 | 0.0083 | 1.039 | 0.931 | 0.063 | 3.301 | |
| SN2014du (95) | UGC 01899 | E/S0(?) | 0.03244 | 0.811 | 1.383 | 0.232 | 1.713 | |
| PSN J13471211-2422171 (98) | ESO 509- G 108 | S0 | 0.01989 | 0.718 | 1.537 | 0.178 | 3.044 | |
| LSQ11pn (101) | 2MASX J05164149+0629376 | S0 | 0.03265 | 0.503 | 2.116 | 0.016 | 3.578 | |
| LSQ12gxj (107) | 2MASX J02525699+0136231 | S0 | 0.0353 | 1.092 | 0.833 | 0.387 | 2.829 | |
| LSQ14ajn (120) | CGCG 068-091 | S0 | 0.02101 | 0.654 | 1.738 | 0.048 | 3.933 | |
| LSQ14asu (121) | 2MASX J11113635-2127597 | S0 | 0.0684 | 0.767 | 1.426 | 0.078 | 4.296 | |
| LSQ14gov (126) | GALEXMSC J040601.67-160139.7 | S0 | 0.08954 | 1.052 | 1.042 | 0.008 | 3.731 | |
| LSQ14jp (128) | 2MASX J12572166-1547411 | S0 | 0.04539 | 0.675 | 1.743 | 0.138 | 3.698 | |
| LSQ14xi (132) | 2MASX J12304088-1346236 | S0 | 0.05074 | 1.142 | 0.911 | 0.324 | 2.797 | |
| PS1-14ra (138) | IC 1044 | S0 | 0.02808 | 0.777 | 1.161 $^{\dagger}$ | 0.112 | 2.931 | |
| PS1-14rx (139) | SDSS J124653.32+144748.4 | S0 * | 0.06695 | 0.846 | 1.153 $^{\dagger}$ | 0.071 | 3.837 | |
| iPTF13ebh (143) | NGC 0890 | S0 | 0.01326 | 0.636 | 1.763 | 0.084 | 3.623 | |
| iPTF14w (146) | UGC 07034 | S0 | 0.01889 | 0.742 | 1.529 | 0.093 | 3.5 | |

| Name | Host | Host type | $z_{hel}$ | $s_{BV}$ | $\Delta\bar{m}_{15}(B)$ | $E(B-V)^{\dagger}$ | $R_V$ Host $^{\dagger}$ | Type |
|---|---|---|---|---|---|---|---|---|
| SNe Ia in the bulge of spiral galaxies | | | | | | | | |
| SN2005ke (15) | NGC 1371 | S | 0.00488 | 0.438 | 1.755 | 0.175 | 1.617 | |
| SN2006ob (23) | UGC 1333 | S | 0.0592 | 0.743 | 1.558 | 0.118 | 2.896 | |
| SN2007bd (26) | UGC 4455 | S | 0.031 | 0.917 | 1.166 | 0.073 | 2.691 | |
| SN2013aj (40) | NGC 5339 | S | 0.00912 | 0.794 | 1.466 | 0.092 | 2.756 | |
| ASAS14me (68) | ESO 113- G 047 | S | 0.0178 | 1.078 | 0.861 | 0.091 | 3.671 | 99aa-B |
| ASAS14my (71) | NGC 3774 | S | 0.0205 | 0.923 | 1.187 | 0.084 | 2.77 | |
| ASAS15bm (77) | LCRS B150313.2-052600 | S | 0.02079 | 0.991 | 0.976 | 0.205 | 2.133 | |
| ASAS15db (80) | NGC 5996 | S | 0.01099 | 0.955 | 1.092 | 0.192 | 2.856 | |
| OGLE-2014-SN-141 (96) | 2MASX J05371898-7543157 | S | 0.0625 | 1.24 | 0.62 | 0.078 | 3.346 | 91T |
| SNe Ia in large elliptical galaxies | | | | | | | | |
| SN2005ir (13) | SDSS J011643.87+004736.9 | E | 0.07631 | 1.03 | 0.883 | 0.105 | 2.718 | |
| SN2011iv (28) | NGC 1404 | E | 0.00649 | 0.699 | 1.744 | 0.073 | 4.307 | |
| SN2012ij (38) | CGCG 097-050 | E | 0.01099 | 0.536 | 1.938 | 0.016 | 3.575 | |
| ASAS15dd (81) | CGCG 107-031 | E | 0.02436 | 0.849 | 1.329 | 0.079 | 4.947 | |
| SN2015bo (97) | NGC 5490 | E | 0.01618 | 0.505 | 1.855 | 0.163 | 2.352 | |
| LSQ12agq (102) | GALEXASC J101741.80-072452.2 | E | 0.06416 | 1.152 | 0.889 † | 0.198 | 2.891 | |
| LSQ12bld (103) | SDSS J134244.72+080531.7 | E | 0.08336 | 0.896 | 1.081 † | 0.184 | 2.112 | |
| LSQ12hzj (108) | 2MASX J09591230-0900095 | E | 0.0334 | 0.969 | 1.047 | 0.03 | 3.375 | 99aa-A |
| LSQ13cwp (109) | 2MASX J04035024-0239275 | E | 0.0666 | 0.944 | 1.188 | 0.146 | 2.349 | |
| LSQ13ry (115) | SDSS J103247.83+041145.5 | E | 0.02984 | 0.881 | 1.245 | 0.019 | 3.589 | |
| LSQ14auy (122) | 2MASX J14281171-0403150 | E | 0.08241 | 1.147 | 0.743 | 0.094 | 4.197 | |
| LSQ14ghv (125) | 2MASX J03234449-3135101 | E | 0.06665 | 0.968 | 1.143 | 0.073 | 4.339 | |
| No host type information | | | | | | | | |
| SN2006gt (21) | 2MASX J00561810-013732 | - | 0.04474 | 0.575 | 1.884 | 0.054 | 3.262 | |
| SN2012ar (31) | 2MASX J16203650-1028061 | - | 0.02824 | 0.807 | 1.392 | 0.091 | 3.813 | |
| SN2012ht (37) | NGC 3447 (GPair) | - | 0.0036 | 0.919 | 1.271 | 0.042 | 3.768 | |
| ASAS15as (74) | SDSS J093916.69+062551.1 | - | 0.02868 | 1.076 | 0.787 | 0.082 | 2.359 | 99aa-B |
| ASAS15be (76) | GALEXASC J025245.83-341850.6 | - | 0.02188 | 1.134 | 0.876 | 0.085 | 2.914 | |
| OGLE-2013-SN-126 (87) | Anonymous | - | 0.05976 | 0.984 | 1.083 | 0.049 | 4.318 | |
| MLS140102:120307-010132 (89) | SDSS J120306.76-010132.4 | Old$^{+}$ | 0.07715 | 1.208 | 0.739 | 0.024 | 3.758 | 03fg |
| OGLE-2014-SN-021 (92) | Anonymous | - | 0.04217 | 0.982 | 1.015 | 0.068 | 3.5 | |
| CSS140914-010107-101840 (93) | Anonymous | - | 0.02998 | 0.96 | 0.998 | 0.007 | 3.559 | |
| LSQ11bk (99) | Anonymous | - | 0.04027 | 1.071 | 0.822 | 0.021 | 3.666 | |
| LSQ13dby (110) | Anonymous | - | 0.09993 | 1.139 | 0.825 | 0.049 | 3.914 | |

| Name | Host | Host type | $z_{hel}$ | $s_{BV}$ | $\Delta\bar{m}_{15}(B)$ | $E(B-V)^{\dagger}$ | $R_V$ Host [†] | Type |
|---|---|---|---|---|---|---|---|---|
| LSQ14ahm (119) | GALEXASC J114122.65-122354.9 | - | 0.04977 | 1.156 | 0.688 | 0.019 | 3.767 | |
| LSQ14fms (123) | 2MASX J00145929-5112380 | - | 0.07795 | 0.848 | 1.277 | 0.25 | 1.782 | |
| LSQ14ie (127) | Anonymous | - | 0.08954 | 1.162 | 0.722 | 0.05 | 3.669 | |
| LSQ15aja (135) | SDSS J170308.90+122741.5 | - | 0.06995 | 1.032 | 0.924 | 0.044 | 4.02 | |
| LSQ15bv (137) | 2MASX J10594717-1649070 | - | 0.0689 | 0.952 | 1.164 [†] | 0.065 | 3.322 | |
| PS15sv (140) | GALEXASC J161311.68+013532.2 | - | 0.03328 | 0.993 | 1.033 | 0.107 | 3.221 | 99aa-B |
| iPTF13anh (142) | SDSS J130650.44+153432.7 | - | 0.06146 | 0.944 | 1.167 | 0.003 | 3.616 | |
| iPTF14fpg (144) | SDSS J002812.09+070940.0 | - | 0.034 | 1.078 | 0.796 | 0.086 | 2.998 | |
| | | | Outliers | | | | | |
| SN2013ao (153) | - | Young | 0.0435 [a] | | 1.00 | 0.09 [a] | | 03fg |
| ASAS15hy (154) | Faint host | Young | 0.0186 | 1.24 [*] | 0.726 | 0.18 [*] | | 03fg |
| SN2012Z (155) | NGC 1309 [b] | S | 0.0071 [b] | | 1.393 | 0.11 [b] | 2.2 | Iax |
| SN2013gr (156) | ESO 114- G7 | S | 0.0074 | | 1.94 [o] | | | Iax |
| SN2014ek (157) | UGC 12850 | S? | 0.023 | | 1.644 | | 0.77 | Iax |
| KISS15m (158) | NGC 4098 | Gpair, S | 0.02432 | 0.425 | 1.729 | 0.101 | 2.134 | 91bg |
| ASAS15go (159) | 2MASX J06113048-1629085 | - | 0.01891 | 1.071 | 0.82 | 0.308 | 2.643 | |
| SN2012bl (160) | ESO 234-019 | S | 0.01869 | 1.08 | 0.779 | 0.108 | 1.927 | |
| OGLE-2014-SN-019 (161) | 2MASX J06134795-6755146 | E | 0.03591 | 0.898 | 1.113 | 0.018 | 3.615 | |

[+] Star-formation near SN (Lu et al. 2021).
[†] Chris Burns (private communication).
[*] (Lu et al. 2021)
[a] (Ashall et al. 2021)
[b] (Stritzinger et al. 2015).
[o] Our $\Delta m_{15}(B)$.

Note: The names of all ASASSN objects are abbreviated in this table as ASAS. For example, ASASSN-14ad is written as ASAS14ad.

## C. COLOR INFORMATION OF THE SN IA AND HOST-GALAXIES

**For orientation and as a consistency check, some additional host-galaxy properties and SNe-colors are briefly discussed.In Tab. A3, previously published data and the 'generic' V-based model fits are shown and compared to observations. For the entire brightness range of SN Ia, the generic synthetic and the intrinsic SNe colors show good agreement and are consistent within the absolute model accuracy of $\approx 0.1...0.2^m$ at $t_V(max)$ (Höflich et al. 1998c) For a detailed discussion of the observations and their uncertainties, see references to the CSP papers in Sect. 1. The spread in $(B-V)$ is consistent with our previous studies [17] (e.g. Figs. 2 and 4 in Hoeflich et al. 2017a).**

[17] which include uncertainties in the rise times and intrinsic diversity to be expected both by different explosion scenarios (see Sect. 2) and variations within each scenario such as details of the ignition and flame propagation (e.g. Domínguez & Höflich 2000).

**In the absence of detailed fits, the discussion of the relation between SN-color and the galactic host will be limited to the effect of the progenitor properties $M_{MS}$ and $\rho_c$. Note that the amount of deflagration burning determines, to first order, the maximum brightness $M_V$, $(B-V)$, and $\Delta m_{15}(V)$.**

**In literature, several empirical relations have been studied between SNe Ia colors and the host galaxies properties (Hashimoto et al. 1995; Sullivan et al. 2006; Childress et al. 2013; Wiseman et al. 2020), particularly for CSP-I SNe Ia (Uddin et al. 2020). Some studies related the SN Ia properties to the morphological type of the host (Wang et al. 1997; Galbany et al. 2012).**

**Our delayed-detonation models show (see also Fig. 3 in Hoeflich et al. 2017a) that a) with decreasing $M_{MS}$ from 7 to 1.5 $M_\odot$, $(B-V)_{max}$ becomes redder by $\approx 0.06^m$, b) brighter in $B$ by $\approx -0.1^m$ , c) $\Delta m_{15}(B)$ increases with otherwise the same model parameters. Taking the shift of the average $M_{MS}$ distribution between spirals and ellipticals (Fig. 18), typical ellipticals should be redder by $\approx 0.04^m$ and brighter by about $\approx -0.05^m$. However, they become even redder but dimmer $\approx 0.2^m$ with increasing production of EC elements, i.e. $\rho_c$ (Fig. 3), as can be expected for low accretion rates, e.g. in old systems. Note, from theory, an additional term to equation 2 for the '$U$' color may be attributed to the metallicity $Z$ and leading to dimmer B with increasing Z (Höflich et al. 1998c). The effect of Z has been studied for the few CSP-I SNe Ia with U colors (Sadler 2012).**

**From observations, the metallicity in ellipticals is increasing with the host mass but with a significant overlap with spirals (Graves & Faber 2010; Li et al. 2018). One may expect that old ellipticals are low $Z$ but they are formed by multiple mergers leading to, intermittently, irregular morphology with episodes of star-formation which may lead to super-solar metallicity in high-mass ellipticals. $Z$ may cause a systematic shift in either direction or a dispersion in $(B-V)$ and $B$. Both $M_{MS}$ and $\rho_c$ provide the trend in color and luminosity decline rate seen in elliptical vs. spirals but $Z$ and mixing may dominate.**

**Tests failed to use host-galaxy properties alternative to the morphology. Using the host mass suffers from the significant overlap in mass between galaxies with and without recent star formation, though a weak indication can be seen for bluer colors in massive galaxies. Alternatively, using the color of the host and passive stellar evolution can probe recent star formation but it is limited to time-scales of $\approx 10-30$ Myrs (Cristallo et al. 2011) which are at the upper end of the stellar mass range for the WD formation. To use statistics, much larger SN samples than 161 objects are needed, and a direct application to the brightness evolution, i.e.differentials, in individual bands may be more stable (see Sect. 8).**

**Table A3**. SNe Ia in our sample with the SN and host names, host masses $M_{host}$ in $10^9 M_\odot$, the reddening $E_{B-V,(host,MW)}$ by the host and the Milky Way $(MW)$, $(B-V)_{CSP}$ observed at maximum, the observed intrinsic, reddening corrected $(B-V)_{o,CSP}$ and synthetic $(B-V)_{gen}$ colors. In the last column, we give the difference between the latter two. Note that the generic color does not include higher-order corrections due to e.g. metallicity, mixing and velocity gradients (see Sect. 3, and text). The numbers in brackets beside the name of the object are their order in Tab. A1.

| Name | Host | $M_{host}$ | $E_{B-V,(host,MW)}$ | $(B-V)_{CSP}$ | $(B-V)_{o,CSP}$ | $(B-V)_{gen}$ | $\Delta(B-V)$ |
|---|---|---|---|---|---|---|---|
| | | | SNe Ia in spiral galaxies | | | | |
| SN2004ef (1) | UGC 12158 | 54.95 | 0.175 , 0.046 | 0.117 | -0.104 | 0.014 | -0.118 |
| SN2004eo (2) | NGC 6928 | 144.54 | 0.124 , 0.093 | 0.025 | -0.192 | -0.004 | -0.188 |
| SN2004ey (3) | UGC 11816 | 11.48 | 0.052 , 0.120 | -0.051 | -0.223 | -0.108 | -0.115 |
| SN2004gu (5) | FGC 175A | 10.47 | 0.197 , 0.022 | 0.145 | -0.074 | -0.007 | -0.067 |
| SN2005am (7) | NGC 2811 | 67.61 | 0.118 , 0.043 | 0.074 | -0.087 | 0.021 | -0.108 |
| SN2005A (8) | NGC 958 | 154.88 | 1.096 , 0.026 | 0.984 | -0.138 | 0.053 | -0.191 |
| SN2005eq (10) | MCG -01-09-006 | 36.31 | 0.135 , 0.063 | 0.035 | -0.163 | -0.057 | -0.106 |
| SN2005hc (11) | MCG +00-06-003 | 39.81 | 0.092 , 0.028 | 0.012 | -0.108 | -0.045 | -0.063 |
| SN2005iq (12) | MCG -03-01-008 | 34.67 | 0.041 , 0.019 | -0.032 | -0.092 | 0.013 | -0.105 |

| Name | Host | $M_{host}$ | $E_{B-V,(host,MW)}$ | $(B-V)_{CSP}$ | $(B-V)_{o,CSP}$ | $(B-V)_{gen}$ | $\Delta(B-V)$ |
|---|---|---|---|---|---|---|---|
| SN2005kc (14) | NGC 7311 | 114.81 | 0.330 , 0.114 | 0.238 | -0.206 | -0.017 | -0.189 |
| SN2005na (18) | UGC3634 | 93.32 | 0.079 , 0.068 | -0.020 | -0.167 | -0.084 | -0.083 |
| SN2006ax (19) | NGC 3663 | 83.17 | 0.046 , 0.041 | -0.072 | -0.159 | 0.011 | -0.170 |
| SN2006bh (20) | NGC 7329 | 26.91 | 0.060 , 0.023 | -0.010 | -0.093 | 0.013 | -0.106 |
| SN2006D (149) | MCG -01-33-34 | 5.75 | 0.150 , 0.039 | 0.098 | -0.091 | 0.057 | -0.148 |
| SN2006X (150) | NGC 4321 | 2.14 | 1.349 , 0.023 | 1.214 | -0.158 | 0.095 | -0.253 |
| SN2007af (24) | NGC 5584 | NA | 0.180 , 0.034 | 0.065 | -0.149 | -0.054 | -0.095 |
| SN2007S (27) | UGC 5378 | 9.55 | 0.468 , 0.022 | 0.366 | -0.124 | -0.118 | -0.006 |
| SN2011jh (29) | NGC 4682 | 12.30 | 0.456 , 0.032 | 0.368 | -0.120 | 0.042 | -0.162 |
| SN2012bo (32) | NGC 4726 | 20.89 | 0.104 , 0.045 | 0.012 | -0.137 | -0.003 | -0.134 |
| SN2012fr (33) | NGC 1365 | NA | 0.074 , 0.018 | 0.013 | -0.079 | -0.139 | 0.060 |
| SN2012G (34) | IC 0803 NED01 | 5.37 | 0.038 , 0.021 | -0.040 | -0.099 | 0.048 | -0.147 |
| SN2012hd (35) | IC 1657 | 27.54 | 0.218 , 0.023 | 0.123 | -0.118 | 0.027 | -0.145 |
| SN2012hr (36) | ESO 121- G 026 | 20.89 | 0.085 , 0.039 | 0.009 | -0.115 | -0.009 | -0.106 |
| SN2013aa (39) | NGC 5643 | 9.77 | 0.035 , 0.146 | -0.078 | -0.259 | -0.042 | -0.217 |
| SN2013bz (41) | 2MASX J13265081-100126 | 9.55 | 0.225 , 0.038 | 0.154 | -0.109 | -0.073 | -0.036 |
| SN2013E (42) | IC 2532 | 9.12 | 0.151 , 0.084 | 0.076 | -0.159 | -0.029 | -0.130 |
| SN2013fy (43) | ESO 287- G 040 | 72.44 | 0.089 , 0.023 | 0.023 | -0.089 | -0.121 | 0.032 |
| SN2013fz (44) | NGC 1578 | 48.98 | 0.099 , 0.012 | 0.005 | -0.106 | -0.111 | 0.005 |
| SN2013gy (45) | NGC 1418 | 1.00 | 0.084 , 0.049 | -0.008 | -0.141 | -0.089 | -0.052 |
| SN2013hh (46) | UGC 06483 | 3.89 | 0.706 , 0.025 | 0.566 | -0.165 | 0.020 | -0.185 |
| SN2013H (47) | ESO 036- G 019 | 31.62 | 0.299 , 0.113 | 0.186 | -0.226 | -0.101 | -0.125 |
| SN2013M (48) | ESO 325- G 043 | 79.43 | 0.110 , 0.072 | 0.023 | -0.159 | -0.120 | -0.039 |
| SN2013U (49) | CGCG 008-023 | 18.62 | 0.204 , 0.025 | 0.168 | -0.061 | -0.090 | 0.029 |
| SN2014ao (50) | NGC 2615 | 37.15 | 0.715 , 0.028 | 0.609 | -0.134 | 0.031 | -0.165 |
| SN2014at (51) | NGC7119 | 177.82 | 0.052 , 0.016 | -0.034 | -0.102 | -0.041 | -0.061 |
| SN2014eg (53) | ESO 154- G 010 | 416.87 | 0.316 , 0.316 | 0.257 | -0.375 | -0.019 | -0.356 |
| SN2015F (55) | NGC 2442 | 1.44 | 0.165 , 0.175 | 0.058 | -0.282 | 0.022 | -0.304 |
| ASAS14ad (57) | KUG 1237+183 | 0.41 | 0.062 , 0.016 | -0.011 | -0.089 | 0.078 | -0.167 |
| ASAS14hp (58) | 2MASX J21303015-703848 | 1.23 | 0.022 , 0.030 | -0.049 | -0.101 | -0.042 | -0.059 |
| ASAS14hu (60) | ESO 058- G 012 | 8.51 | 0.036 , 0.061 | -0.045 | -0.142 | -0.040 | -0.102 |
| ASAS14jc (61) | 2MASX J07353554-624609 | 2.88 | 0.538 , 0.135 | 0.437 | -0.236 | 0.052 | -0.288 |
| ASAS14jg (62) | 2MASX J23331223-603420 | 0.57 | 0.010 , 0.013 | 0.049 | 0.026 | -0.132 | 0.158 |
| ASAS14kd (63) | 2MASX J22532475+044758 | 13.80 | 0.292 , 0.292 | 0.234 | -0.350 | -0.061 | -0.289 |
| ASAS14kq (64) | 2MASX J23451480-294700 | 0.81 | 0.059 , 0.016 | -0.021 | -0.096 | 0.231 | -0.327 |
| ASAS14lp (65) | NGC 4666 | 54.95 | 0.351 , 0.021 | 0.243 | -0.129 | -0.132 | 0.003 |
| ASAS14lw (67) | GALEXASC J010647.95-46 | .0022 | 0.050 , 0.018 | -0.012 | -0.080 | -0.005 | -0.075 |
| ASAS15aj (72) | NGC 3449 | 158.49 | 0.208 , 0.066 | 0.103 | -0.171 | 0.013 | -0.184 |
| ASAS15al (73) | GALEXASC J045749.46-21 | 0.12 | 0.124 , 0.029 | 0.055 | -0.098 | 0.106 | -0.204 |

| Name | Host | $M_{host}$ | $E_{B-V,(host,MW)}$ | $(B-V)_{CSP}$ | $(B-V)_{o,CSP}$ | $(B-V)_{gen}$ | $\Delta(B-V)$ |
|---|---|---|---|---|---|---|---|
| ASAS15ba (75) | SDSS J140455.12+085514 | 0.10 | 0.086 , 0.022 | 0.008 | -0.100 | -0.112 | 0.012 |
| ASAS15cd (78) | CGCG 064-017 | 8.91 | 0.046 , 0.026 | -0.021 | -0.093 | -0.064 | -0.029 |
| ASAS15gr (84) | ESO 366- G 015 | 0.87 | 0.077 , 0.082 | -0.036 | -0.195 | -0.118 | -0.077 |
| PSN J03055989 +0432382 (88) | SDSS J030559.63+043246 | NA | 0.069 , 0.149 | -0.006 | -0.224 | 0.029 | -0.253 |
| MASTER OT J093953.18+165516.4 (90) | CGCG 092-024 | 11.74 | 0.034 , 0.024 | -0.004 | -0.062 | -0.087 | 0.025 |
| CSS130303:105206-133424 (91) | GALEXASC J105206.27-13 | 0.22 | 0.020 , 0.048 | -0.023 | -0.091 | -0.136 | 0.045 |
| LSQ11ot (100) | CGCG 421-013 | 14.79 | 0.474 , 0.158 | 0.320 | -0.312 | 0.014 | -0.326 |
| LSQ12fxd (105) | ESO 487- G 004 | 19.05 | 0.083 , 0.022 | -0.006 | -0.111 | 0.025 | -0.136 |
| LSQ12gdj (106) | ESO 472- G 007 | 0.46 | 0.029 , 0.029 | -0.031 | -0.089 | 0.008 | -0.097 |
| LSQ13dpm (112) | GALEXASC J102908.61-17 | 1.48 | 0.107 , 0.053 | -0.010 | -0.170 | 0.013 | -0.183 |
| LSQ13dsm (113) | APMUKS(BJ) B033105.19- | 1.02 | 0.068 , 0.009 | -0.016 | -0.093 | -0.059 | -0.034 |
| LSQ13lq (114) | SDSS J134410.77+030345 | 0.28 | 0.033 , 0.022 | -0.071 | -0.126 | 0.245 | -0.371 |
| LSQ13vy (116) | 2MASX J16065563+030004 | 14.79 | 0.175 , 0.068 | 0.076 | -0.167 | -0.076 | -0.091 |
| LSQ14age (117) | GALEXASC J132408.58-13 | 0.26 | 0.046 , 0.058 | -0.009 | -0.113 | 0.135 | -0.248 |
| LSQ14ahc (118) | 2MASX J13434760-325438 | 1.34 | 0.008 , 0.046 | -0.029 | -0.083 | 0.041 | -0.124 |
| LSQ14foj (124) | GALEXASC J002634.59-32 | 0.28 | 0.153 , 0.012 | 0.050 | -0.115 | -0.044 | -0.071 |
| LSQ14mc (129) | SDSS J090213.35+170335 | 1.48 | 0.062 , 0.021 | -0.027 | -0.110 | -0.104 | -0.006 |
| LSQ14q (130) | SDSS J085357.19+171942 | 2.09 | 0.050 , 0.020 | -0.046 | -0.116 | -0.008 | -0.108 |
| LSQ14wp (131) | SDSS J101405.83+064032 | 0.03 | 0.023 , 0.021 | -0.037 | -0.081 | 0.0385 | -0.119 |
| LSQ15aae (133) | 2MASX J16301506+055551 | 18.62 | 0.138 , 0.062 | 0.077 | -0.123 | 0.043 | -0.166 |
| LSQ15agh (134) | 2MASX J10525434+233551 | 7.41 | 0.055 , 0.022 | -0.032 | -0.109 | 0.100 | -0.209 |
| LSQ15alq (136) | ESO 508- G 016 | 25.11 | 0.072 , 0.067 | -0.002 | -0.141 | -0.012 | -0.129 |
| iPTF11pbp (141) | NGC 7674 | 75.85 | 0.152 , 0.051 | 0.055 | -0.148 | -0.125 | -0.023 |
| iPTF11pra (151) | NGC 881 | 47.86 | 0.436 , 0.024 | 0.920 | 0.460 | 0.495 | -0.035 |
| iPTF14aje (152) | SDSS J152512.43-014840 | 60.25 | 0.654 , 0.152 | 0.608 | -0.198 | 0.118 | -0.316 |
| iPTF14gnl (145) | LCSB S0066P | 11.48 | 0.055 , 0.027 | -0.055 | -0.137 | 0.058 | -0.195 |
| iPTF14yw (147) | NGC 3861 | 17.37 | 0.011 , 0.026 | -0.038 | -0.075 | 0.059 | -0.134 |
| iPTF14yy (148) | SDSS J122608.78+095847 | 2.57 | 0.356 , 0.020 | 0.256 | -0.120 | -0.057 | -0.063 |
| SNe Ia in low-mass elliptical galaxies | | | | | | | |
| OGLE-2014-SN-107 (94) | APMUKS(BJ) B004021.02- | 0.15 | 0.130 , 0.017 | 0.122 | -0.025 | -0.092 | 0.067 |

| Name | Host | $M_{host}$ | $E_{B-V,(host,MW)}$ | $(B-V)_{CSP}$ | $(B-V)_{o,CSP}$ | $(B-V)_{gen}$ | $\Delta(B-V)$ |
|---|---|---|---|---|---|---|---|
| LSQ12fuk (104) | GALEXASC J045815.88-16 | 0.22 | 0.100 , 0.071 | -0.008 | -0.179 | -0.025 | -0.154 |
| LSQ13dhj (111) | GALEXMSC J021234.60-37 | 0.42 | 0.163 , 0.013 | 0.060 | -0.116 | 0.092 | -0.208 |
| | | | SNe Ia in S0 galaxies | | | | |
| SN2004gs (4) | MCG +03-22-020 | 42.65 | 0.231 , 0.026 | 0.202 | -0.055 | 0.117 | -0.172 |
| SN2005al (6) | NGC 5304 | 30.90 | 0.014 , 0.048 | -0.051 | -0.113 | 0.060 | -0.173 |
| SN2005el (9) | NGC 1819 | 69.18 | 0.016 , 0.098 | -0.073 | -0.187 | -0.001 | -0.186 |
| SN2005ki (16) | NGC 3332 | 63.09 | 0.037 , 0.027 | -0.018 | -0.082 | 0.104 | -0.186 |
| SN2005M (17) | NGC 2930 | 0.63 | 0.077 , 0.027 | 0.018 | -0.086 | 0.084 | -0.170 |
| SN2006kf (22) | UGC 2829 | 66.07 | 0.083 , 0.210 | 0.026 | -0.267 | 0.081 | -0.348 |
| SN2007ba (25) | UGC 9798 | 97.72 | 0.175 , 0.032 | 0.328 | 0.121 | 0.062 | 0.059 |
| SN2011jn (30) | 2MASX J12571157-172434 | 223.87 | 0.071 , 0.059 | 0.121 | -0.009 | 1.228 | -1.237 |
| SN2014dn (52) | IC 2060 | 44.66 | 0.228 , 0.016 | 0.653 | 0.409 | 0.344 | 0.065 |
| SN2014I (54) | ESO 487-G36 | 79.43 | 0.059 , 0.021 | 0.012 | -0.068 | -0.027 | -0.041 |
| SNhunt281 (56) | NGC 5839 | 5.62 | 0.073 , 0.045 | 0.045 | -0.073 | 0.043 | -0.116 |
| ASAS14hr (59) | 2MASX J01504127-143103 | 21.87 | 0.129 , 0.013 | 0.061 | -0.081 | 0.104 | -0.185 |
| ASAS14lt (66) | IC 0299 | 43.65 | 0.050 , 0.047 | -0.033 | -0.130 | -0.049 | -0.081 |
| ASAS14mf (69) | GALEXASC J000454.54-32 | 0.85 | 0.097 , 0.013 | -0.012 | -0.122 | 0.029 | -0.151 |
| ASAS14mw (70) | AM 0139-655 NED02 | 47.86 | 0.046 , 0.018 | -0.034 | -0.098 | 0.041 | -0.139 |
| ASAS15da (79) | 2MASX J05235106-244220 | 25.11 | 0.012 , 0.030 | -0.132 | -0.174 | 0.093 | -0.267 |
| ASAS15eb (82) | ESO 561- G 012 | 91.20 | 0.013 , 0.165 | -0.129 | -0.307 | 0.091 | -0.398 |
| ASAS15ga (83) | NGC 4866 | 23.98 | 0.218 , 0.024 | 0.431 | 0.189 | NA | NA |
| ASAS15hf (85) | ESO 375- G 041 | 3.63 | 0.127 , 0.085 | 0.009 | -0.203 | -0.108 | -0.095 |
| ASAS15hx (86) | GALEXASC J134316.80-31 | 0.03 | 0.063 , 0.042 | -0.030 | -0.135 | -0.026 | -0.109 |
| SN2014du (95) | UGC 01899 | 64.56 | 0.232 , 0.095 | 0.232 | -0.095 | 0.034 | -0.129 |
| PSN J13471211-2422171 (98) | ESO 509- G 108 | 89.12 | 0.178 , 0.064 | 0.140 | -0.102 | 0.015 | -0.117 |
| LSQ11pn (101) | 2MASX J05164149+062937 | 50.11 | 0.016 , 0.146 | 0.287 | 0.125 | 0.258 | -0.133 |
| LSQ12gxj (107) | 2MASX J02525699+013623 | 1.51 | 0.387 , 0.057 | 0.264 | -0.180 | -0.035 | -0.145 |
| LSQ14ajn (120) | CGCG 068-091 | 12.88 | 0.048 , 0.019 | 0.075 | 0.008 | 0.033 | -0.025 |
| LSQ14asu (121) | 2MASX J11113635-212759 | 46.77 | 0.078 , 0.037 | 0.004 | -0.111 | -0.007 | -0.104 |
| LSQ14gov (126) | GALEXMSC J040601.67-16 | 1.05 | 0.008 , 0.036 | -0.104 | -0.148 | 0.041 | -0.189 |
| LSQ14jp (128) | 2MASX J12572166-154741 | 34.67 | 0.138 , 0.053 | 0.123 | -0.068 | 0.097 | -0.165 |
| LSQ14xi (132) | 2MASX J12304088-134623 | 79.43 | 0.324 , 0.037 | 0.218 | -0.143 | -0.073 | -0.070 |

| Name | Host | $M_{host}$ | $E_{B-V,(host,MW)}$ | $(B-V)_{CSP}$ | $(B-V)_{o,CSP}$ | $(B-V)_{gen}$ | $\Delta(B-V)$ |
|---|---|---|---|---|---|---|---|
| PS1-14ra (138) | IC 1044 | 44.66 | 0.112 , 0.023 | 0.060 | -0.075 | 0.032 | -0.107 |
| PS1-14rx (139) | SDSS J124653.32+144748 | 7.76 | 0.071 , 0.027 | -0.022 | -0.120 | 0.049 | -0.169 |
| iPTF13ebh (143) | NGC 0890 | 72.44 | 0.084 , 0.067 | 0.123 | -0.028 | 0.086 | -0.114 |
| iPTF14w (146) | UGC 07034 | 1.38 | 0.093 , 0.021 | 0.056 | -0.058 | 0.073 | -0.131 |
| SNe Ia in the bulge of spiral galaxies | | | | | | | |
| SN2005ke (15) | NGC 1371 | 26.91 | 0.175 , 0.020 | 0.670 | 0.475 | 0.110 | 0.365 |
| SN2006ob (23) | UGC 1333 | 181.97 | 0.118 , 0.029 | 0.111 | -0.036 | 0.083 | -0.119 |
| SN2007bd (26) | UGC 4455 | 66.07 | 0.073 , 0.029 | -0.001 | -0.103 | -0.036 | -0.067 |
| SN2013aj (40) | NGC 5339 | 9.77 | 0.092 , 0.032 | 0.052 | -0.072 | 0.061 | -0.133 |
| ASAS14me (68) | ESO 113- G 047 | 1.12 | 0.091 , 0.018 | 0.007 | -0.102 | -0.041 | -0.061 |
| ASAS14my (71) | NGC 3774 | 20.41 | 0.084 , 0.032 | 0.015 | -0.101 | -0.106 | 0.005 |
| ASAS15bm (77) | LCRS B150313.2-052600 | 10.47 | 0.205 , 0.069 | 0.127 | -0.147 | -0.101 | -0.046 |
| ASAS15db (80) | NGC 5996 | 12.88 | 0.192 , 0.029 | 0.086 | -0.135 | -0.024 | -0.111 |
| OGLE-2014-SN-141 (96) | 2MASX J05371898-754315 | 30.90 | 0.078 , 0.077 | 0.035 | -0.120 | -0.011 | -0.109 |
| SNe Ia in large elliptical galaxies | | | | | | | |
| SN2005ir (13) | SDSS J011643.87+004736 | 1.69 | 0.105 , 0.027 | 0.050 | -0.082 | 0.039 | -0.121 |
| SN2011iv (28) | NGC 1404 | NA | 0.073 , 0.010 | 0.277 | 0.194 | 0.030 | 0.164 |
| SN2012ij (38) | CGCG 097-050 | 8.71 | 0.016 , 0.024 | 0.257 | 0.217 | -0.004 | 0.221 |
| ASAS15dd (81) | CGCG 107-031 | 25.70 | 0.079 , 0.046 | -0.007 | -0.132 | 0.032 | -0.164 |
| SN2015bo (97) | NGC 5490 | 138.04 | 0.163 , 0.023 | 0.454 | 0.268 | 0.305 | -0.037 |
| LSQ12agq (102) | GALEXASC J101741.80-07 | 1.99 | 0.198 , 0.037 | 0.089 | -0.146 | -0.029 | -0.117 |
| LSQ12bld (103) | SDSS J134244.72+080531 | 35.48 | 0.184 , 0.023 | 0.078 | -0.129 | -0.020 | -0.109 |
| LSQ12hzj (108) | 2MASX J09591230-090009 | 2.23 | 0.030 , 0.057 | -0.069 | -0.156 | -0.040 | -0.116 |
| LSQ13cwp (109) | 2MASX J04035024-023927 | 34.67 | 0.146 , 0.133 | 0.088 | -0.191 | -0.084 | -0.107 |
| LSQ13ry (115) | SDSS J103247.83+041145 | 4.36 | 0.019 , 0.040 | -0.048 | -0.107 | 0.039 | -0.146 |
| LSQ14auy (122) | 2MASX J14281171-040315 | 54.95 | 0.094 , 0.067 | -0.012 | -0.173 | -0.008 | -0.165 |
| LSQ14ghv (125) | 2MASX J03234449-313510 | 15.13 | 0.073 , 0.009 | -0.016 | -0.098 | -0.087 | -0.011 |
| No host type information | | | | | | | |
| SN2006gt (21) | 2MASX J00561810-013732 | 8.51 | 0.054 , 0.032 | 0.249 | 0.163 | 0.214 | -0.051 |
| SN2012ar (31) | 2MASX J16203650-102806 | 87.09 | 0.091 , 0.202 | 0.033 | -0.260 | 0.029 | -0.289 |
| SN2012ht (37) | NGC 3447 (GPair) | 0.01 | 0.042 , 0.025 | -0.016 | -0.083 | 0.008 | -0.091 |

| Name | Host | $M_{host}$ | $E_{B-V,(host,MW)}$ | $(B-V)_{CSP}$ | $(B-V)_{o,CSP}$ | $(B-V)_{gen}$ | $\Delta(B-V)$ |
|---|---|---|---|---|---|---|---|
| ASAS15as (74) | SDSS J093916.69+062551 | 0.04 | 0.082 , 0.040 | 0.029 | -0.093 | 0.190 | -0.283 |
| ASAS15be (76) | GALEXASC J025245.83-34 | 0.06 | 0.085 , 0.017 | 0.019 | -0.083 | -0.077 | -0.006 |
| OGLE-2013-SN-126 (87) | Anonymous | NA | 0.049 , 0.029 | -0.055 | -0.133 | 0.131 | -0.264 |
| MLS140102:120307-010132 (89) | SDSS J120306.76-010132 | NA | 0.024 , 0.021 | -0.061 | -0.106 | -0.133 | 0.027 |
| OGLE-2014-SN-021 (92) | Anonymous | 1.17 | 0.068 , 0.093 | -0.021 | -0.182 | -0.073 | -0.109 |
| CSS140914-010107-101840 (93) | Anonymous | NA | 0.007 , 0.026 | -0.071 | -0.104 | -0.061 | -0.043 |
| LSQ11bk (99) | Anonymous | 0.00 | 0.021 , 0.094 | -0.078 | -0.193 | 0.002 | -0.195 |
| LSQ13dby (110) | Anonymous | NA | 0.049 , 0.009 | -0.040 | -0.098 | -0.144 | 0.046 |
| LSQ14ahm (119) | GALEXASC J114122.65-12 | 0.16 | 0.019 , 0.027 | -0.051 | -0.097 | 0.027 | 0.124 |
| LSQ14fms (123) | 2MASX J00145929-511238 | 12.88 | 0.250 , 0.014 | 0.171 | -0.093 | -0.008 | -0.085 |
| LSQ14ie (127) | Anonymous | 0.12 | 0.050 , 0.072 | -0.010 | -0.132 | 0.0273 | -0.159 |
| LSQ15aja (135) | SDSS J170308.90+122741 | 0.03 | 0.044 , 0.055 | -0.087 | -0.186 | 0.020 | -0.206 |
| LSQ15bv (137) | 2MASX J10594717-164907 | 36.30 | 0.065 , 0.044 | -0.001 | -0.110 | -0.100 | -0.010 |
| PS15sv (140) | GALEXASC J161311.68+01 | 0.20 | 0.107 , 0.078 | 0.011 | -0.174 | -0.073 | -0.101 |
| iPTF13anh (142) | SDSS J130650.44+153432 | 0.03 | 0.003 , 0.022 | -0.170 | -0.195 | 0.018 | -0.213 |
| iPTF14fpg (144) | SDSS J002812.09+070940 | NA | 0.086 , 0.026 | 0.010 | -0.102 | 0.090 | -0.192 |
| Outliers | | | | | | | |
| SN2013ao (153) | NA | NA | NA , 0.034 | NA | NA | NA | NA |
| ASAS15hy (154) | Faint host | NA | NA , 0.13 | NA | NA | NA | NA |
| SN2012Z (155) | NGC 1309 [b] | NA | NA , 0.034 | NA | NA | NA | NA |
| SN2013gr (156) | ESO 114- G7 | NA | NA , NA | NA | NA | NA | NA |
| SN2014ek (157) | UGC 12850 | NA | NA , NA | NA | NA | NA | NA |
| KISS15m (158) | NGC 4098 | 6.02 | 0.101 , 0.028 | 0.622 | 0.420 | NA | NA |
| ASAS15go (159) | 2MASX J06113048-162908 | 15.13 | 0.308 , 0.137 | 0.224 | -0.392 | NA | NA |
| SN2012bl (160) | ESO 234-019 | 12.30 | 0.108 , 0.03 | 0.044 | -0.172 | NA | NA |
| OGLE-2014-SN-019 (161 ) | 2MASX J06134795-675514 | 120.22 | 0.018 , 0.052 | -0.090 | -0.126 | -0.022 | -0.104 |